\documentclass[reqno, a4paper]{amsart} 
\usepackage{latexsym}
\usepackage[english]{babel}
\usepackage{fancyhdr}
\usepackage[mathscr]{eucal}
\usepackage{amsmath}
\usepackage{mathrsfs}
\usepackage{amsthm}
\usepackage{amsfonts}
\usepackage{amssymb}
\usepackage{amscd}
\usepackage{bbm}
\usepackage{graphicx}
\usepackage{graphics}
\usepackage{latexsym}
\usepackage{color}
\usepackage[dvipsnames]{xcolor}

\usepackage{ascmac}

\theoremstyle{plain}
\newtheorem{theorem}{Theorem}[section]
\newtheorem{remark}[theorem]{Remark}
\newtheorem{lemma}[theorem]{Lemma}
\newtheorem{corollary}[theorem]{Corollary}
\newtheorem{proposition}[theorem]{Proposition}

\theoremstyle{definition}
\newtheorem{definition}[theorem]{Definition}

\newcommand {\absleq} {{\leq_{|\, \cdot\, |}\, }}

\def\ta {{\tilde{a}}}
\def\Kg {{\mathcal L}}
\def\Lg {{\mathcal L}}

\def\Xg {{\mathcal X}}
\def\Yg {{\mathcal Y}}

\def\tL {{\tilde{L}}}

\numberwithin{equation}{section}
\def\HL {{L^2}}
\def\Lam{{\Lambda}}
\def\tchi{{\tilde{\chi}}}

\def\ben{\begin{enumerate}}
\def\een{\end{enumerate}}
\def\bgdf{\begin{definition}}
\def\eddf{\end{definition}}
\def\bglm{\begin{lemma}}
\def\edlm{\end{lemma}}
\def\bgpf{\begin{proof}}
\def\edpf{\end{proof}}
\def\bgth{\begin{theorem}}
\def\edth{\end{theorem}}
\def\bgcor{\begin{corollary}}
\def\edcor{\end{corollary}}
\def\bgprop{\begin{proposition}}
\def\edprop{\end{proposition}}
\def\bgrm{\begin{remark}}
\def\edrm{\end{remark}}

\def\lbeq(#1){\label{eqn:#1}}
\def\refeq(#1){{\rm (\ref{eqn:#1})}}
\def\refeqs(#1,#2){{\rm (\ref{eqn:#1}) and (\ref{eqn:#2})}}
\def\refeqss(#1,#2,#3){{\rm (\ref{eqn:#1}),\ (\ref{eqn:#2}) and (\ref{eqn:#3})}}
\def\refeqsss(#1,#2,#3,#4){{\rm (\ref{eqn:#1}),\ (\ref{eqn:#2}),\ 
(\ref{eqn:#3}) and (\ref{eqn:#4})}}
\def\lbth(#1){\label{th:#1}}
\def\refth(#1){{\rm Theorem \ref{th:#1}}}
\def\refths(#1,#2){{\rm Theorems \ref{th:#1} and \ref{th:#2}}}
\def\refthb(#1){{\bf Theorem \ref{th:#1}}}
\def\lblm(#1){\label{lm:#1}}
\def\reflm(#1){{\rm Lemma \ref{lm:#1}}}
\def\reflms(#1,#2){{\rm Lemmas \ref{lm:#1} and \ref{lm:#2}}}
\def\reflmss(#1,#2,#3){{\rm Lemmas \ref{lm:#1}, \ref{lm:#2} and \ref{lm:#3}}}
\def\reflmsss(#1,#2,#3,#4){{\rm Lemmas \ref{lm:#1},\, \ref{lm:#2},\, \ref{lm:#3} and \ref{lm:#4}}}
\def\reflmb(#1){{\bf Lemma \ref{lm:#1}}}
\def\lbprop(#1){\label{prp:#1}}
\def\refprop(#1){{\rm Proposition \ref{prp:#1}}}
\def\refprops(#1,#2,#3,#4){{\rm Propositions \ref{prp:#1},\, \ref{prp:#2},
\, \ref{prp:#3} \, and \ref{prp:#4}}}
\def\refpropb(#1){{\bf Proposition \ref{prp:#1}.}}
\def\lbcor(#1){\label{cor:#1}}
\def\refcor(#1){{\rm Corollary \ref{cor:#1}}}
\def\refcors(#1,#2){{\rm Corollaries \ref{cor:#1} and \ref{cor:#2}}}
\def\lbrm(#1){\label{rm:#1}}
\def\refrm(#1){{\rm Remark \ref{rm:#1}}}
\def\lbass(#1){\label{ass:#1}}
\def\refass(#1){{\rm Assumption \ref{ass:#1}}}
\def\lbdf(#1){\label{df:#1}}
\def\refdf(#1){{\rm Definition \ref{df:#1}}}
\def\refdfs(#1,#2){{\rm Definitions \ref{def:#1} and \ref{def:#2}}}
\def\lbsec(#1){\label{s:#1}}
\def\refsec(#1){{\rm \S\ref{s:#1}}}
\def\lbsubsec(#1){\label{ss:#1}}
\def\refsubsec(#1){{\rm \S\ref{ss:#1}}}

\def\Ag{{\mathcal A}}

\def\Dg{{\mathcal D}}
\def\Fg{{\mathcal F}}
\def\Gg{{\mathcal G}}
\def\Mg{{\mathcal M}}
\def\Ng{{\mathcal N}}

\def\Og{{\mathcal O}}

\newcommand{\lam}{\lambda}

\def\Bb{{\bf B}}

\def\ph{{\varphi}}
\def\tph{{\tilde{\varphi}}}
\def\bqn{\begin{equation}}
\def\eqn{\end{equation}}
\def\C{{\mathbb C}}

 \def\Cb{{\overline{\mathbb C}}}

\def\R{{\mathbb R}}
\def\Rg {{\mathcal R}}
\def\a{\alpha}
\def\b{\beta}
\def\c{\gamma}
\def\Ga{\Gamma}
\def\d{\delta}
\def\Dg{{\mathcal D}}
\def\Eg{{\mathcal E}}
\def\Sg{{\mathcal S}}

\def\p{\psi}
\def\tp{\tilde{\psi}}

\def\ep{\varepsilon}

\def\th{\theta}

\def\m{\mu}
\def\n{\nu}
\def\r{\rho}
\def\s{\sigma}
\def\t{\tau}
\def\w{\omega}
\def\W{\Omega}
\def\Hg {{\mathcal H}}

\def\la{\langle}
\def\ra{\rangle}

\def\lap{\Delta}
\def\ax{{\la x \ra}}
\def\ay{{\la y \ra}}
\def\az{{\la z \ra}}
\def\pa{{\partial}}

\def\br{\begin{array}}
\def\er{\end{array}}

\def\Ker{\rm Ker\,}
\def\rank{{\rm rank\,}}

\begin{document}

\title[$L^p$-boundedness of wave operators in $\R^4$]
{The $L^p$-boundedness of wave operators for four dimensional 
Schr\"odinger operators with threshold resonances}

\author{Kenji \textsc{Yajima}}
\address{Department of Mathematics \\ Gakushuin University 
\\ 1-5-1 Mejiro \\ Toshima-ku \\ Tokyo 171-8588 (Japan)}
\footnote{Supported by JSPS grant in aid for scientific research No. 19K03589}


\subjclass[2020]{Primary 47A40; Secondary 81Q10}
\keywords{$L^p$-boundedness, Wave operators, Schr\"odinger operator}




\begin{abstract} 
We prove that the low energy parts of the wave operators $W_\pm$ for 
Schr\"odinger operators $H = -\lap + V(x)$ on $\R^4$ are bounded in 
$ L^p(\R^4)$ for $1<p\leq 2$ and are unbounded for $2<p\leq \infty$ if 
$H$ has resonances at the threshold. 
If $H$ has eigenfunctions only at the threshold, it has recently been 
proved that they are bounded in $L^p(\R^4)$ for $1\leq p<4$ in general and 
for $1\leq p<\infty$ if all threshold eigenfunctions $\ph$ satisfy 
$\int_{\R^4}x_j V(x) \ph(x)dx=0$ for $1\leq j\leq 4$. 
We prove in this case that they are unbounded in $L^p(\R^4)$ for $4<p<\infty$ 
unless the latter condition is satisfied. 
It is long known that 
the high energy parts are bounded in $L^p(\R^4)$ for all 
$1\leq p\leq \infty$ and that the same holds for $W_\pm$ 
if $H$ has no eigenfunctions nor resonances at the threshold. 
\end{abstract}

\maketitle


\section{Introduction and theorem}
Consider Schr\"odinger operators $H = -\lap + V$ on $\R^d$, 
$d=1,2, \dots$ with real potentials $V(x)$ which satisfy 
$V(x)=O(|x|^{-\d})$, $\d>2$ near infinity in a suitable norm. 
Then, $H$ is selfadjoint in $L^2(\R^d)$ and its spectrum consists 
of finite number of non-positive eigenvalues 
and absolutely continuous (AC for short) part $[0,\infty)$.  
Let $H_0 = -\lap $ be the free 
Schr\"odinger operator. Then, wave operators $W_{\pm}$ 
defined by the strong limits in $L^2(\R^d)$ 
\[
W_{\pm}= \lim_{t \to \pm\infty} e^{itH} e^{-itH_0}
\]
exist and are complete in the sense ${\textrm{Image}}\, W_{\pm} = L^2_{ac}(H)$, 
the AC subspace of $L^2(\R^d)$ for $H$ (\cite{Ag, Ku}). 
Let $P_{ac}(H)$ be the projection onto $L^2_{ac}(H)$. Wave operators  
satisfy the intertwining property: For Borel functions $f$ on $\R$
\bqn \lbeq(inter) 
f(H)P_{ac}(H) = W_\pm f(H_0) W_{\pm}^\ast, 
\eqn 
and, more generally, 
for Fourier and generalized Fourier multiplies $F(D)= \Fg^\ast M_F \Fg$ 
and $F(D_\pm)= \Fg_\pm^\ast M_F \Fg_\pm$, $M_F$ being multiplications with 
Borel functions $F(\xi)$ on $\R^d$, 
\bqn \lbeq(inter-1)
F(D_\pm)P_{ac}(H) = W_\pm F(D) W_\pm^\ast,  
\eqn 
where $\Fg$ and $\Fg_\pm$ are the Fourier and generalized 
Fourier transforms defined by   
\[
\Fg u(\xi) = \frac1{(2\pi)^{d/2}}
\int_{\R^d} e^{-ix\xi}u(x)dx,  \quad 
\Fg_\pm u(\xi) = \frac1{(2\pi)^{d/2}}
\int_{\R^d} \overline{\ph_{\pm}(x,\xi)}u(x)dx
\]
by using scattering eigenfunctions $\ph_{\pm}(\cdot,\xi)$ of $H$ with eigenvalues 
$|\xi|^2$ (see \cite{Ya}, \S 1). 
It follows that, if $W_\pm$ are bounded in $L^p(\R^d)$, then  
$L^p$-mapping properties of $f(H)P_{ac}(H)$ and $F(D_\pm)P_{ac}(H)$ 
may be deduced from the corresponding ones 
of $f(H_0)$ and $F(D)$ respectively.
 
Thus the problem of $L^p$-boundedness of wave operators has attracted  
many authors' interest and many results have been obtained under various 
assumptions on $V$: If $H$ has no zero energy 
eigenfunctions nor resonances (see the remark 
below \refdf(GT-1) for the definition), 
$W_\pm$ are bounded in $L^p(\R^d)$ for all 
$1\leq p \leq \infty$ when $d \geq 3$ and for $1<p<\infty$ in general 
when $d=1,2$ (the cases $p=1, \infty$ for $d=2$ are unknown). 
When $H$ has zero energy eigenfunctions or resonances, the range of $p$ for which 
$W_\pm$ are bounded in $L^p(\R^d)$ generally shrinks and the results depend on 
$d$ and the decay properties of zero energy eigenfunctions 
(see the introduction of \cite{GG,Ya-2dim} and the referenece therein for 
more details). These results are from the perspective of the author 
rather satisfactory when $d\not =4$ except those for end points $p=1, \infty$ 
and for $p$'s where $W_\pm$ change the $L^p$ properties and 
for sharper conditions on $V$ (see e.g. recent paper \cite{Weder} for 
the end point results for $d=1$).

In this paper we are concerned with $W_\pm$ on $\R^4$. Let  
$\chi_{\leq{a}}\in C_0^\infty(\R)$ and 
$\chi_{\geq{a}}\in C^\infty(\R)$ be such that 
\[
\chi_{\leq{a}}(\lam)=\left\{\br{ll} 1, \ \ & |\lam|\leq a \\[2pt] 
0, \ \ & |\lam|\geq 2a \er \right.\,, \quad 
\chi_{\geq{a}}(\lam)= 1 - \chi_{\leq{a}}(\lam)
\]
and define the high and the low energy parts of $W_{\pm}$ by 
$W_{\pm, \geq a}= W_{\pm}\chi_{\geq{a}}(|D|)$ and  
$W_{\pm, \leq a}= W_{\pm}\chi_{\leq{a}}(|D|)$ respectively.
For the high energy part the following result is long known: 

\bgth[\cite{Ya,Ya-4dim}]  \lbth(high) 
Suppose either that 
$\Fg (\ax^{2\s} V) \in L^{\frac32}(\R^4)$ 
for some $\s>2/3$ and $|V(x)|\leq C \ax^{-\d}$ for some $\d>7$ 
or that, for $|\alpha|\leq 1$, 
$\|D^\alpha V(y)\|_{L^p(|x-y|<1)}\leq C \ax^{-\d}$ 
for some $p>2$ and $\d>7$. Then, $W_{\pm}\chi_{\geq{a}}(|D|)$ 
are bounded in $L^p(\R^4)$ for all $1\leq p \leq \infty$ for arbitray $a>0$. 
If $H$ has no zero energy eigenfunctions nor resonances, the same  
holds for  $W_{\pm}$ themselves. 
\edth 

We recall that some smoothness condition on $V$ is necessary in \refth(high) 
because of Goldberg-Visan's conter-example for dispersive estimates on 
$e^{-itH}$ (\cite{GV}). 
For the high energy parts we have nothing to add to \refth(high) 
and we shall exclusively study the low energy part in this paper. 

When $H$ has zero energy eigenfunctions but not resonances 
Goldberg and Green (\cite{GG}, see also \cite{J-Y-4}) have  
recently proved that $W_{\pm,\leq a}$ 
are bounded in $L^p(\R^4)$ for $1\leq p <4 $ 
and, for $1\leq p <\infty$ if the eigenfunctions $\ph$ satisfy 
$\int_{\R^4} x_j V(x) \ph(x) dx=0$ for $1\leq j \leq 4$. However, if 
$H$ has zero energy resonances 
$W_{\pm, \leq a}$ must be unbounded in $L^p$ for $2 < p \leq \infty$ 
because of Murata's result (\cite{Mu}) that 
$(e^{-itH}P_{ac}(H)u, v)_{L^2}$ for some $u,v \in C_0^\infty(\R^4)$ can decay 
as slowly as $C(\log t)^{-1}$ as $t\to \infty$. 
In this paper, we show when $H$ has zero energy resonances 
that $W_\pm$ are nevertheless bounded in $L^p(\R^4)$ 
for $1 < p \leq 2$ and give a direct proof of the unboundedness 
for $2<p \leq \infty$. When $H$ has zero energy eigenfunctions only,  
we prove that $W_{\pm}$ are unbounded in $L^p$ for $4<p \leq \infty$ unless 
$\int_{\R^4} x_j V(x) \ph(x) dx=0$ for $1\leq j \leq 4$, which 
supplements Goldberg-Green's result mentioned above. We 
give a new proof of some known results on $W_\pm$. 
 
For stating the main theorem we need introduce some notation. 
We define 
\[
U(x) = {\textrm{sign}}\, V (x)=\left\{\br{cl} 1, & \ V(x)\geq 0, \\[2pt]
-1 & V(x)<0, \er \right. 
\quad 
v(x) = |V(x)|^{1/2}, 
\quad w(x)= U(x) v(x).
\]
Integral operators $T$ and their 
kernels $T(x,y)$ are often identified and 
we say operator $T(x,y)$ or write $|T|\leq S(x,y)$ for 
$|T(x,y)|\leq S(x,y)$ for example. Multiplication operator by $F$ 
is denoted by $M_F$, however, we often write $F$ for $M_F$. 
Define 
\bqn \lbeq(M0)
\Mg_0 = M_U + M_v N_0 M_v, \quad 
N_0 u(x) = \frac1{4\pi^2}\int_{\R^4} \frac{u(y)}{|x-y|^2} dy\,.
\eqn  
$\|u\|_p$ is the norm of $L^p(\R^4)$ and $\|u\|=\|u\|_2$;
$L^p_w(\R^4)= L^{p,\infty}(\R^4)$ is the weak-$L^p$ space and 
its norm is denoted by $\|u\|_{p,\infty}$. 
$\hat{u}=\Fg{u}$ and $\check{u}=\Fg^\ast u$.
For a Banach space $\Xg$, $\Bb(\Xg)$ is the Banach space of bounded operators 
in $\Xg$ with norm $\|T\|_{\Bb(\Xg)}$.

\bglm \lblm(null)  Suppose that $V \in L^2$. Then: \\[2pt]
{\textrm {(1)}} The operator  $M_v N_0 M_v$ is compact and selfadjoint 
in $L^2$. \\[2pt] 
{\textrm{(2)}} If $\Ker \Mg_0 \not= \{0\}$, 
$0$ is isolated eigenvalue and $\dim\Ker \Mg_0<\infty$. \\[2pt]
{\textrm (3)} The number of negative eigenvalues of $H= -\lap + V$ is finite.   
\edlm 
\bgpf (1) Denote $B_0=M_v N_0 M_v$, 
$B_0 u (x)= (4\pi)^{-1} v(x)(|\,\cdot\,|^{-2} \ast (vu))(x)$. 
By applying H\"older's, generalized Young's and again H\"older's inequalities 
in this order we have 
\bqn \lbeq(D0-est)
\|B_0 u\| \leq C \|v\|_4 \||x|^{-2}\ast (vu)\|_{4} 
\leq C\|v\|_4 \||x|^{-2}\|_{2,\infty} \|vu\|_{\frac43}
\leq C\|v\|_4^2 \||x|^{-2}\|_{2,\infty} \|u\|
\eqn 
and $B_0$ is bounded selfadjoint in $\HL$. To show that it is a 
compact operator, take a sequence $v_n \in C_0^\infty(\R^4)$, 
$n=1,2, \dots$ such that $\|v_n - v \|_4 \to 0 $ as $n\to \infty$ 
and define $B_n= M_{v_n} N_0 M_{v_n}$. 
Then \refeq(D0-est) implies 
$\|B_n - B_0\|_{\Bb(L^2)} \leq C (\|v_n\|_4+\|v\|_4)\|v_n-v\|_4 \to 0$ 
as $n \to \infty$ and it suffices to prove that 
$B_0$ is compact if $v \in C_0^\infty(\R^4)$. 
Let 
\[
\tilde{B}_m u(x)= \frac1{4\pi^2} \int_{\R^4} 
\frac{v(x)v(y)u(y)}{|x-y|^2 +m^{-1}}dy, 
\quad  m=1,2, \dots . 
\] 
Then, $\tilde{B}_m$ is of Hilbert-Schmidt type since 
for any $1<p < 2$ and $q=2p/(2p-1)$ 
\[
\int_{\R^4} \frac{|v(x)v(y)|^2}{(|x-y|^2 +m^{-1})^2}dx dy
\leq \|v^2\|_{q}^2 \|(|x|^2 +m^{-1})^{-2}\|_{p,\infty}\,. 
\]
Set $f_m(x) = |x|^{-2}- (|x|^2 +m^{-1})^{-1}$. We have 
$\|f_m\|_{r,\infty} \leq C \|f_m\|_r = C m^{1-\frac2{r}} \to 0 $ 
as $m\to \infty$ for any $1<r<2$ and, as in \refeq(D0-est), for $p=2r/(r-1)$  
\[
\|(\tilde{B}_m- B_0)u\| \leq C\|v\|_p \|f_m \ast (vu)\|_{2r} 
\leq C \|v\|_p^2 \|f_m\|_{r,\infty}\|u\| \,.
\]
Hence $\|\tilde{B}_m- B_0\|_{\Bb(L^2)} \to 0$ and 
$B_0$ is compact.

\noindent 
(2)  Since $\s(U)\subset \{-1,1\}$ and $B_0$ is compact, the stability theorem 
for the essential spectrum implies $\s_{ess}(\Mg_0)\subset \{-1,1\}$ 
and $0$ is possibly be an isolated eigenvalue of 
$\Mg_0$ with finite multiplicity. 

\noindent 
(3) Cwikel-Lieb-Rozenbljum's theorem (\cite{RS3}) implies (3).  
\edpf 

The following definition can be found in \cite{GT}. 
$\la f, g \ra = \int_{\R^4} f(x) \overline{g(x)}dx$ 
and we denote the rank one operator $f(x) \mapsto \la f, b\ra a(x)$ 
by $|a\ra \la b|$ or $a \otimes b$ indiscriminately. 
$P=\|V\|_1^{-1}(v\otimes v)$ is the orthogonal projection onto 
the space $\C{v}=\{\a{v}\colon \a \in \C\}$.  

\bgdf \lbdf(GT-1) {\textrm{(1)}} 
We say $H= -\lap + V$ is regular at zero if 
if $\Mg_0$ is invertible in $\HL$ and, otherwise $H$ is singular at zero. \\[2pt]
{\textrm{(2)}} Assume that $H$ is singular at zero and let 
$S_1$ be the orthogonal projection onto $\Ker\, \Mg_0$. 
We say $H$ has singularities of 
the first kind if $S_1P{S_1}\vert_{S_1 \HL}$ is invertible 
in $S_1\HL$, of the second kind if $S_1P{S_1}\vert_{S_1 \HL}=0$ 
and of the third kind if $S_1P{S_1}\vert_{S_1 \HL}$  
is singular but $S_1P{S_1}\vert_{S_1 \HL}\not=0$.  
\eddf
We denote $D_0 = (\Mg_0 +S_1)^{-1}$ which exists by virtue of \reflm(null). 
Let $S_2$ be the projection onto $\Ker S_1P{S_1}\vert_{S_1 \HL}$ 
in $S_1\HL$. If $\ph \in S_1\HL$, then $u(x) = (N_0v\ph)(x)$ is 
a distributional solution of $(-\lap + V(x))u(x)=0$ which satisfies 
$|u(x)|\leq C \ax^{-2}$ and $\ph \mapsto u$ is an isomorphism between 
$S_1\HL$ and $\{u\colon |u(x)|\leq C \ax^{-2}, (-\lap + V(x))u=0\}$  
(cf. \reflm(7-12) below, see also \cite{GT,EGG}); 
$u= N_0v\ph\in L^2(\R^2)$ and it becomes a zero energy eigenfunction of $H$ 
if and only if $\la \ph, v \ra=0$, or $\ph\in S_2\HL$;  
$u$ is called (threshold) resonance otherwise. Thus, $H$ is regular at zero 
if $H$ has no zero energy eigenfunctions nor resonances, 
$H$ has singularities of the first 
kind if $H$ has zero energy resonances only, of the second kind if zero energy 
eigenfunctions only and of the third kind if both zero energy 
eigenfunctions and resonances. For $d \geq 3$, 
distribution solutions of $(-\lap + V(x))u(x)=0$ such that 
$u(x) \to0$ as $|x|\to \infty$ satisfy 
$|u(x)|\leq C \ax^{2-d}$. Such solutions are called threshold 
resonances if $u \not\in L^2(\R^d)$. 
Thus, no threshold resonaces exist when $d\geq 5$. 
\bgth \lbth(main-theorem) 
{\textrm{(1)}} Assume that $\ax^3 V \in (L^1 \cap L^4)$ 
and $H$ is regular at zero. Then, for arbitrary $a>0$, $W_{\pm,\leq a}$ 
are bounded in $L^p$ for all $1<p<\infty$.
 
\noindent 
{\textrm{(2)}} Suppose that $H$ is singular at zero and $a>0$  
is arbitrary. 
\ben 
\item[{\textrm {(a)} }]  Assume $\ax^{3+\ep} V \in (L^1 \cap L^4)$ 
for an $\ep>0$. Let singularities be of the first kind.  Then, 
$W_{\pm, \leq{a}}$ are bounded in $L^p$ for $1<p\leq 2$ and are 
unbounded for $2<p<\infty$.  
\item[{\textrm {(b)}}] Assume $\ax^{4+\ep} V \in (L^1 \cap L^4)$ 
for an $\ep>0$. Let singularities be of the second kind. Then, 
$W_{\pm,\leq {a}}$ are bounded in $L^p$ for 
$1<p<4$. They are bounded in 
$L^p$ for $1<p<\infty$ if all zero energy eigenfunctions $u$ 
satisfy $\la V, x_j u\ra=0$, $j=1,\dots,4$ and are unbounded  
for $4<p\leq \infty$ otherwise. 
\item[{\textrm{(c)}}] Assume $\ax^{4+\ep} V \in (L^1 \cap L^4)$ 
for an $\ep>0$. Let singularities be of the third kind, then 
$W_{\pm, \leq {a}}$ are bounded in $L^p$ for $1<p\leq 2$ and are 
unbounded in $L^p$ for $2<p\leq \infty$. 
\een
\edth 

The rest of the paper is devoted to the proof of \refth(main-theorem). 
Statements (1) and (2b) are known for wider ranges of $p$ 
under slightly different conditions 
(cf. \refth(high) and \cite{GG}) and we give a new proof for them. 
We explain here the basic 
strategy of the proof, introducing some more notation and displaying the plan of 
the paper. We arbitrarily take and fix $\Lambda>0$ and consider $a$ such that  
$0<a<\Lambda/2$. 

We prove \refth(main-theorem) for $W_{+}$ only. 
$W_{-}(x,y)= \overline{W_{+}(x,y)}$ and statements for $W_{-}$ 
instantly follow from the ones for $W_{+}$. 
$\C^{+}=\{z\in \C \colon \Im z>0\}$ and 
$\Cb^{+}=\C^{+} \cup \R$.  
$G_0(\lam)=(H_0-\lam^2)^{-1}$ for $\lam \in \C^{+}$. 
It is well known (cf. e.g. \cite{KRS, KY,IS}) that for $V \in \HL$   
\bqn \lbeq(Q0-def)
Q_0(\lam)= M_v G_0(\lam)M_v, \quad \lam \in \C^{+}
\eqn 
has a bounded closure which is compact and, which we denote again 
by $Q_0(\lam)$.  $\C^{+}\ni \lam \mapsto Q_0(\lam)\in \Bb(\HL)$ 
is holomorphic, uniformly bounded, 
$Q_0(\lam)\to 0$ as $\Im \lam \to \infty$ and has continuous extension to 
$\Cb^{+} \setminus \{0\}$. Let  
\bqn \lbeq(M-def)
\Mg(\lam)=M_U + Q_0(\lam), \quad \lam \in \Cb^{+}\setminus\{0\}.  
\eqn 
then, the 
analytic Fredholm theory implies that $\Mg(\lam)$, $\lam\in \C^{+}$, 
is invertible outside a discrete set $\Eg(H)$,  
$\C^{+}\ni \lam \mapsto \Mg(\lam)^{-1}\in \Bb(\HL)$ is meromorphic.  
$\Eg(H)$ is the set of 
$\lam\in i\R^{+}$ such that $\lam^2<0$ is an eigenvalue of $H$ and   
\bqn \lbeq(reso-eq)
(H-\lam^2)^{-1} = G_0(\lam) - G_0(\lam)M_v \Mg(\lam)^{-1}M_v G_0(\lam), 
\quad \lam\in \C^{+}\setminus\Eg(H) .
\eqn 
For $\lam\in \R\setminus\{0\}$, $\Mg(\lam)$ is invertible unless 
$\lam^2$ is an eigenvalue of $H$ (\cite{IS}) and the absence of 
positive eigenvalues (cf. \cite{IJ,KT}) implies $\Mg(\lam)^{-1}$ 
exists for all $\lam \in \R \setminus \{0\}$. Thus, $\Mg(\lam)^{-1}$ has 
continuous extension to $\R \setminus \{0\}$. 

For $\lam>0$, 
``the spectral projection'' of $H_0$ for the energy $\lam^2$ is defined by 
$\Pi({\lam})u(x) = {i\pi}^{-1}(G_0(\lam)-G_0(-\lam))u(x)$ or 
\bqn \lbeq(Pi-def)
\Pi({\lam})u(x) =\frac{\lam^2} {4\pi^2}
\int_{{\mathbb S}^3} e^{i\lam \w x}\hat{u}(\lam{\w})d\w 
= (\Pi({\lam})\tau_{-x} u)(0), 
\eqn 
where $\t_y$ is the translation by $y\in \R^4$.
It is evident from \refeq(Pi-def) that, for continuous functions 
$f(\lam)$ on $(0,\infty)$,  
\bqn \lbeq(mult-1)
f(\lam) \Pi(\lam ) u = \Pi(\lam) f(|D|) u, \quad \lam>0\,.
\eqn 

\bgdf \lbdf(Mikhlin) 
{\textrm{(1)}} We say $T$ is a good operator if $T$ is bounded in 
$L^p$ for all $1<p<\infty$.

\noindent 
{\textrm{(2)}} A function $f(\lam)$ of $0<\lam<\Lam$ is called  
Mikhlin multiplier if it satisfies 
$|f^{(j)}(\lam)|\leq C_j \lam^{-j}$ for 
$0\leq j \leq 3$ and $0<\lam<\Lam$.  
\eddf 
\noindent 
If $f(\lam)$ is a Mikhlin multiplier, then  
$\chi_{\leq a}(|D|)f(|D|)$ is a good operator for $2a<\Lam$ (\cite{Stein}).

We use the function space 
\bqn 
\Dg_\ast= \{u \in \Sg(\R^4) \ | \ \hat u \in C_0^\infty(\R^4 \setminus\{0\})\}  
\eqn 
which is dense in $L^p$ for any $1\leq p<\infty$ (cf. \cite{CMY}). 
The proof of \refth(main-theorem) is based on the stationary 
representation of $W_{+}$ (\cite{Ku,Ku-1,RS3}): If $V \in L^{2}$, then  
\bqn \lbeq(sta-0)
W_{+}u(x)= u(x) -\int_0^\infty 
(G_0(-\lam)M_v\Mg(\lam)^{-1}M_v \Pi({\lam})u)(x) \lam d\lam\,, 
\quad u \in \Dg_\ast.
\eqn 
By virtue of \refeq(mult-1) it suffices to prove \refth(main-theorem) for   
\bqn \lbeq(sta-1)
\W_{\leq {a}}u(x) \equiv \int_0^\infty 
(G_0(-\lam)M_v\Mg(\lam)^{-1}M_v \Pi({\lam})u)(x) \chi_{\leq{a}}(\lam)\lam d\lam\,.\eqn 
Let $\Ng(\lam)= M_v\Mg(\lam)^{-1}M_v$ and 
$\Ng(\lam,x,y)$ be its integral kernel. Then, by 
\refeq(Pi-def) 
\bqn \lbeq(L-repre-W)
\W_{\leq a}u(x)=\int_{0}^{\infty}\left(
\int_{\R^4\times \R^4}\Gg_{-\lam}(x-z)\Ng(\lam,z,y)
(\Pi(\lam)\t_{-y} u)(0)dy dz\right) 
\chi_{\leq a}(\lam)\lam d\lam. 
\eqn 
\bgdf \lbdf(Wtilde)
For $\tilde\Ng(\lam,z,y)$, define $\W_{\leq a}(\tilde\Ng)$ 
by \refeq(L-repre-W) with $\tilde\Ng(\lam,z,y)$ in 
place of $\Ng(\lam,z,y)$. We say $\tilde\Ng(\lam,z,y)$ 
is a good producer if $\W_{\leq a}(\tilde\Ng)$ is a good operator. 
\eddf

In \S 2, after proving some properties of $\Pi(\lam)$, 
we record some results on $\Mg(\lam,x,y)$ which will be used for 
studying $\Ng(\lam,x,y)$ in later sections. For shortening formulas, 
we write $\Lg_1$ and $\Lg_{2,1}$ for Banach spaces 
$L^1(\R^4 , L^1(\R^4))$ and $L^2(\R^4 , L^1(\R^4))$ respectively.

\begin{definition} \lbdf(Lg-def) The Banch space 
$\Lg_1 \cap \Lg_{2,1}$ with the norm 
$\|L\|_{\Lg}{=} \|L\|_{\Lg_1}+ \|L\|_{\Lg_{2,1}}$ is denoted by $\Lg$. 
It also denotes the Banch space of integral operators with the kernels 
$L(x,y) \in \Lg$. 
\eddf 

For a function $T(\lam)$ of $\lam \in (0,\Lam)$, 
$T^{(j)}(\lam)$ is the $j$-th derivative of $T(\lam)$. 
\bgdf
Let $\Yg$ be Banach space and $f(\lam)>0$ a function on $(0,\Lam)$. Then 
we say $T(\lam)=\Og^{(j)}_{\Yg}(f(\lam))$ if $T(\lam)$ is 
$\Yg$-valued function of $\lam \in (0,\Lam)$ which is 
$C^{j-1}$, $T^{(j-1)}(\lam)$ is absolutely continuous and it satisfies 
\bqn \lbeq(T-kk)
\|\pa_\lam^k T(\lam)\|_{\Yg}\leq C_k \lam^{-k}f(\lam), \  
0\leq k \leq j, \ \lam\in (0,\Lam).
\eqn 
\eddf 

It will be shown in later sections 
that $\Ng(\lam,x,y)$ is in general of the form 
\bqn \lbeq(Nlam-xy)
\Ng(\lam,x,y)= \sum_{j=0}^2 \sum_{j\ell}^{\ell_j} 
\lam^{j-2}(\log\,\lam)^\ell \m_{j\ell}(\lam) L_{j\ell}(x,y) + \Rg(\lam,x,y),
\eqn 
where $\m_{j\ell}(\lam)$ are Mikhlin multipliers, $L_{j\ell} \in \Lg$ 
and $\Rg(\lam)\in \Og^{(3)}_{\Lg}(\lam^2(\log\lam)^2)$. 
Section 3 is technically the main section where $L^p$ mapping 
properties of operators 
$\W_{\leq a}^{(j,\ell)}(\m_{j\ell}(\lam) L_{j\ell})$ are studied 
which are defined by the right of 
\refeq(L-repre-W) by substituting 
$\lam^{j-2}(\log\,\lam)^\ell \m_{j\ell}(\lam) L_{j\ell}(x,y)$ 
for $\Ng(\lam,x,y)$, $j=0,1,2$ and $\ell=0,1, \dots$. 
After changing the order of integrations it becomes 
\bqn \lbeq(Wjell)
\int_{\R^4\times \R^4} L_{j\ell}(z,y)\tau_z 
\left( \int_{0}^{\infty}\Gg_{-\lam}(x)(\Pi(\lam)\tau_{-y} u)(0) 
\chi_{\leq a}(\lam)\lam^{j-1} (\log\,\lam)^\ell d\lam \right) dy dz\,.
\eqn 
Then, by defining  $K_{a}^{(j,\ell)}$ by 
\begin{align} 
& K_{a}^{(j,\ell)} u(x)= \int_0^\infty \Gg_{-\lam}(x)
(\Pi({\lam})u)(0)
\chi^{(j,\ell)}_{\leq {a}}(\lam)\lam^{-1}d\lam\,, \lbeq(K-op-low) \\
& \hspace{1cm} \chi^{(j,\ell)}_{\leq {a}}(\lam) = \lam^j 
(\log\,\lam)^{\ell}\chi_{\leq a}(\lam)\,.
\lbeq(chijl)
\end{align}
and by using \refeq(mult-1), we express  \refeq(Wjell) in the form 
\bqn 
\W_{\leq a}^{(j,\ell)}(\m_{j\ell}(\lam) L_{j\ell})u
= \int_{\R^4\times \R^4} L_{j\ell}(z,y)\tau_z K_{a}^{(j,\ell)} \tau_{-y}  
\m_{j\ell}(|D|) u dy dz.  
\eqn 
Notice that $\W_{\leq a}(L)= \W_{\leq a}^{(2,0)}(L)$. 
$K_a$ denotes $K_{a}^{(2,0)}$. 
We let $\tchi_{\leq a}(\lam)$ be another cut off function  such that 
$\chi_{\leq a}(\lam)\tchi_{\leq a}(\lam) = \chi_{\leq a}(\lam)$ 
and define 
\bqn \lbeq(K-op-a)
K_{a,\leq }^{(j,\ell)}=\tchi_{\leq a}(|D|)K_{a}^{(j,\ell)}, \quad 
K_{a,\geq }^{(j,\ell)}=\tchi_{\leq a}(|D|)K_{a}^{(j,\ell)}\,
\eqn 
so that $K_{a}^{(j,\ell)}=K_{a,\leq }^{(j,\ell)}+ K_{a,\geq }^{(j,\ell)}$. 
We show in \reflm(Kgeq) and \reflm(Kleq) that:  

\noindent 
(1) 
$K_{a,\geq}^{(j,\ell)} \equiv 
(4\pi^2)^{-1} \widehat{\m_{1,a}} \otimes \widehat{\n_{a}^{(j,\ell)}}$ 
modulo a good operator, 
where $\widehat{{\m}_{1,a}}$ is smooth for $x\not=0$, rapidly 
decreasing as $|x|\to \infty$ and $\widehat{{\m}_{1,a}}(x)=|x|^{-2}+ O(1)$ 
as $|x|\to 0$; $\widehat{\n_{a}^{(j,\ell)}}$ is smooth and 
$|\pa_y^\a \widehat{{\n}^{(j,\ell)}_{a}}(y)|\leq C_{\a} 
\la \log\, \ay \ra^{\ell}\ay^{-(2+j+|\a|)}$.  

\noindent 
(2) There exist singular integral operators $T_1$ and $T_2$ which are 
bounded in $L^p$ for $2<p<\infty$ and $1<p<2$ respectively and  
\bqn \lbeq(KT)
K^{(j,\ell)}_{a,\leq}u=
\widehat{{\tchi}_{\leq a}} \ast \big(|x|^{-2}\otimes |y|^{-2} 
-T_{1} \big) (\widehat{\chi^{(j,\ell)}_{\leq {a}}} \ast u)
=\widehat{{\tchi}_{\leq{a}}} \ast 
T_{2} (\widehat{\chi^{(j,\ell)}_{\leq {a}}} \ast u)\,.
\eqn
In particular, 
$K_{a,\leq}^{(0,0)}$, $K_{a,\leq}^{(1,\ell)}$ and $K_{a,\leq}^{(2,\ell)}$ 
are bounded in $L^p$ for $1<p<2$, for $1<p<4$ and 
for $1<p<\infty$ respectively, $\ell=0,1, \dots$. 

By using (1) and (2), we then show in \reflm(Funda) that 
$\W_{\leq a}^{(j,\ell)}(\m_{j\ell}(\lam) L_{j\ell})$ 
is bounded in $L^p$ for $1<p<2$ if $(j,\ell)=(0,0)$, for $1<p<4$ 
if $j=1$ and for $1<p<\infty$ if $j=2$ and $\ell=0,1, \dots$. 
At the end of \S 3 we show \refprop(R-theo) that 
$\W_{\leq a}(\tilde\Ng)$ is a good operator if 
$\tilde\Ng(\lam,x,,y)$ is an $\Lg$-valued function of 
$\lam\in (0,\Lam)$ which is $C^1$, $\pa_{\lam}\tilde{\Ng}(\lam)$ 
is absolutely continuous  and 
$\int_0^{\Lam} \|\tilde{\Ng}^{''}(\lam)\|_{\Lg}d\lam<\infty$, 
which partly generalizes \reflm(Funda) and which proves that 
$\Rg(\lam,x,y)$ is a good producer. 

In \S 4, we show that $\Ng(\lam,x,y)$ 
is an $\Lg$-valued $C^3$ function of $\lam$ on any closed interval 
$[b,a]$, $b<a$ if $\ax^3 V \in (L^1\cap L^4)(\R^4)$. 
Then \refprop(R-theo) implies 
that $\W_{\leq a} \chi_{\geq b}(|D|)$ is a good operator. This   
reduces the proof of \refth(main-theorem) for $a>0$ as small as we wish.

In \S 5 we prove \refth(main-theorem)\, (1) for $H$ which is regular at zero 
by showing 
$\Ng(\lam,x,y)\in \Lg + \Og^{(3)}_{\Lg}(\lam^{2+\ep})$ for an $\ep>0$. 
\refprop(R-theo) then implies $\W_{\leq a}$ is a good operator.  

We begin to study the case that $H$ is singular at zero in \S 6. 
After stating Feshabach formula and the lemma due to Jensen-Nenciu (\reflm(JN)) 
which will be used many times in later sections, we prepare a few lemma 
to study $\Ng(\lam,x,y)$ for small $\lam\in (0,a)$. 
In particular, we show that $\ph \in S_1\HL$ satisfies 
$|\ph(x)|\leq C v(x)\ax^{-2}$ and prove the statement below \refdf(GT-1) 
on the relation between the type of 
singularities of $H$ at zero and the existence/absence of zero energy 
eigenfunctions and resonances. 

In sections 7,8 and 9 we prove statements (2a), (2b) and (2c) of 
\refth(main-theorem) respectively. 
It will be shown in all cases that modulo a good producer  
$\Ng(\lam,x,y)$ is equal to an operator valued function 
in the finite dimensional subspace $S_1\HL$ which, however, 
has singularities which can be as strong as $\lam^{-2}$ at $\lam=0$.  

In \S 7, we prove \refth(main-theorem) (2a) when 
$H$ has singularity of the first kind. 
Then, $\dim S_1 \HL=1$ and modulo a good producer 
$\Ng(\lam,z,y) \equiv \lam^{-2}\m(\lam) (v\ph)(z)(v\ph)(y)$, 
where $\m(\lam)$ is Mikhlin multiplier and 
$\ph$ is the normalized basis vector of $S_1 \HL$. 
Then, modulo a good operator and with  $L(x,y)=(v\ph)(x)(v\ph)(y)\in \Lg$ 
\bqn \lbeq(intro-8)
\W_{\leq a}u(x)= \W_{\leq{a}}^{(0,0)}(\mu(\lam)L)u(x) 
= \int_{\R^4 \times \R^4} L(y,z)(\t_z K_a^{(0,0)})\t_{-y} \mu(|D|)u(x) 
dy dz
\eqn 
and \reflm(Funda) implies that $W_{\leq a}$ is bounded 
in $L^p(\R^4)$ for $1<p<2$. To prove the negative part of 
\refth(main-theorem) (2a), we shall invoke the explict formula   
$\m(\lam)= c_1 (\log \lam)^{-1} (1+ o(\log \lam)^{-1})$ with $c_1\not=0$ 
and the first representation of \refeq(KT) for $K_{\leq{a}}^{(0,0)}$ 
which implies that $\tchi_{\leq{a}}(|D|)\W_{\leq{a}}$ is equal 
modulo a bounded operator in $L^p$ for $2<p<\infty$ to the rank one operator 
$(F \otimes G)\m(|D|)$ where 
$F =(|x|^{-2}\ast v\ph) \ast \widehat{{\tchi}_{\leq{a}}}$ 
and $G =(|x|^{-2}\ast v\ph) \ast \widehat{{\chi}_{\leq{a}}}$ 
are in $L^p$ for any $2<p<\infty$ but the linear functional  
$\ell(u)=\la G,\m(|D|)u\ra$ is unbounded on $L^p$ unless 
$\la v, \ph\ra =0$.  

In \S 8, we prove \refth(main-theorem)\, (2b). If $S_1P S_1=0$, then 
$S_1=S_2$. We take an orthonormal basis 
$\{\ph_1, \dots, \ph_n\}$ of $S_1\HL$. They satisfy the vanishing property. 
$\la v, \ph_j\ra=0$, $j=1, \dots, n$. We then show that, modulo a good producer, 
$\Ng(\lam) \equiv \lam^{-2} \sum_{j,k=1}^n c_{jk}(v\ph_j) \otimes (v\ph_k)$ 
with a non-singular matrix $(c_{jk})$ and modulo a good operator 
$\W_{\leq a}$ is a linear combination of operators which are given by 
the right of \refeq(intro-8) 
with $\m(\lam)=1$ and 
$(v\ph_j) \otimes (v\ph_k)$ in place of $(v\ph) \otimes (v\ph)$. 
Here, however, the vanishing property  $\la v, \ph_j\ra=0$ implies 
$|(|\cdot|^{-2}\ast (v\ph_j))(x)|\leq C\ax^{-3}$ and 
$\W_{\leq a}$ becomes bounded in $L^p$ for $1<p<4$ by virtue of \refeq(KT). 
If $\la v,x_k \ph_j\ra=0$, $k=1, \dots, 4$ is satisfied in addition, then 
$|(|\cdot|^{-2}\ast (v\ph))(x)|\leq C\ax^{-4}$ and 
$\W_{\leq a}$ becomes bounded in $L^p$ for all $1<p<\infty$. 
The proof for the negative part is similar as in the case of 
the first kind. 

In the final \S 9, we study the case that 
$H$ has singularity of the third kind. Then  
$S_1= S_2\oplus (S_1 \ominus S_2)$, 
${\rank} (S_1 \ominus S_2)=1$ and the basis vectors 
$\ph_1$ of $(S_1 \ominus S_2)\HL$ and $\{\ph_2, \dots, \ph_n\}$ of 
$S_2\HL$ satisfy $\la \ph_1, v \ra \not=0$ and 
$\la \ph_j, v \ra =0$, $j=2, \dots, n$. 
We then show that modulo a good producer 
$\Ng(\lam) \equiv \lam^{-2} \sum_{j,k=2}^n d_{jk}(v\ph_j \otimes v\ph_k) 
+ \lam^{-2}\b(\lam)v(\ph_1 \oplus \tilde \ph)
\otimes v(\ph_1 \oplus \tilde \ph)$ 
where $d_{jk}$ are constants and $\b(\lam)$ is Mikhlin multiplier. 
Thus, $\W_{\leq a}$ is a sum 
of the operators which are studied in \S 7 and \S 8 
and \refth(main-theorem)\ (2c) follows.

\section{Preliminaries,  $\Mg(\lam)$ for small $\lam>0$.} 

Recall that $\Pi(\lam)= (i\pi)^{-1}(G_0(\lam) - G_0(-\lam))$, $\lam>0$. 

\bglm \lblm(spec-proj) $\Pi(\lam)$ satisfies \refeq(Pi-def) 
and \refeq(mult-1) for continuous functions $f(\lam)$. 
\edlm 
\bgpf 
The Fouier inversion formula 
implies that $\Pi({\lam})u(x)$ is equal to 
\begin{align*} 
& \lim_{\ep \downarrow 0} \frac{\ep}{2\pi^3}\int_{\R^4} 
((\xi^2-\lam^2)^2+\ep^2)^{-1}
\Fg(\tau_{-x} u)(\xi)d\xi \\
&= \lim_{\ep \downarrow 0}\frac{1}{4\pi^2} 
\frac{\ep}{\pi}\int_0^\infty \frac{1}{(\r-\lam^2)^2+\ep^2}\left(
\int_{{\mathbb S}^3}
\Fg(\tau_{-x} u)(\sqrt{\r}\w)d\w \right) \r{d\r}. 
\end{align*}
and the first of \refeq(Pi-def) follows from which other statements are obvious.  
\edpf 

It is well known (e.g.\cite{ST,DLMF}) that $G_0(\lam)$, 
$\lam\in \Cb^{+}\setminus \{0\}$ is the convolution with   
\begin{align} 
\Gg_{\lam}(x)& = 
\frac{i}{4}\left(\frac{\lam}{2\pi |x|}\right) H^{(1)}_1 (\lam|x|) \lbeq(Green) 
\\
& = \frac1{4\pi^2|x|^2}+ 
\frac{\lam^2}{4\pi}\sum_{n=0}^\infty 
\left(g(\lam)+\frac{c_n}{2\pi}-\frac{\log |x|}{2\pi}\right)
\frac{(-\lam^2|x|^2/4)^n}{n!(n+1)!}\,,  \lbeq(Hankel-1)
\end{align} 
where $H^{(1)}_{1}(z)$ is the Hankel function, 
$g(z)$ is the principal branch of 
\bqn 
g(z)= -\frac1{2\pi}\log\left(\frac{z}{2}\right)-\frac{\c}{2\pi} + \frac{i}{4}, 
\eqn 
$\c$ is Euler's constant, $c_0 = 1/2$  and  
$c_n= 1+ \cdots+ {n}^{-1}+ 2^{-1}(n+1)^{-1}$ for $n=1,2,\dots$. 
$\Gg_{\lam}(x)$ has the integral representation (\cite{DLMF}), 
with the brach of $z^\frac12$ such that $z^\frac12>0$ for $z>0$,
\begin{equation}\lbeq(re-i0) 
\Gg_{\lam}(x)= \frac{e^{i\lam|x|}}{2(2\pi)^{\frac{3}2}
\Ga \left(\frac{3}{2}\right) |x|^{2}} 
\int_0^\infty e^{-t} t^{\frac{1}2}
\left(\frac{t}2 -i\lam|x| \right)^{\frac{1}2}dt\,.
\end{equation} 
Eqns. \refeq(Hankel-1) and \refeq(re-i0) imply that 
$\Gg_{\lam}(x)$ is smooth in $\lam>0$ and 
$x\in \R^4\setminus\{0\}$, bounded by $C|x|^{-3/2}$ as $|x|\to \infty$ 
and by $C|x|^{-2}$ near $x=0$.

We need investigate $\Mg(\lam)^{-1}$ for small $\lam>0$. 
We begin by studying $\Mg(\lam)$ which is defined by \refeq(M-def). 
$a\absleq b$ means $|a|\leq |b|$.

\bgdf \lbdf(Og)  
Let functions be defined for $\lam\in (0,\Lam)$ and $j=0,1,2,3$.
For a function $K(\lam,x,y)$ and 
an operator valued function $T(\lam)$, we say that 
$T(\lam)=\Og^{(j)}(K(\lam,x,y))$ if the kernel $T(\lam,x,y)$ 
is of class $C^j$ in $\lam$ and, for 
$0\leq k \leq j$,   
\bqn \lbeq(T-k)
|\pa_\lam ^k T(\lam,x,y)|\leq C_k \lam^{-k} |K(\lam,x,y)|,  
\ (\lam,x,y)\in (0,\Lam)\times \R^4 \times \R^4.
\eqn 
\eddf

\bglm \lblm(G-bound)
{\textrm{(1)}} Let $\lam|x|\geq 1$. For $j=0,1, \dots$, 
\bqn 
\Gg_{\lam}(x)\absleq C\lam^2,\ \ 
\pa_\lam \Gg_{\lam}(x)\absleq C\lam , \ \
\pa_\lam^{2+j} \Gg_{\lam}(x)\absleq 
C\lam^\frac12 |x|^{\frac12+j}.  \quad  \lbeq(large-1)
\eqn 
{\textrm{(2)}} Let $(\lam, x)\in (\Lam^{-1},\Lam)\times \R^4$, $\Lam>1$. Then,
\begin{gather} 
\Gg_{-\lam}(x) \absleq C \ax^{1/2}{|x|^{-2}}, \ \ 
\pa_\lam \Gg_{-\lam}(x) \absleq C {\ax^{1/2}}|x|^{-1}, \lbeq(G-bound)
 \\
\pa_\lam^{2+j} \Gg_{-\lam}(x) \absleq C (\ax^{1/2+j}+ |\log |x||). \ \  
j=0,1, \dots. \lbeq(G-bound+)
\end{gather}
\edlm 
\bgpf (1) We rewrite \refeq(re-i0) in the form  
\bqn  \lbeq(large) 
\Gg_{\lam}(x)= \frac{\lam^{\frac12}e^{i\lam|x|}}
{2^{\frac{3}{2}}\pi^2 |x|^{\frac32}}F(\lam |x|), \quad 
F(\r)= \int_0^\infty e^{-t} t^{\frac{1}2}
\left(\frac{t}{2\r}-i\right)^{\frac{1}2}dt\,.
\eqn 
\refeq(large-1) follows since 
$|(d/d\r)^k F(\r)|\leq C_{k}$, $k=0,1, \dots$  for $\r \geq 1$. 

\noindent 
(2) We use \refeq(re-i0). Then, \refeq(G-bound) is obvious. 
By Leibniz' rule, 
\bqn 
\pa_\lam^{j+2}\Gg_{-\lam}(x) = \sum_{k=0}^{j+2}
C_{jk}(-i)^{2+j}e^{-i\lam|x|}|x|^{j}\int_{0}^\infty 
e^{-t}t^{\frac12} \left(\frac{t}{2}-i\lam|x|\right)^{\frac12-k}dt\,.
\eqn 
For $\Lam^{-1} <\lam <\Lam$, the term with $k=0$ is bounded by 
$C |x|^j \ax^{\frac12}$,
$k=1$ by $C|x|^j$ since $({t}/2-i\lam|x|)^{-\frac12}\absleq C(t/2)^{-\frac12}$, 
those with $k=2, \dots, j+1$ by 
\[
C|x|^{j+1-k}\int_{0}^\infty 
e^{-t}t^{\frac12} \left(\frac{t}{2}-i\lam|x|\right)^{-\frac12}dt 
\leq C |x|^{j+1-k}
\]
since $({t}/{2}-i\lam|x|)^{\frac12-k}\absleq C|x|^{1-k}t^{-\frac12}$, 
and with $k=j+2$ by 
\[
C \int_{0}^\infty 
e^{-t}t^{\frac12} \left(\frac{t}{2}-i\lam|x|\right)^{-\frac32} dt\, 
\]
which is bounded by $C$ for $|x|\geq 1/100$ and 
by $C \la \log |x| \ra$ for $|x|<1/100$. We obtain \refeq(G-bound+). 
\end{proof}

The following lemma is Lemma 2.4 of \cite{EGG} with a slightly 
different assumption. Define 
\bqn \lbeq(hj-def)
g_1(\lam) = \frac{\|V\|_1 }{4\pi}\left(g(\lam)+ \frac{1}{4\pi}\right), \ 
h_{j}(\lam) = \lam^{j}\la \log \lam \ra, \quad j \geq 0.
\eqn 
Recall that $\Mg_0$ is defined by \refeq(M0). We define $\Mg_1$ by 
\bqn \lbeq(M1)
\Mg_1 u(x) = - \frac{1}{8\pi^2}\int_{\R^4} v(x)(\log |x-y|)v(y) u(y)dy\,.
\eqn

\bglm \lblm(M0-3) Let $0<\lam\leq \Lambda$ with an arbitrary 
but fixed $\Lambda$. Let $j=0,1$. Then: 

\noindent 
{\textrm {(1)}} $\Mg(\lam)= \Mg_0  + \tilde{\Mg}_1(\lam)$ 
with $\tilde{\Mg}_1(\lam)$ satisfying 
\bqn 
\tilde{\Mg}_1(\lam,x,y)= \Og^{(2+j)}(h_{2}(\lam)
v(x)(|x-y|^{j+\frac12}+ \la \log|x-y|\ra\ra) v(y)), \quad j=0,1\,. \lbeq(M-1a)
\eqn 

\noindent 
{\textrm {(2)}}  
$\tilde{\Mg}_1(\lam)= \lam^2 g_1(\lam) P + \lam^2 \Mg_1+ \tilde{\Mg}_2(\lam)$ 
with $\tilde{\Mg}_2(\lam)$ satisfying for any $0\leq \ep_1\leq 2$ 
\bqn 
\tilde{\Mg}_2(\lam)= 
\Og^{(2+j)}(h_{2+\ep_1}(\lam)v(x)|x-y|^{\ep_1}
(\la x-y\ra^{\max \left(j+{\frac12}-\ep_1, 0 \right)} +\la \log|x-y| \ra) v(y)).
\lbeq(M-2a)
\eqn 

\noindent 
{\textrm {(3)}} Let $\ep$ be such that $0\leq \ep\leq 2$. 
Suppose $\ax^{\d}V \in (L^1 \cap L^2)$ for some 
$\d>\max(1+2j,2\ep)$. Then for any $0\leq \ep_1 \leq  \ep$ 
\begin{align} 
& \tilde{\Mg}_1(\lam) = \Og^{(2+j)}_{\Hg_2}(h_{2}(\lam)),
 \quad 
\tilde{\Mg}_2(\lam) = \Og^{(2+j)}_{\Hg_2}(h_{2+\ep_1}(\lam)), \lbeq(HS-est-1) \\
& M_v \tilde{\Mg}_1(\lam),\ \ \tilde{\Mg}_1(\lam)M_v, \ \ 
M_v\tilde{\Mg}_1(\lam)M_v \in \Og^{(2+j)}_{\Hg_2}(h_{2}(\lam)). 
\lbeq(HS-est-2)
\end{align}
\edlm 

\bgpf We prove \refeqs(M-1a,M-2a) only. Other statements 
follow from them instantly since $||\log |x-y||\leq C_\d \ax^\d \ay^\d$ 
for any $0<\d$ if $|x-y|\geq 1$ and since 
\[
\int_{|x-y|\leq 1}|v(x)|^2 |v(y)|^2 |\log |x-y||dx dy <\infty 
\]
if $V \in (L^1 \cap L^2)$. 
Substituting \refeq(Hankel-1) for $\Gg_\lam(x)$, we obtain for $0<\lam |x-y|<1$
\begin{align*} 
\pa^j \tilde \Mg_1(\lam)& = \Og^{(3)}(\lam^{2-j}\la \log \lam \ra 
v(x)\la \log|x-y| \ra v(y)) , \\
\pa^j \tilde \Mg_2(\lam)& = \Og^{(3)}(\lam^{4-j}\la \log \lam \ra 
v(x)|x-y|^2 \la \log|x-y| \ra v(y)) \\
& =\Og^{(3)}(\lam^{2+\ep_1 -j}\la \log \lam \ra 
v(x)|x-y|^{\ep_1} \la \log|x-y| \ra v(y)), \quad 0\leq \ep_1 \leq 2.
\end{align*} 
Then, \refeqs(M-1a,M-2a) for $\lam|x-y|\leq 1$ follow. 

For $\lam |x-y|\geq 1$ we observe that for any real $k \geq 0$ 
and $j=0,1, \dots$ 
\begin{align}
& v(x) N_0(x-y)v(y)= \Og^{(3)}(\lam^k v(x)|x-y|^{k-2}v(y))\,.\lbeq(M-2b) \\
& g_1(\lam)\lam^2 P(x,y) =\Og^{(3)}(\lam^{2+k}\la \log\lam\ra 
v(x)|x-y|^{k}v(y))\,. \lbeq(M-3b) \\
& \lam^2 \Mg_1(x,y)= 
\Og^{(3)}(\lam^{2+k} v(x)|x-y|^k v(y) |\log|x-y|)\,. \lbeq(M-4b)
\end{align}
We have 
\[
\tilde{\Mg}_1(\lam,x,y) = v(x)(\Gg_\lam(x-y)- N_0(x-y))v(y).
\]
Then, \refeq(large-1) and \refeq(M-2b) 
with $k=2$ imply \refeq(M-1a) for $\lam|x-y|\geq 1$. Likewise 
\[
\tilde{\Mg_2}(\lam)= M_v (G_0(\lam)-N_0 - g_1(\lam)\lam^2 \Mg_1-\lam^2 \Mg_2)
M_v\,.
\]
We apply \refeq(large-1), 
\refeq(M-2b) with $k=2+\ep_1$, 
\refeq(M-3b) with $k=\ep_1$ and  
\refeq(M-4b) with $k=\ep_1$ to obtain \refeq(M-2a) for $\lam|x-y|\geq 1$. 
\edpf

\section{Integral operators $K_{a}^{(j,\ell)}$} 

In this section we study $L^p$ mapping properties of  
$K_a^{(j,\ell)}$. We first prove that the change of the order of 
integration which lead to \refeq(Wjell) is possible. 
Define for integral operator 
$\tilde{\Ng}(\lam)=\tilde{\Ng}(\lam,x,y)$ and 
$\chi^{(j,\ell)}_{\leq {a}}$ of \refeq(chijl) that 
\bqn \lbeq(Wjl)
\W^{(j,\ell)}_{\leq {a}}(\tilde{\Ng}(\lam))u(x)\equiv \int_0^\infty 
(G_0(-\lam)f(\lam) \tilde{\Ng}(\lam) \Pi({\lam})u)
(x) \chi^{(j,\ell)}_{\leq {a}}(\lam)\lam^{-1}d\lam\,.
\eqn 
Do not confuse $\W^{(j,\ell)}_{\leq {a}}$ with 
$\tilde{\W}^{(j,\ell)}_{{a},\leq}$ or 
$\tilde{\W}^{(j,\ell)}_{{a},\geq}$ which will appear shortly.

\bglm \lblm(WL) Let $L\in \Lg_1$ and Mikhlin multiplier $f(\lam)$. Then,  
\bqn  
\W^{(j,\ell)}_{\leq {a}}(f(\lam)L)u(x)
= \iint_{\R^4\times \R^4} 
L(z,y)(\tau_z K_{a}^{(j,\ell)} \tau_{-y} f(|D|)u)(x)dz dy, \quad u \in \Dg_{\ast} 
\lbeq(WL)  
\eqn 
for almost all $x \in \R^4$ and for $j=0,1,2$ and $\ell=0,1, \dots$\,. 
\edlm 
\bgpf We may assume $f(\lam)=1$ by virtue of \refeq(mult-1). Then, 
$\W^{(j,\ell)}_{\leq a}(L)u(x)$ is equal to  
\bqn \lbeq(iil)
\int_0^\infty \left(
\iint_{\R^4\times \R^4} 
\Gg_{-\lam}(x-z)L(z,y)(\Pi(\lam)u)(y) 
\chi^{(j,\ell)}_{\leq {a}}(\lam)\lam^{-1}dy dz
\right) d\lam\, .
\eqn 
We prove that 
$|\Gg_{-\lam}(x-z)L(z,y)(\Pi(\lam)u)(y)\chi^{(j,\ell)}_{\leq{a}}(\lam)\lam^{-1}|$ 
is integrable with respect to 
$(x,y,z,\lam)\in B_R \times \R^4\times \R^4 \times (0,\infty)$ 
for any $R>0$, where $B_R=\{x \in \R^4 \colon |x|\leq R\}$. 
Then Fubini's theorem guaratees that for almost all $x\in \R^4$ 
the order of integrations in \refeq(iil) may be freely changed.  
However, this follows easily because, for $u \in \Dg_\ast$, 
(i) $(\Pi(\lam)\tau_y u)(0)=0$ for $\lam$ outside a compact interval 
of $(0,\infty)$, say $[\a,\b]$, 
(ii) $|(\Pi(\lam)\tau_y u)(0)|\leq C \ay^{-3/2}$ uniformly for $\lam\in [\a,\b]$ 
which is well known, (iii) \refeq(large) implies 
\[
\sup_{(\lam,z)\in [\a,\b]\times \R^4} 
\int_{B_R} \int_{\a}^\b | \Gg_{-\lam}(x-z)| d\lam dx \leq C<\infty
\]
and (iv) $L(y,z) \in L^1(\R^4 \times \R^4)$. This completes the proof. 
\edpf 

\subsection{High energy part of $K_{a}^{(j,\ell)}$} 

Let $\tchi_{\leq}(\lam)\in C_0^\infty(\R)$ be another cut offfunction such that 
$\tchi_{\leq}(\lam) = 1$ for $|\lam|\leq 2$ and $\tchi_{\leq}(\lam) = 0$ for 
$|\lam|>4$ and let $\tchi_{\geq}(\lam) = 1- \tchi_{\leq}(\lam)$ . 
Then, define $\tchi_{\leq a}(\lam)=\tchi_{\leq}(\lam/a)$ and  
$\tchi_{\geq a}(\lam)=\tchi_{\geq}(\lam/a)$  for $a>0$ so that 
$\chi_{\leq a}(\lam)\tchi_{\leq a}(\lam) = \chi_{\leq a}(\lam)$.  
Define as in \refeq(K-op-a)  
\[
K^{(j,\ell)}_{a,\leq}=\tchi_{\leq {a}}(|D|)K_{a}^{(j,\ell)}, 
\quad 
K^{(j,\ell)}_{a,\geq}=\tchi_{\geq{a}}(|D|)K_{a}^{(j,\ell)}\,.
\]
We first study the high energy (output) part $K^{(j,\ell)}_{a,\geq}$.
To prove the following lemma we use the well known identity 
\bqn 
\lbeq(Bessel-def) 
\int_{{\mathbb S}^3}e^{i{\r}x\w}d\w = 4\pi^2  \frac{J_1(\r|x|)}{\r|x|}
\eqn 
for the Bessel function and the asymptotic formula  
\begin{multline} J_1(\r)= \sqrt{\frac{\pi}{2}} 
\Big((\r^{-\frac12}+O(\r^{-\frac52})) \cos(\r-\tfrac{3\pi}{4})  \\
+  (\tfrac38\r^{-\frac32}+ O(\r^{-\frac72})
\sin(\r-\tfrac{3\pi}{4})\Big)\,, \quad \r \to \infty.
\lbeq(Bessel-asymp)
\end{multline}

\bglm \lblm(mu-nu) 

\textrm{(1)} Let 
$\m_{1,a}(\xi)= |\xi|^{-2}\tchi_{\geq{a}}(|\xi|)$ and 
$\m_{2,a}(\xi)= |\xi|^{-4}\tchi_{\geq{a}}(|\xi|)$. Then,  
$\widehat{{\m}_{1,a}}(x)$ and $\widehat{{\m}_{2,a}}(x)$ 
are $C^\infty$ in $\R^4\setminus\{0\}$, 
they are rapidly decreasing as $|x|\to \infty$ and as $|x|\to 0$ 
\bqn \lbeq(mu-1-2)
\widehat{{\m}_{1,a}}(x)=\frac{1}{|x|^{2}}+ O(1), 
\quad \widehat{{\m}_{2,a}}(x) = \frac12 \log |x| + O(1). 
\eqn 

\noindent 
\textrm{(2)} Let 
$\n^{(j,\ell)}_{a}(\xi)= |\xi|^{-2}\chi^{(j,\ell)}_{\leq {a}}(|\xi|)$ 
Then, 
$\widehat{{\n}^{(j,\ell)}_{a}}(y)$ is $C^\infty$ on $\R^4$ and, for large $|y|$, 
it satisfies for any $\a$ 
\bqn \lbeq(F-1)
|\pa_y^\a \widehat{{\n}^{(j,\ell)}_{a}}(y)|\leq C_{\a} 
\frac{\la \log\, \ay \ra^{\ell}}{\ay^{2+j+|\a|}}\,, \quad 
j=0,1,2, \ \ell=0, 1, \dots\,.
\eqn 
For $j=0, 1$, we have the lower bound 
\bqn \lbeq(F-1a)
|\widehat{{\n}^{(j,\ell)}_{a}}(y)|\geq  C  
\frac{\la \log\, \ay \ra^{\ell}}{\ay^{2+j}}\,, \quad \ell=0,1, \dots. 
\eqn 
\edlm 
\bgpf It suffices to prove the lemma 
when $a=1$ since differences of ${\m}_{1,a}$, 
${\m}_{2,a}$ and ${\n}_{a}^{(j,\ell)}$ for different $a$'s  
are of class $C_0^\infty(\R^4)$. We omit the index $a=1$ in the proof. 

\noindent
(1) Integration by parts shows that 
$\widehat{\m_{1}}(x), \widehat{\m_{2}}(x)$ are smooth 
outside $x=0$ and are rapidly decreasing as $|x|\to \infty$. 
Since $\m_{1}(\xi)-|\xi|^{-2}= - \tchi_{\leq}(\xi)|\xi|^{-2}$ is a 
compactly supported integrable function, 
$\widehat{\m_{1}}(x) -|x|^{-2}\in C^\infty(\R^4)$ 
and the first of \refeq(mu-1-2) follows.  
By integrating by the spherical variable first and by using 
\refeq(Bessel-def), we obtain   
\[
\widehat{\m_{2}}(x) = \frac {1}{|x|} \int_0^\infty J_1(\r|x|)
\frac{\tchi_{\geq}(\r)}{\r^2} d\r 
=  \int_0^\infty J_1(\r)
\frac{\tchi_{\geq }(\r/|x|)}{\r^2} d\r. 
\] 
Then \refeq(Bessel-asymp) implies 
that the integral over $[1,\infty)$ contributes only by $O(1)$ 
and,  
$J_1(\r) = (\r/2)+ O(|\r|^3)$ as $\r\to 0$ implies that 
modulo $O(1)$ as $|x|\to 0$, 
\[
\hat\m_{2}(x) \equiv  
\frac12 \int_{|x|}^1 \frac{d\r}{\r} 
+ \frac12 \int_{|x|}^{2|x|} \frac{\tchi_{\geq}(\r/|x|)-1}{\r} d\r
= \frac12 \log |x| + O(1) .
\]
(2) For $j=0,1$ we use the well known identity for the Fourier transforms of 
four dimensional homegenous distributions (cf. \cite{Grafakos-1}, Theorem 2.4.6): 
\bqn 
\Fg (|\xi|^{-z})(x) = C_z |x|^{z-4}, \quad C_z= (2\pi^2)^{z-2}\Ga(z/2)^{-1}
\Ga((4-z)/2)
\eqn 
and $C_z$ is holomorphic for $0<\Re{z}<4$. It follows by differentiating $\ell$ 
times by $z$ that 
\bqn \lbeq(p-log-Fourier)
\Fg (|\xi|^{-z}(\log |\xi|)^\ell )(x) = 
\sum_{k=0}^\ell \begin{pmatrix} \ell \\ k \end{pmatrix}
\frac{d^{\ell-k} C_z}
{dz^{\ell-k}} 
|x|^{z-4}(\log |x|)^k. 
\eqn 
Since $\hat{\chi}(x)\in \Sg(\R^4)$ and $\int_{\R^4}\hat{\chi}(x)dx = 4\pi^2$, 
\refeqs(F-1,F-1a) for $j=0,1$ follow from \refeq(p-log-Fourier). 
For $j=2$ we avoid using \refeq(p-log-Fourier) because of the obvious reason. 
We apply \refeq(Bessel-def) which implies   
\bqn 
\n^{(2,\ell)}(y)
= \int_{0}^\infty \left(\frac{J_1(\lam|y|)}{\lam|y|}\right)
\chi(\lam)(\log\,\lam)^{\ell}\lam^{3}d\lam. \lbeq(F-2)
\eqn 
We may assume $|y|\geq 100$. 
Since $J_1(\r)= (\r/2)+ O(\r^3)$ for $\r\leq 1$, the integral \refeq(F-2) 
over $(0,1/|y|)$ is estimated in modulus by 
$C\la \log\, \ay \ra^{\ell} |y|^{-4}$. The integral over $[1/y,\infty)$ 
is equal to by change of variable by 
\bqn \lbeq(j=2)
\frac{1}{|y|^{4}} \int_{1}^\infty J_1(\r)
\chi(\r/|y|)(\log |\r/|y||)^{\ell}\r^{2}d\r\,.
\eqn 
We substitute \refeq(Bessel-asymp) for $J_1(\r)$ and apply 
integration by parts $3$ times which implies that \refeq(j=2) 
is also bounded in modulus by $C\la \log\, \ay \ra^{\ell} |y|^{-4}$. 
This completes the proof. 
\edpf 

Hereafter we shall often write $\chi_{\leq a}(\xi)$ and $\chi_{\geq a}(\xi)$ for 
$\chi_{\leq a}(|\xi|)$ and $\chi_{\geq a}(|\xi|)$ respectively and 
$\widehat{{\chi}_{\leq a}}(x)$ and 
$\widehat{{\chi}_{\geq a}}(x)$ for their Fourier transforms. Similar 
convention will be used for $\tchi_{\leq a}$ and $\tchi_{\geq a}$.

\bglm \lblm(Kgeq) Modulo a good operator 
$K_{a,\geq}^{(j,\ell)}$, $j=0,1,2, \ell=0,1, \dots$ 
satisfies 
\bqn 
K_{a,\geq}^{(j,\ell)}\equiv 
\frac1{4\pi^2} \widehat{\m_{1,a}} \otimes \widehat{{\n}_{a}^{(j,\ell)}}, \quad 
\m_{1,a}(\xi)= \frac{\tchi_{\geq{a}}(|\xi|)}{|\xi|^2}, \ \ 
{\n}_{a}^{(j,\ell)}(\eta)= 
\frac{\chi^{(j,\ell)}_{\leq {a}}(|\eta|)}{|\eta|^2}.
\eqn 
\edlm 
\bgpf In view of \reflm(mu-nu) it suffices to prove    
\begin{align}
K_{a,\geq}^{(j,\ell)}= (4\pi^2)^{-1}(
\widehat{\m_{1,a}}\otimes \widehat{{\n}^{(j,\ell)}_{a}}+ 
\widehat{\m_{2,a}}\otimes \widehat{{\n}_{a}^{(j+2,\ell)}}) + R_{a}^{(j,\ell)}, 
\lbeq(Kgeq) \\
|R^{(j,\ell)}_{a}(x,y)|\leq C_{N}\ax^{-N} \ay^{-5-j}, 
\quad N=0,1, \dots\,. \lbeq(Kgeq-add)
\end{align}
Notice first $\m_{j,a}$, $j=1,2$ are even. 
We have from \refeq(K-op-low) 
\[
K_{a,\geq}^{(j,\ell)}u(x) = \int_0^\infty \tchi_{\geq {a}}(|D|)\Gg_{-\lam}(x)
(\Pi({\lam})u)(0)\chi^{(j,\ell)}_{\leq {a}}(\lam)\lam^{-1}d\lam\,.
\]
Since $|\xi|^2-\lam^2\geq 2a^2$ if 
$\tchi_{\geq {a}}(|\xi|)\chi_{\leq a}(\lam)\not=0$, we have 
for $\lam$ such that $\chi^{(j,\ell)}_{\leq {a}}(\lam)\not=0$  
\[
\tchi_{\geq {a}}(|D|)\Gg_{-\lam}(x)= \frac1{4\pi^2}
\int_{\R^4}\frac{e^{ix\xi}\tchi_{\geq {a}}(|\xi|)}{|\xi|^2-\lam^2}d\xi 
\]
and, integration by parts 
implies that $\tchi_{\geq {a}}(|D|)\Gg_{-\lam}(x)$ is $C^\infty$ outside $x=0$. 
It follows by recalling \refeq(Pi-def) and by setting $\eta= \lam\w$ that 
\begin{align*}
K^{(j,\ell)}_{a,\geq}u(x)& = \frac1{(2\pi)^4}
\int_0^\infty 
\left(\int_{\R^4}
\frac{e^{ix\xi}\tchi_{\geq {a}}(|\xi|)}{|\xi|^2-\lam^2}d\xi\right) 
\left(\lam^2 \int_{{\mathbb S}^3} \hat{u}(\lam\w)d\w\right)
\chi^{(j,\ell)}_{\leq {a}}(\lam)\lam^{-1}d\lam \\
&= 
\frac1{(2\pi)^4}
\iint_{\R^4\times \R^4}
\frac{e^{ix\xi}\tchi_{\geq {a}}(|\xi|)}{(|\xi|^2-|\eta|^2)|\eta|^2} 
\hat{u}(\eta)\chi^{(j,\ell)}_{\leq {a}}(|\eta|)d\xi d\eta \,.
\end{align*}
Thus, the integral kernel of $K^{(j,\ell)}_{a,\geq}$ is given by 
the oscillatory integral 
\bqn \lbeq(Kje)
K^{(j,\ell)}_{a,\geq}(x,y)= \frac1{(2\pi)^6}\int_{\R^4 \times \R^4}
\frac{e^{ix\xi-iy\eta}\tchi_{\geq {a}}(|\xi|)
\chi^{(j,\ell)}_{\leq {a}}(|\eta|)}{(|\xi|^2-|\eta|^2)|\eta|^{2}}d\xi d\eta\,.
\eqn 
Substituting    
\bqn \lbeq(expand)
\frac1{|\xi|^2-|\eta|^2}= \frac1{|\xi|^2}+ \frac{|\eta|^2}{|\xi|^4}+ 
\frac{|\eta|^4}{|\xi|^4(|\xi|^2-|\eta|^2)} 
\eqn
in \refeq(Kje) implies \refeq(Kgeq) if we set   
\[
R^{(j,\ell)}_{a}(x,y)  = \frac1{(2\pi)^6}\int_{\R^4 \times \R^4}
\frac{e^{ix\xi-iy\eta}\tchi_{\geq {a}}(|\xi|)
\chi^{(j+2,\ell)}_{\leq a}(|\eta|)}{|\xi|^4 (|\xi|^2-|\eta|^2)}d\xi d\eta\,.
\]
The amplitude function is smooth for $\eta\not=0$ and it satisfies  
for any multi-indices $\a,\b$ 
\[
\left|
\pa_\xi^\a \pa_\eta^\beta \left(
\frac{\tchi_{\geq {a}}(|\xi|)\chi^{(j+2,\ell)}_{\leq a}(|\eta|)}
{|\xi|^4(|\xi|^2-|\eta|^2)} \right)
\right| 
\leq C_{N\a\b\ep}\la \xi\ra^{-6-|\a|}\la \eta \ra^{-N}
|\eta|^{2+j-|\b|-\ep}, \quad \eta \not=0
\]
for any $N=0,1, \dots$ and small $\ep>0$. 
Then, integration by parts implies \refeq(Kgeq-add).  
\edpf

\subsection{Low energy part $K_{a,\leq}^{(j,\ell)}$}
We next study the high energu output paart 
$K_{a,\leq}^{(j,\ell)}$. We define for $\ep>0$
\bqn 
T_{1,\ep}(x,y)= \frac{1}{(x^2-y^2-i\ep)y^2}\,, \quad 
T_{2,\ep}(x,y)= \frac{-1}{x^2(x^2-y^2-i\ep)}\,.
\eqn 

\bglm \lblm(T) 
\ben
\item[\textrm{(1)}]  $T_{1,\ep}$ is bounded in $L^p$ for $2<p<\infty$ 
and $T_{2,\ep}$ for $1<p<2$.   
\item[\textrm {(2)}]  $T_{1,\ep}$ and $T_{2,\ep}$ are strongly convergent as 
$\ep \to 0$ in $L^p$ for $2<p<\infty$ and $1<p<2$  respectively 
and we denote the limit operators by $T_1$ and $T_2$.
\een
\edlm 
\bgpf Let $Mu(r)= |{\mathbb S}^3|^{-1}\int_{{\mathbb S}^3} u(r\w) d\w $. 
$T_{1,\ep}u(x)$ is spherical symmetric and we write  
$T_{1,\ep} u(\r) = T_{1,\ep} u(x)$ for $|x|=\r$. We have 
\[ 
T_{1,\ep} u(\r)= |{\mathbb S}^3| 
\int_{\R^4} \frac{Mu(r)}{(\r^2-r^2-i\ep)}rdr
\]
and by changing variables
\[
T_{1,\ep} u(\sqrt{\r})= \frac{|{\mathbb S}^3|}{2}
\int_{0}^\infty \frac{Mu(\sqrt{r})}{(\r-r-i\ep)}dr.
\] 
It is well-known that one dimensional integral operator with 
kernel $(\r-r-i\ep)^{-1}$ is uniformly bounded in $L^p(\R)$ 
for $1<p<\infty$ and converges to the Hilbert transform and 
the weight $\r$ is one-dimensional $(A)_p$ weight for $p>2$ 
(see \cite{Stein}, Chapter 5.6.4. p. 218). 
It follows by the weighted 
inequality for the Hilbert transform applied to 
$Mu(\sqrt{r})$ (extended to $\R$ by setting $Mu(\sqrt{r})=0$ for $r<0$) 
and by H\"older's inequality that for $2<p<\infty$ 
\begin{align}
& \int_{\R^4} |T_{1,\ep}u(x)|^p dx = 
\frac{|{\mathbb S}^3|}{2} \int_0^\infty |T_{1,\ep}u(\sqrt{\r})|^p \r d\r 
\lbeq(ti-est)
\\
& \leq C_1  \int_0^\infty |Mu(\sqrt{r})|^p rdr 
= C_2 \int_0^\infty |Mu({r})|^p r^3 dr \leq C \|u\|_p^p. \notag 
\end{align}
The result for the Hilbert-transform recalled above and the application of 
\refeq(ti-est) to $T_{1,\ep_m}- T_{1,\ep_n}$ imply that 
$T_{1,\ep}u$ is convergent in $L^p(\R^4)$ as $\ep \to 0$ for any $u \in L^p$. 
This proves lemma for $T_{1,\ep}$. Similar argument with the weight 
$\r^{1-p}$ in place of $\r$ implies the same for $T_{2,\ep}$.  
Note that $T_{2,\ep}$ is the adjoint of $T_{1,-\ep}$.  
\edpf

Define for $j=0,1,2$, $\ell=0,1, \dots$ and $\ep>0$ 
\bqn \lbeq(L-02)
K_{a,\leq,\ep}^{(j,\ell)}(x,y)=\frac1{(2\pi)^3}
\int_{\R^8}\frac{e^{ix\xi-iy\eta}\tchi_{\leq{a}}(|\xi|)
\chi_{\leq a}^{(j,\ell)}(|\eta|)}
{(|\xi|^2-|\eta|^2 + i\ep)|\eta|^{2}} d\xi d\eta. \\
\eqn
Since we have uniformly with respect to $(\lam, x)$ in compact sets  
of $(0,\infty)\times \R^4$ that 
\[
\tchi_{\leq a}(|D|)\Gg_{-\lam}(x) = \lim_{\ep \downarrow 0}
\frac1{4\pi^2}\int_{\R^4} \frac{e^{ix\xi}\tchi_{\leq a}(|\xi|)}
{|\xi|^2-\lam^2 -i\ep}d\xi ,
\]
\refeq(Bessel-def) and the uniform convergence theorem imply 
$K_{\ep,\leq}^{(j,\ell)}(x,y)$ converges to 
\[
K_{a,\leq}^{(j,\ell)}(x,y)= 
\frac{2pi}{|y|}
\int_0^\infty 
\chi_{\leq a}(|D|)\Gg_{-\lam}(x)
J_1({\lam}|y|)
\chi_{\leq a}^{(j,\ell)}
(|\eta|)d\lam , 
\quad x,y \in \R^4.
\]
We make this statement more precise:

\bglm \lblm(K-funda)
$K_{a,\leq,\ep}^{(j,\ell)}(x,y)$ converges as ${\ep \to 0}$ uniformly to 
$K_{a,\leq}^{(j,\ell)}(x,y)$ along with derivatives of all order. 
$K_{a,\leq}^{(j,\ell)}(x,y)\in C^\infty(\R^4 \times \R^4)$ 
and for $s,t\geq 0$ with $s+t=2$
\bqn  \lbeq(L-02a)
|K_{a,\leq}^{(j,\ell)}(x,y)|\leq C \ax^{-t}\ay^{-s}, 
\quad (x,y) \in \R^4 \times \R^4\,.
\eqn 
For $u \in \Dg_\ast$, we have \refeq(KT) for 
$K_{a,\leq}^{(j,\ell)}u(x)$, viz.  
\bqn \lbeq(KT-2)
K^{(j,\ell)}_{a,\leq}u=
\widehat{{\tchi}_{\leq a}} \ast \big(|x|^{-2}\otimes |y|^{-2} 
-T_{1} \big) (\widehat{\chi^{(j,\ell)}_{\leq {a}}} \ast u)
=\hat{\tchi}_{\leq{a}} \ast 
T_{2} (\widehat{\chi^{(j,\ell)}_{\leq {a}}} \ast u)\,.
\eqn 
\edlm 
\bgpf 
We prove the lemma for $j=\ell=0$. Modifications for other cases 
are obvious. Substituting  
$(\xi^2-\eta^2 + i\ep)^{-1}= -(i/2) \int_0^\infty 
e^{it(\xi^2-\eta^2 +i\ep)/2}dt$ in \refeq(L-02) yields that  
\bqn 
K_{a,\leq,\ep}^{(0,0)}(x,y){=} 
-i\pi  \int_{0}^\infty e^{-\ep{t}/2} u(t,x) v(t,y) dt, \lbeq(r-1) 
\eqn 
where $u(t,x)$ and $v(t,y)$ are solutions of Schr\"odinger equations:
\begin{align*} 
& u(t,x) = \frac1{(2\pi)^2}
\int_{\R^4}e^{ix\xi+i|\xi|^2t/2}\tchi_{\leq{a}}(|\xi|)d\xi, \\
& 
v(t,y)= \frac1{(2\pi)^2} 
\int_{\R^4}e^{iy\eta- i|\eta|^2t/2}\chi_{\leq a}(|\eta|)\frac{d\eta}{|\eta|^2} 
\end{align*}
and $\|\pa_x^\a u(t,x)\|_{L^\infty}\leq C_{\a} \la t \ra^{-2}$ 
and $\|\pa_y^\a v(t,y)\|_{L^\infty}\leq C_{\a}$ since 
$|\eta|^{-2+|\a|}\chi_{\leq a}(|\eta|)\in L^1$.   
It follows that 
$|\pa_x^\a \pa_y^\b u(t,x) v(t,y)|\leq C_{\a\b} \la t \ra^{-2}$, 
$|\pa_x^\a \pa_y^\b K_{a,\leq,\ep}^{(0,0)}(x,y)|\leq C$ uniformly 
for $\ep>0$ and 
$\pa_x^\a \pa_y^\b K_{a,\leq,\ep}^{(0,0)}(x,y)$ 
converges uniformly as $\ep \to 0$. 
Since $K_{a,\leq,\ep}^{(0,0)}(x,y)$ converges to $K_{a,\leq}^{(0,0)}(x,y)$ 
everywhere as remarked above the lemma,  
$K_{a,\leq}^{(0,0)}(x,y)\in C^\infty(\R^4 \times \R^4)$. 
Moreover, the limit 
is unchanged if we replace the factor $e^{-\ep{t}/2}$ by $e^{-\ep/{t}}$ 
in \refeq(r-1). 
Thus, we replace $e^{-\ep{t}/2}$ by $e^{-\ep/{t}}$ in \refeq(r-1) 
and prove \refeq(L-02a) and \refeq(KT-2). 
Define 
\[
v_1(t,x)= 
\frac1{(2\pi)^2} \int_{\R^4}e^{iy\eta- i|\eta|^2t/2}\chi_{\leq a}(|\eta|)d\eta 
\]
We have  
$i{\pa}_{t} v(t,y) = (1/2) v_1(t,y)$ and   
$v(0,y)= ({1}/{4\pi^2})(|\,\cdot\,|^{-2} \ast \widehat{\chi_{\leq{a}}})(y)$ 
since $\chi_{\leq{a}}$ is even. Hence, 
\bqn 
\lbeq(r-4a)
v(t,y) = \frac{1}{4\pi^2}(|\,\cdot\,|^{-2} \ast \widehat{\chi_{\leq{a}}})(y)
 - \frac{i}2 \int_0^t v_1(s,y) ds, 
\eqn 
which we substitute for $v(t,y)$ in \refeq(r-1) with $e^{-\ep{t}}$ 
being replaced by $e^{-t/\ep}$. Thus, uniformly along with all  
derivatives we have 
\bqn  
K_{a,\leq}^{(0,0)}(x,y)= \lim_{\ep \to 0}(F_{\ep,1}(x,y)+ F_{\ep,2}(x,y))
\eqn 
where $F_{\ep,1}(x,y)$ is equal to 
\bqn \lbeq(final-1)
-\pi{i}\int_{0}^\infty e^{-\ep/{t}} 
\left(\int_{\R^4}\frac{e^{-iz^2/2t}}{(-2\pi{it})^2}\widehat{\tchi_{\leq{a}}}(x-z)
dz\int_{\R^4}\frac{\widehat{\chi_{\leq{a}}}(y-w)}{4\pi^2|w|^2} dw\right) dt
\eqn 
and $F_{\ep,2}(x,y)$  
\[ 
-\frac{\pi}{2}\int_{0}^\infty e^{-\ep/t} 
\left(\int_{\R^4}
\frac{e^{-iz^2/2t}}{(-2\pi{it})^2} \widehat{\tchi_{\leq{a}}}(x-z) dz 
\int_0^t \int_{\R^4}
\frac{e^{iw^2/2s}}{(2\pi{is})^2}\widehat{\chi_{\leq{a}}}(y-w)ds dw\right)dt\,.
\]
Change $t$ to $t^{-1}$ and integrate with respect to $t$ first.   
\refeq(final-1) becomes 
\bqn \lbeq(final-1a)
F_{\ep,1}(x,y)= \frac{1}{8\pi^3}\int_{\R^4\times \R^4}
\frac{\widehat{\tchi_{\leq{a}}}(x-z)\widehat{\chi_{\leq{a}}}(y-w)}
{(z^2-2i\ep)w^2}dw dz. 
\eqn 
It is obvious that $F_{\ep,1}(x,y)\in C^\infty(\R^4 \times \R^4)$ and 
\bqn \lbeq(bound-F1)
|F_{\ep,1}(x,y)| \leq C\ax^{-2}\ay^{-2} \ \ 
\mbox{uniformly with respect to $\ep>0$}
\eqn 
and, along with derivatives of all order, 
\bqn \lbeq(final-1b)
\lim_{\ep \downarrow 0} F_{\ep,1}(x,y)= \frac{1}{8\pi^3}\int_{\R^4\times \R^4}
\frac{\widehat{\tchi_{\leq{a}}}(x-z)\widehat{\chi_{\leq{a}}}(y-w)}
{z^2 w^2}dw dz.
\eqn 
Changing $t$ by $1/t$ and then $s$ by $1/s$, we see that 
$F_{\ep,2}(x,y)$ is equal to 
\[
-\frac{\pi}{2}\int_{0}^\infty e^{-\ep{t}} 
\left(\int_{\R^4}
\frac{e^{-iz^2t/2}}{(2i\pi)^2} \widehat{\tchi_{\leq{a}}}(x-z) dz 
\int_{t}^\infty  \int_{\R^4}
 \frac{e^{iw^2s/2}}{(2i\pi)^2}\widehat{\chi_{\leq{a}}}(y-w)dw ds \right)dt\,.
\]
We compute the $s$-integral explicitly:   
\begin{multline} \lbeq(N-1)
\lim_{N\to \infty} \int_{t}^N \int_{\R^4}
e^{iw^2s/2}\widehat{\chi_{\leq{a}}}(|y-w|)dw ds \\
= -\int_{\R^4}
\frac{e^{iw^2t/2}}{iw^2/2}\widehat{\chi_{\leq{a}}}(|y-w|)dw 
+ \lim_{N\to \infty}\int_{\R^4}
\frac{e^{iw^2N/2}}{iw^2/2}\widehat{\chi_{\leq{a}}}(|y-w|)dw\,.
\end{multline}
We show the second integral on the right is bounded by 
$C_\ell N^{-1}\ay^{-\ell}$ 
for any $\ell=1,2, \dots$ and the limit vanishes. 
We insert $1= \chi_{\leq a}(|w|)+ \chi_{\geq{a}}(|w|)$.
Then integrating by parts $\ell$ times by using the identity
$N^{-1}L e^{iw^2N/2}= e^{iw^2N/2}$ for 
$L= -i|w|^{-2}(w \cdot \nabla_w)$, we obtain 
\begin{multline}
\int_{\R^4}
\frac{e^{iw^2N/2}}{iw^2/2}\widehat{\chi_{\leq{a}}}(|y-w|)\chi_{\geq{a}}(|w|)dw \\
= \frac1{N^\ell}\int_{\R^4}
\frac{e^{iw^2N/2}}{iw^2/2} 
(^{t}L)^\ell(\widehat{\chi_{\leq{a}}}(|y-w|)\chi_{\geq{a}}(|w|))
dw 
\absleq \frac{C}{N^\ell \ay^{\ell+2}},  
\end{multline}
where 
$^{t}L$ is the transpose of $L$ and we used  
$|w|^{-\ell-2}\widehat{\chi_{\leq{a}}}(y-w)\chi_{\geq{a}}(|w|)\leq C_k \ay^{-\ell-2}$
in the last step. Recall that $a\absleq b$ means $|a|\leq |b|$. 
For the integral containing $\chi_{\leq a}(|w|)$ we obtain 
by integrating by parts and by using estimate 
$\widehat{\chi_{\leq{a}}}(y-r\w)\chi_{\leq a}(r\w)\absleq C_\ell \ay^{-\ell}$ that 
\begin{align*}
& \int_{\R^4}
\frac{e^{iw^2N/2}}{iw^2/2}\widehat{\chi_{\leq{a}}}(y-w)\chi_{\leq a}(|w|)dw \\
& =- 2i \int_{0}^\infty e^{ir^2N/2}
\left(\int_{{\mathbb S}^3}\widehat{\chi_{\leq{a}}}(y-r\w)\chi_{\leq a}(r\w)d\w\right) rdr  \\
& = \frac{2}{N}\left(\widehat{\chi_{\leq{a}}}(y)
+ \int_{0}^\infty e^{ir^2N/2} \frac{d}{dr}
\left(\int_{{\mathbb S}^3}\widehat{\chi_{\leq{a}}}(y-r\w)\chi_{\leq a}(r\w)d\w\right) dr
\right) \leq \frac{C_{\ell}}{N \ay^\ell}. 
\end{align*}
In this way we have shown that $F_{\ep,2}(x,y)$ is equal to 
\bqn \lbeq(F-2ep-pre)
\frac{1}{16\pi^3 i} \int_{0}^\infty e^{-\ep{t}} 
\left(\int_{\R^4} e^{-iz^2t/2}\widehat{\tchi_{\leq{a}}}(x-z) dz 
\int_{\R^4}\frac{e^{iw^2t/2}}{\w^2}\widehat{\chi_{\leq{a}}}(y-w)dw\right)dt\,.
\eqn 
The first integral in the parentheses is a Gaussian integral and is 
bounded in modulus by $C \la t \ra^{-2}$ and the second obviously 
by $C \ay^{-2}$. It follows that 
\bqn 
|F_{\ep,2}(x,y)|\leq C\ay^{-2} \ \mbox{uniformly with respect to $\ep>0$}
\lbeq(bound-F2)
\eqn  
which together with \refeq(bound-F1) implies that 
\bqn \lbeq(Kleq-b1)
K_{a,\leq}^{(0,0)}(x,y) \absleq C \ay^{-2}.
\eqn 
Integrating by $t$ first in \refeq(F-2ep-pre) yields that 
\bqn 
F_{\ep,2}(x,y) = -\frac{1}{8\pi^3} \int_{\R^4\times\R^4} 
\frac{\widehat{\tchi_{\leq{a}}}(x-z)\widehat{\chi_{\leq{a}}}(y-w)}
{(z^2 - w^2 -i\ep)w^2} dzdw. \lbeq(F-2ep)
\eqn  
Thus, \refeq(bound-F1), \refeq(final-1b), \refeq(bound-F2) and \refeq(F-2ep) 
jointly imply 
\[
K_{a,\leq}^{(0,0)} u(x) = \big(
\widehat{{\tchi}_{\leq{a}}} \ast  
(|x|^{-2}\otimes |y|^{-2}- T_{1}) \big) 
(\widehat{\chi^{(0,0)}} \ast u)(x).
\]

Equations \refeq(final-1a) and \refeq(F-2ep) imply that 
\bqn 
K_{a,\leq}^{(0,0)}(x,y)= \lim_{\ep\downarrow 0}
\frac{-1}{8\pi^3}\int_{\R^4\times \R^4}
\frac{\widehat{\tchi_{\leq{a}}}(x-z)\widehat{\chi_{\leq{a}}}(y-w)}
{(z^2-i\ep)(z^2 - w^2 -i\ep)} dzdw\,.  \lbeq(F-3a)
\eqn 
The function under the limit sign is equal to \refeq(F-2ep) with the factor 
$w^2$ in the denominator being replaced by $z^2-i\ep$, hence is equal to  
\[
\int_{0}^\infty e^{-\ep{t}} 
\left(\int_{\R^4}
\frac{e^{iw^2t/2}}{(2i\pi)^2} \widehat{\tchi_{\leq{a}}}(y-w) dw 
\int_{t}^\infty  \int_{\R^4}
 \frac{e^{-i(z^2-i\ep)s/2}}{(2i\pi)^2}
\widehat{\chi_{\leq{a}}}(x-z)dz ds \right)dt\,.
\] 
This is equal to $\overline{F_{\ep,2}(y,x)}$ but with a slight modification, 
viz. with extra harmless damping factor $e^{-s\ep}$. Thus, 
by repeating the argument after \refeq(N-1), we obtain 
$K_{a,\leq}^{(0,0)}(x,y) \absleq C \ax^{-2}$ which proves \refeq(L-02a). 
Since 
\[
\left| \frac{2i\ep}{(x^2-i\ep)}\right| \leq 1 \ \ 
\mbox{and} \ \ \lim_{\ep \to 0}\frac{i\ep}{(x^2-i\ep)}=0 \ \ 
\mbox{if} \ \ x \not=0 ,
\]
Lebesgue's dominated convergence theorem 
and \reflm(T) (2) imply that 
\[
-\int_{\R^4}\frac{u(y)dy}{(x^2-i\ep)(x^2 - y^2 -i\ep)}- 
T_{2,\ep}u(x) 
= \frac{i\ep}{x^2-i\ep}T_{2,\ep}u(x)
\]
converges to $0$ in $L^p$ if $1<p<2$.
This implies the second of \refeq(KT-2). 
\edpf 

\bglm \lblm(Kleq) 
\ben 
\item[\textrm{(1)}] $K_{a,\leq}^{(0,0)}$ is bounded in $L^p$
for all $1<p<2$, $K_{a,\leq}^{(1,\ell)}$ for all $1<p<4$ 
and $K_{a,\leq}^{(2,\ell)}$ for all $1<p<\infty$, $\ell=0,1, \dots$. 

\item[\textrm{(2)}] $K_{a,\leq}^{(0,0)}$ is unbounded in $L^p$
if $2<p<\infty$ and $K_{a,\leq}^{(1,\ell)}$ if $4<p<\infty$, 
$\ell=0,1,\dots$.  
\een
\edlm 

The restriction to $\ell=0$ for $K_{a,\leq}^{(0,\ell)}$ is 
made since the operator 
$u \mapsto \widehat{\chi^{(0,\ell)}}\ast u $ is unbounded in $L^p$ 
for any $1,p<\infty$ if $\ell \geq 1$. This will, however, not cause any problem 
for us since the singularity of resolvent $G(\lam)$ as $\lam \to 0$ is 
never stronger than $\lam^{-2}$ by the selfadjointness.

\bgpf (i) Since $T_2$ is bounded in $L^p$ for $1<p<2$  and 
convolution operators by $\widehat{\tchi_{\leq {a}}}$ 
and $\widehat{\chi^{(j,\ell)}_{\leq {a}}}$ are good operators 
unless $(j,\ell)=(0,\ell)$, $\ell\geq 1$,  
$K_{a,\leq}^{(0,0)}$, $K_{a,\leq}^{(1,\ell)}$ and 
$K_{a,\leq}^{(2,\ell)}$, $\ell=0,1, \dots$
are bounded in $L^p$ for $1<p<2$. 

(ii) Since $T_1$  is bounded in $L^p$ for all $2<p<\infty$, $K_{a}^{(j,\ell)}$ 
is bounded in $L^p(\R^4)$ for $2<p<\infty$ for which the rank one operator  
\[
T_{jk}= (\widehat{{\tchi}_{\leq a}} \ast |x|^{-2})\otimes 
(\widehat{\chi^{(j,\ell)}_{\leq {a}}}\ast |y|^{-2})
\]
in the right of \refeq(KT) is bounded in $L^p$. Let 
$F^{(j,\ell)}(x)= \widehat{\chi^{(j,\ell)}_{\leq {a}}}\ast |x|^{-2}$.
$F^{(j,\ell)}$ is smooth and it satisfies \refeqs(F-1,F-1a). Moreover 
\refeq(p-log-Fourier) implies 
\bqn \lbeq(Fjell01)
F^{(j,\ell)}(x)= C |x|^{-(2+j)}(\log |x|)^\ell(1+ O(\log|x|^{-1})), 
\quad j=0,1.  
\eqn 
In particular we have $\widehat{{\tchi}_{\leq a}} \ast |x|^{-2} \in L^p(\R^4)$ 
for $2<p<\infty$. Hence $K_{a,\leq}^{(j,\ell)}$ is bounded in $L^p$ 
for a $p>2$ if and only if the linear functional 
\[
h(u)= \int_{\R^4}
(\widehat{\chi^{(j,\ell)}_{\leq {a}}}\ast |y|^{-2})u(y) dy 
\] 
is bounded in $L^p$ or $F^{(j,\ell)}(y)\in L^q(\R^4)$ for $q=p/(p-1)$. 
Then, \refeq(F-1) implies 
$F^{(2,\ell)}\in L^q$ for all $1<q$ or $1<p<\infty$ and \refeq(Fjell01) implies 
$F^{(0,\ell)} \in L^q$ if and only if $q>2$ or $1<p<2$ and  
$F^{(1,\ell)}\in L^q$ if and only if $d>4/3$ or $1<p<4$. 
Combining (i) and (ii) by interpolation, we obtain the lemma.  
\edpf 

Recall \refdf(Lg-def) that $\Lg=\Lg_1 \cap \Lg_{2,1}$,  
$\Lg_1= L^1(\R^4 \cap \R^4)$ and $\Lg_{2,1}= L^2(\R^4, L^1(\R^4))$.   

\bglm \lblm(Funda)  Let $L(x,y) \in \Lg$ and $f$ be Mikhlin multiplier. 
Then, 
\bqn \lbeq(Wjlest)
\|\W^{(j,\ell)}_{\leq{a}}(f(\lam)L)u\|_p \leq C\|L\|_{\Lg}\|u\|_p
\eqn 
for $1< p <\infty$ if $j=2$ and for $1< p <4$ if $j=1$ 
for any $\ell=0,1, \dots$. Estimate \refeq(Wjlest) holds also 
for $(j,\ell)=(0,0)$ for $1<p<2$. 
The same holds if $L$ is replaced by the multiplication $M_F$ 
with $F \in (L^1 \cap L^2)$.
\edlm
\bgpf 
We may assume $f(\lam)=1$ thanks to \refeq(mult-1). 
Let $\tilde{\W}^{(j,\ell)}_{a,\leq }(L)$ 
and $\tilde{\W}^{(j,\ell)}_{a, \geq}(L)$ respectively be defined by 
\refeq(WL) with $K^{(j,\ell)}_{a,\leq}$ and 
$K^{(j,\ell)}_{a,\geq}$ in place $K_{a}^{(j,\ell)}$ so that  
$\W^{(j,\ell)}_{\leq{a}}(L)=
\W^{(j,\ell)}_{a,\leq}(L) + \W^{(j,\ell)}_{a,\geq}(L)$.  
$\tilde{\W}^{(j,\ell)}_{a,\leq}(L)$ satisfies the lemma 
since 
\[
\|\tilde{\W}^{(j,\ell)}_{a,\leq}(L)u\|_p \leq \iint_{\R^4\times \R^4} 
|L(z,y)| \|\tau_z K^{(j,\ell)}_{a,\leq} \tau_{-y} u \|_p dz dy
\leq C\|L\|_{\Lg_1}\|u\|_p
\]
by Minkowski's inequality and \reflm(Kleq).  

\reflm(Kgeq) and the next \reflm(L-add-1) imply that 
$\|\tilde{\W}^{(j,\ell)}_{a,\geq}(L)u\|_p\leq C \|u\|_p$ 
for the same range of $p$'s and \reflm(Funda) follows.  
\edpf 

\bglm \lblm(L-add-1) 
Let $L\in \Lg$ and $G^{(j,\ell)}$ be the integral operator defined by the kernel 
\[
G^{(j,\ell)}(x,y)= \iint \widehat{{\m}_1}(x-z)L(z,w)
\widehat{{\n}^{(j,\ell)}}(w-y) dz dw , \quad \ell=0,1, \dots. 
\]
Then, $\|G^{(j,\ell)}u\|_p \leq C\|L\|_{\Ng}\|u\|_p$ 
for any $1<p<2$ if $j=0$, for any $1<p<4$ if 
$j=1$ and for any $1 < p<\infty$ if $j=2$. 
\edlm 
\bgpf  Let $q=p/(p-1)$ be the conjugate exponent of $p$. Then 
\reflm(mu-nu)\,(2) implies that 
$\widehat{{\n}^{(j,\ell)}} \in L^q$ for $q>2$ if $j=0$, for 
$q>4/3$ if $j=1$ and for $q>1$ if $j=2$ and H\"older's inequality implies 
\[ 
|G^{(j,\ell)}u(x)|
\leq I(x) \|\widehat{\n}^{(j,\ell)}\|_q \|u\|_p\, , 
\quad 
I(x)= \int |\widehat{{\m}_1}(x-z)| \left(\int_{\R^4} |L(z,w)| dz\right) dw\,.
\] 
By virtue of \refeq(mu-1-2) $\widehat{{\m}_1}\in L^s(\R^4)$ for any $1\leq s<2$ 
and we evidently have $\|I \|_1 \leq \|L\|_{\Lg_1}\|\widehat{{\m}_1}\|_1$. 
Moreover, 
Young's inequality implies 
$\|I\|^p \leq \|\widehat{{\m}_1}\|_s \|L\|_{\Lg_{2,1}}$ 
for $1/p= 1/s-1/2$ and when $s$ runs over $1<s<2$, $p$ does over $(2,\infty)$. 
Hence $\|I\|_p\leq C \|L\|_{\Lg}$ for all $1\leq p<\infty$  by interpolation. 
\edpf 

For genuinely $\lam$-dependent $\tilde{\Ng}(\lam,x,y)$  
we have the following proposition. 
Note that it is not quite a generalization of \reflm(Funda) since 
$\m^{(2)}(\lam)$ is in general {\it not} integrable for Mikhlin multipliers 
$\m(\lam)$.  

\bgprop \lbprop(R-theo) 
Let $f(\lam)$ be Mikhlin multiplier
and $F(\lam)$ an $\Lg$-valued function of $\lam\in(0,\Lam)$ 
of class $C^1$ such that $F'(\lam)$ is absolutely continuous and 
$\int_0^\Lam \|F''(\lam)\|_{\Lg}d\lam <\infty$.  
Let $\W_{\leq a}^{(j,\ell)}(f(\lam)F(\lam))$, 
$j=0,1,2$ and $\ell=0,1, \dots$, 
be defined by \refeq(Wjl) with $f(\lam)F(\lam,x,y)$ in place of 
$\tilde{\Ng}(\lam,x,y)$. Then, 
$\W_{\leq a}^{(j,\ell)}(f(\lam)F(\lam))$ is bounded in $L^p(\R^4)$ 
for the same range of $p$'s as in \reflm(Funda).  
\edprop
\bgpf We may asume $f(\lam)=1$ as previously. By Taylor's formula we have 
\[
F(\lam)= F(0) + \lam F'(0)+ R(\lam), \quad 
R(\lam)= \int_0^\lam (\lam-\r) F''(\r) d\r,
\]
which we substitute for $\tilde{\Ng}(\lam)$ in \refeq(Wjl). 
Since $F(0)$ and $F'(0) \in \Lg$, \reflm(Funda) implies that 
they produce bounded operators in $L^p(\R^4)$ for the desired $p$'s. 
Changing the order of 
integrations and inserting $\tchi_{\leq{a}}(\r)$ 
which satisfies  $\tchi_{\leq{a}}(\r)=1$ if  
$(\lam-\rho)_{+}\chi_{\leq a}(\lam)\not=0$, we express the operator 
produced by $R(\lam)$ as 
\bqn \lbeq(R-2a)
\int_0^\infty 
\left(
\int_0^\infty (\lam-\rho)_{+} 
(G_0(-\lam)F''(\r)\Pi({\lam})u)(x) \chi_{\leq {a}}^{(j,\ell)}(\lam)
\lam^{-1} d\lam 
\right)\tchi_{\leq{a}}(\r) d\r.
\eqn 
Evidently $(\lam-\r)_{+} = (\lam-\r) - (\r-\lam)_{+}$. 
If $(\lam-\r)_{+}$ is replaced by $(\r-\lam)_{+}=\r(1-\lam/\r)_{+}$, 
\refeq(R-2a) becomes by virtue of \refeq(mult-1) 
\bqn 
\int_0^\infty 
\left(\int_{\R^8} 
F''(\r,y,z) (\tau_y K^{(j,\ell)}_{a}\tau_{-z} (1-|D|/\r)_{+} u)(x) dy dz
\right) \r \tchi_{\leq{a}}(\r) d\r.  \lbeq(R-3)
\eqn 
Here $\sup_{\r>0}\|(1-|D|/\r)_{+}\|_{\Bb(L^p)}\leq C<\infty $ 
for any $1< p < \infty$ since the Fourier transform of 
$(1-|\xi|^2)_{+}$ in $\R^4$ is integrable (\cite{Stein}, p.389), 
$(1-|\xi|)_{+}= (1+|\xi|)^{-1}(1-|\xi|^2)_{+}$ and 
$(1+|\xi|)^{-1}$ is 4-dimensional Mikhlin multiplier. 
It follows by \reflm(Funda) that 
\[
\|\refeq(R-3)\|_{L^p} \leq C
\int_0^\infty \|F''(\r)\|_{\Lg}\|u\|_p \r^2 \tchi(\r) d\r \leq C \|u\|_p\,
\]
for $p$'s of \reflm(Funda).   
The term $\lam+\r$ produces
the sum over $(k_1,k_2)=(1,0), (0,1)$ of 
\bqn 
\int_0^\infty 
\left(
\int_{\R^4\times \R^4} 
F''(\r,z,y)(\tau_z {K}^{(j+k_1,\ell)}_{a}\tau_{-y} u)(x) 
dy dz \right)\r^{k_2}\tchi(\r) d\r 
\eqn 
which have the same or better $L^p$-properties than \refeq(R-3)  
since ${K}^{(j,\ell)}_{a}$ enjoys the better properties 
for the larger $j$ and 
$\r^{k_2}\tchi(\r)$ for $k_2 \geq 0$ is no worse than $\tchi(\r)$. 
\edpf 

\section{$W_{+}$ for intermediate energy} 
For $0<b<a<\infty$, we define $\chi_{b\leq a}(\lam) = 
\chi_{\geq b}(\lam)\chi_{\leq a}(\lam)$. 
In this section we prove the following theorem on $W_{+}\chi_{b\leq a}(|D|)$ 
which reduces the proof of \refth(main-theorem) for an arbitarily small $a.0$. 

\bgth\lbth(intermediate)  Suppose that 
$\ax^3 V \in L^1 \cap L^4$. Then, for any $0<b<a<\infty$, 
$W_{+}\chi_{b\leq a}(|D|)$ is bounded in $L^p$ for any 
$1<p<\infty$. 
\edth 

Recall $\Ng(\lam)=M_v \Mg(\lam)^{-1} M_v$. 
It suffice to prove \refth(intermediate) for  
\bqn  
\W_{b\leq a}u(x) = \int_0^\infty G_0(-\lam)\Ng(\lam)
\Pi(\lam)u(x)\chi_{b\leq a}(\lam) \lam d\lam \,.
\eqn 
For the proof we need some lemmas. Recall \reflm(G-bound).

\bgdf[\cite{EGG}] For an integral operator $K(x,y)$ on $\R^4$, 
$|K|$ is the operator with kernel $|K(x,y)|$.
$K$ is said to be absolutely bounded if $|K|$ is bounded in $L^2$. 
$\Ag$ is the Banach space of all absolutely bounded integral operators 
in $L^2$ with the norm $\||A|\|_{\Bb(L^2)}$. $\Ag$ will often denote an 
element of $\Ag$. As a convention, $M_F\in \Ag$ if $F \in L^\infty$. 
\eddf

\bglm \lblm(important) {\textrm (1)}
Let $K_1, K_2\in \Ag$ and $v\in \HL$. We have  
\bqn \lbeq(7-4)
\int_{\R^4} |(K_1 K_2 M_v)(x,y)|dy \leq 
(|K_1K_2||v|)(x) \leq (|K_1||K_2||v|)(x). 
\eqn 
{\textrm (2)} If $K\in \Ag$ and $F \in \HL$, then 
$\|(K M_F)(x,y)\|_{\Lg_{2,1}}\leq \||K|\|_{\Bb(\HL)}\|F\|_2$.
\edlm
\bgpf (1) It is evident that the left side is bounded as 
\[
\int_{\R^4} \left|\int_{\R^4} K_1(x,z)K_2(z,y)v(y)dz\right|dy   
\leq \int_{\R^4} |(K_1 K_2)(x,y)||v(y)|dy  
\]
and \refeq(7-4) follows. (2) is obvious.    
\edpf 

Recall that $v=|V|^{\frac12}, w=Uv$,  
$\Hg_2$ is the space of Hilbert-Schmidt  operators in $\HL$, 
$\Hg_2$ often denotes an element of $\Hg_2$ 
and $a\absleq b$ means $|a|\leq |b|$.

\bglm \lblm(HS) 
Suppose $\m, \n \in L^{2p}\cap L^{2q}$ 
for some $1\leq q <2< p \leq \infty$ 
such that $p^{-1}+ q^{-1}=1$. Let 
$N_0u(x) =(4\pi^2)^{-1}\int_{\R^4}|x-y|^{-2}u(y) dy$. 
Then, $(M_{\m} N_0 M_{\n})^2\in \Hg_2$. 
\edlm 
\bgpf We estimate the integral kernel of $(M_{\m} N_0 M_{\n})^2$ 
by using H\"older's inequality by 
\begin{multline*} 
\frac1{(2\pi)^4}\int_{\R^4}\frac{\m(x)\n(z)\m(z)\n(y)}{|x-z|^2|z-y|^2} dz \\
\absleq \frac{\|\m\n\|_p}{(2\pi)^4}   
\left(\int_{\R^4}\frac{|\m(x)\n(y)|dz}{|x-z|^{2q}|z-y|^{2q}} 
\right)^{\frac1{q}} \leq 
C \|\m\|_{2p}\|\n\|_{2p} \frac{|\m(x)\n(y)|}{|x-y|^{4/p}}\,,
\end{multline*}
where we used $2q<4$ but $4q>4$. 
Since $|x|^{-\frac{8}{p}}\in L_{w}^{\frac{p}{2}}$, $p/2>1$ and 
$2/p+ 2/q=2$,  
\[
\iint_{\R^4\times \R^4}\frac{|\m(x)\n(y)|^2}{|x-y|^{{8}/{p}}}dx dy 
\leq C \|\m\|_{2q}^2 \|\n\|_{2q}^2 \||x|^{-\frac{8}{p}}\|_{\frac{p}2,\infty}
\]
by the Hardy-Littlewood-Sobolev (HLS) inequality. This proves the lemma. 
\edpf 

\bglm\lblm(M0-1) 
\textrm{(1)} Let $p,q>1$ satisfy $1/p+ 1/q=3/2$. Then 
$\|M_f N_0 M_g\|_{\Lg_1} \leq C\|f\|_{p}\|g\|_{q}$. \\[3pt]
\textrm{(2)} Let $p>2>q$ satisfy $1/p+ 1/q=1$. then 
$\|M_f N_0 M_g\|_{\Lg_{2,1}} \leq C\|f\|_{p}\|g\|_{q}$.\\[3pt]
\textrm{(3)} Let $2<p,q \leq \infty$ satisfy $1/p+ 1/q=1/2$. Then 
$\|M_f N_0 M_g\|_{\Bb(L^2)} \leq C \|f\|_{p}\|g\|_{q}$.
\edlm 
\bgpf  (1) follows from the HLS inequality (\cite{LL},p.98). 
By the weak Young's inequalty (\cite{LL},p.99), 
$\||x|^{-2}\ast g \|_{\frac1{q}-\frac12}\leq  
C_q \||x|^{-2}\|_{2,\infty} \|g\|_{q}$. Then, H\"older's inequality 
implies (2). We likewise have 
$\||x|^{-2}\ast (g u) \|_{\frac1{q}}\leq  
C_q \||x|^{-2}\|_{2,\infty} \|g\|_{q}\|u\|_2$ and 
H\"older's inequality implies (3).
\edpf 

\bglm \lblm(Q-estimate) Define $Q(\lam)= M_w G_0(\lam)M_v$ and 
$I=(\Lam^{-1}, \Lam)$ for $\Lam>1$. Then: 
\ben 
\item[\textrm{(1)}] $Q(\lam)$ and $M_v Q(\lam)$ are $\Ag$-valued 
$C^2$ function of $\lam\in I$. 

\item[\textrm{(2)}] $Q^2(\lam), Q'(\lam)$ and $Q''(\lam)$ 
are $\Hg_2$-valued continuous on $\lam \in I$. 
\een
\edlm 
\bgpf By the first of the bound \refeq(G-bound) and 
$\la x-y\ra^{1/2}\leq \ax^{\frac12}\ay^{\frac12}$  
\bqn \lbeq(5-8)
|Q^{(j)}(\lam)(x,y)|
\leq C \frac{v(x)\ax^{1/2}v(y)\ay^{1/2}}{|x-y|^{2-j}}, \quad j=0, 1. 
\eqn 
It follow by H\"older's and the HLS inequalities 
\begin{align*}
& \||Q(\lam)|u\|_2 \leq C \|v(x)\ax^{1/2}\|_4
\||x|^{-2}\|_{2,\infty}\|v(y)\ax^{1/2}\|_4  \|u\|_2. \\
& \||M_v Q(\lam)|u\|_2 \leq C \|V(x)\ax^{1/2}\|_4
\||x|^{-2}\|_{2,\infty}\|v(y)\ax^{1/2}\|_4  \|u\|_2. \\
& \||Q'(\lam)|u\|_2 \leq C \|v(x)\ax^{1/2}\|_{\frac83}
\||x|^{-1}\|_{4,\infty}\|v(y)\ax^{1/2}\|_{\frac83}  \|u\|_2. \\
& \||M_v Q'(\lam)|u\|_2 \leq C \|V(x)\ax^{1/2}\|_{\frac83}
\||x|^{-1}\|_{4,\infty}\|v(y)\ax^{1/2}\|_{\frac83}  \|u\|_2.\lbeq(5-11)
\end{align*}
Thus, $\||Q(\lam)-Q(\mu)|\|_{\Bb(L^2)}
+ \||M_v (Q(\lam)-Q(\m))|\|_{\Bb(L^2)}\leq C |\lam -\m|$ 
and $Q(\lam)$ and $M_v Q(\lam)$ are $\Ag$-valued continuous. 
For $|Q^{''}(\lam)| |u|(x)$ is bounded 
by virtue of \refeq(G-bound+) by a constant times  
\[
\int_{\R^4}\ax^{\frac12}v(x) \ay^{\frac12}v(y) u(y)dy 
+ \int_{|x-y|\leq 1} v(x) |\log |x-y| | v(y) u(y) dy.
\]
It follows by H\"older's inequality that 
\begin{align*}
& \||Q^{''}(\lam)|\|_{\Bb(L^2)} \leq 
C(\|\ax^{\frac12}v\|_2^2 + 
\|V\|_1^\frac12 \|V\|_2^\frac12 \|\log |x|\|_{L^2(|x|<1)}). \\
& \|M_v |Q^{''}(\lam)|\|_{\Bb(L^2)} \|\leq 
C(\|\ax^{\frac12}v\|_2 \|\|\ax^{\frac12}V\|_2  + 
\|V\|_2^\frac12 \|V\|_2^\frac12 \|\log |x|\|_{L^2(|x|<1)}). 
\end{align*}
Then repeating the proof of the continuity of  
$Q(\lam)$ and $M_v Q(\lam)$ presented above, we obtain statement (1). 

\noindent 
(2) Since $\ax^\frac12 v \in L^4 \cap L^{\frac{8}{3}}$, \refeq(5-8) and 
 \reflm(HS) imply $Q(\lam)^2 \in \Hg_2$. By HLS inequality  
\[
\|Q'(\lam)\|_{HS}^2 \leq C \int_{\R^4 \times \R^4}
\frac{\ax V(x) \ay V(y)}{|x-y|^2}dxdy 
\leq C \|\ax V\|_{\frac43}\||x|^{-2}\|_{2,\infty}\,.
\]
We have by using \refeq(G-bound+) and H\"older's inequality that  
\begin{align*}
\|Q^{''}(\lam)\|_{HS} 
& \leq C \|v\ax^{1/2}\|_2^2 
+ C \left(\int_{|x-y|\leq 1}(v(x)|\log |x-y|v(y))^2 dxy \right)^{1/2} \\
& \leq C(\|\ax V\|_1 + \|V\|_2 \|(\log |x|)^2 \|_1 ).
\end{align*}
Statement (2) follows as in (1). 
\edpf

The obvious modification of \refeq(L-repre-W) implies 
\bqn \lbeq(lll)
\W_{b\leq a}u(x)=\int_{0}^{\infty}\left(
\int_{\R^4\times \R^4}\Gg_{-\lam}(x-z)\Ng(\lam,z,y)
(\Pi(\lam)\t_{-y} u)(0)dy dz\right) 
\chi_{b\leq a}(\lam)\lam d\lam  
\eqn 
and \refth(intermediate) follows from \refprop(R-theo) and the following lemma. 

\bglm Suppose $\ax^3 V \in (L^1 \cap L^4)$. Then, 
$\Ng(\lam)$ is $\Lg$-valued $C^2$ function on $(\Lam^{-1},\Lam)$ for $\Lam>1$, 
\edlm 
\bgpf Let $Q(\lam)= M_w G_0(\lam)M_v$ so that 
$\Ng(\lam)= M_v (1+Q(\lam))^{-1}M_w$. We often omit the variable $\lam$ of 
$Q(\lam)$. By virtue of \reflm(Q-estimate), 
$(1+Q)^{-1}= 1-Q + Q^2(1+Q)^{-1}$ is $\Ag$-valued continuous. 
We have $\Ng(\lam)= 
M_V - M_v Q(\lam)(1+Q(\lam))^{-1}M_w$ and $M_V \in \Lg$ is evident. 
We define $Y(\lam)= M_v Q(\lam)(1+Q(\lam))^{-1}M_w$ and 
prove the lemma for $Y(\lam)$.  

\noindent 
(i) Define $g(\lam)= |Q(\lam)(1+Q(\lam))^{-1}||w|$ 
and $h(\lam)= |(1+Q(\lam))^{-1}||w|$ (recall that $|K|$ is the 
integral operator with the kernel $|K(x,y)|$). Since $w\in L^2$, 
\reflm(Q-estimate) implies $g(\lam)$ and $h(\lam)$ are $L^2$-valued 
continuous functions. By virtue of \reflm(important)  
\bqn \lbeq(Y-lam)
\int_{\R^4} |Y(\lam, x,y)|dy \leq v(x)g(\lam,x) \ \mbox{and} \  
\int_{\R^4} |Y(\lam, x,y)|dy \leq (|M_v Q(\lam)| h(\lam))(x) .
\eqn 
Since $v g(\lam)\in L^1$ and  $|M_v Q(\lam)| h(\lam)\in L^2$, 
we obtain that $Y(\lam) \in \Lg$ with bounded $\|Y(\lam)\|_{\Lg}$. 

\noindent 
(ii) $Y'(\lam)= - M_v (1+ Q)^{-1} Q'(\lam)(1+Q)^{-1}M_w$. 
Since $(1+ Q)^{-1} Q'(\lam)(1+Q)^{-1}$ is $\Hg_2$-valued continuous 
by \reflm(Q-estimate), $Y'(\lam)$ is $\Lg_1$-valued continuous. 
\bqn 
Y'(\lam)=-M_v Q'(\lam)(1+Q)^{-1}M_w 
+ M_v Q(1+ Q)^{-1} Q'(\lam)(1+Q)^{-1}M_w\,.
\lbeq(N-1a)
\eqn 
The first on the right of \refeq(N-1a) is  
equal to $Y(\lam)$ with $Q(\lam)$ replaced by $-Q'(\lam)$ 
and the latter can play the role of the former in \refeq(Y-lam) 
and $M_v Q'(\lam)(1+Q)^{-1}M_w$ is $\Ng$-valued continuous. The 
second is equal to $Y(\lam)$ with $M_w$ replaced by 
$Q'(\lam)(1+Q)^{-1}M_w$ and we repreat the argument of (i) 
by setting $g(\lam)$ and $h(\lam)$ as in (i) with 
$|Q'(\lam)(1+Q)^{-1}||w|$ in place of $|w|$ and obtain that 
$M_v Q(1+ Q)^{-1} Q'(\lam)(1+Q)^{-1}M_w$ is also $\Lg_{2,1}$-valued 
bounded. Thus, $\Ng'(\lam)$ is $\Lg$-valued bounded and hence  
$\Ng(\lam)$ is $\Lg$-valued continuous. 

\noindent 
(iii) By further differentiating $Y(\lam)$ and repeating the argument of 
(i) and (ii) by using \reflm(Q-estimate), we obtain the lemma. 
We omit the repetitous details.  
\edpf 

\section{The case $H$ is regular at zero} 
In this section we prove \refth(main-theorem) (2a). 
If $H$ is regular at zero, we have \refth(high) which allows 
end points $p=1,\infty$ and which, however, assumes 
the differentiability condition on $V$. We do not assume it here, however, 
exclude the end points.   
By virtue of \refth(intermediate) we may assume $a>0$ is 
arbitrarily small. We assume 
$\ax^{3+\ep} V \in L^1 \cap L^4$ for a fixed $\ep>0$ throughout this section. 

\bglm \lblm(M0-2) Suppose that $\Mg_0$ is invertible in $\HL$. Then:
\ben 
\item[\textrm{(1)}] $\Mg_0^{-1}\in \Ag$, 
$\Mg_0^{-1}- U =- M_wN_0 M_w + \Hg_2$ and 
$\Mg_0^{-1}- U = M_wN_0 M_v \cdot \Ag$.
\item[\textrm{(2)}] $(M_v\Mg_0^{-1}M_v)(x,y) \in \Lg$. 
\een  
\edlm 
\bgpf We have 
\bqn \lbeq(Mg0)
\Mg_0^{-1} = U -M_wN_0M_w+ (M_wN_0 M_v)^2 \Mg_0^{-1}.
\eqn   
Since $M_wN_0M_w \in \Ag$ by \refeq(D0-est) and $(M_wN_0 M_v)^2 \in \Hg_2$  
by \reflm(HS), we obtain statement (1). Multiplying \refeq(Mg0) by $M_v$ 
from both sides yields 
$M_v \Mg_0^{-1}M_v  = M_V - M_{V}N_0 M_{V}  + 
M_v (M_wN_0 M_v)^2 \Mg_0^{-1} M_v$. It is evident that 
$M_V \in \Lg$; $M_{V}N_0 M_{V} \in \Lg$ by \reflm(M0-1) 
since $V \in L^{\frac43} \cap L^2$. 
Let $X=M_v (M_wN_0 M_v)^2 \Mg_0^{-1} M_v$. Then 
$X\in \Lg_1$ because $(M_wN_0 M_v)^2 \Mg_0^{-1}\in \Hg_2$. 
Express as $X = (M_V N_0 M_v)(M_wN_0 M_v) \Mg_0^{-1} M_v$.
Let $f=|(M_wN_0 M_v) \Mg_0^{-1}| |v|$. Then $f \in L^2(\R^4)$ and 
\refeq(7-4) implies 
$\int_{\R^4} |X(x,y)|dy \leq (|M_V N_0 M_v||f|)(x)$. 
Since $v, V \in L^4$, $M_V N_0 M_v\in {\Bb(L^2)}$ by \reflm(M0-1) (3),  
$\int_{\R^4} |X(x,y)|dy \in L^2(\R^4)$ and $X \in \Lg_{2,1}$. 
Thus, $X\in \Lg$ and statement (2) follows. 
\edpf 

By virtue of \refprop(R-theo) the following lemma,
which is actually more than necessary, proves 
\refth(main-theorem) (2a).  

\bglm \lblm(M-R) Suppose $H$ is regular at zero, then 
$\Ng(\lam) = \Lg+ \Og^{(3)}_{\Lg}(h_2(\lam))$ for $0<\lam<a$ 
when $a>0$ is sufficiently small.  
\edlm 
\bgpf Recall that $\Mg(\lam) = \Mg_0 + \tilde{\Mg}_1(\lam)$. 
\reflms(M0-3,M0-2) imply that 
$\tilde{\Mg}_1(\lam)\Mg_0^{-1}$ is of class $\Og^{(3)}_{\Hg}(h_2(\lam))$. 
Hence, $1+ \tilde{\Mg}_1(\lam)\Mg_0^{-1}$ is invertible for small $\lam>0$ 
and $\Mg(\lam)^{-1}= \Mg^{-1}_0 - \Mg^{-1}_0 \Sg(\lam)$ with   
$S(\lam)= \tilde{\Mg}_1(\lam)\Mg_0^{-1}(1+ \tilde{\Mg}_1(\lam)\Mg_0^{-1})^{-1}$.  
Then  
\[
\Ng(\lam)= M_v\Mg_0^{-1}M_v- X_1(\lam), \quad  
X_1(\lam)= M_v\Mg_0^{-1}S(\lam)M_v. 
\]
We have $M_v\Mg_0^{-1}M_v \in \Lg$ by \reflm(M0-2) (2); since 
$\tilde{\Mg}_1(\lam)\in \Og^{(3)}_{\Hg_2}(h_2(\lam))$ by \refeq(HS-est-1), 
$S(\lam)\in \Og^{(3)}_{\Hg_2}(h_2(\lam))$ and 
$X_1(\lam)\in \Og^{(3)}_{\Lg_1}(h_2(\lam))$. \reflm(M0-1)(2) implies 
\bqn \lbeq(X1-def)
X_1(\lam)  = M_wS(\lam)M_v + M_V N_0 M_v \Ag S(\lam)M_v  \,.  
\eqn 
Since $M_w S(\lam)\in \Og^{(3)}_{\Hg_2}(h_2(\lam))$
by virtue of \refeq(HS-est-2) and $v \in L^2$, 
\reflm(important) (2) implies 
$M_wS(\lam)M_v\in \Og^{(3)}_{\Lg_{2,1}}(h_2(\lam))$. 
Since $\Ag S(\lam)v \in \Og^{(3)}_{L^2}(h_2(\lam))$ and 
$M_VN_0 M_v \in \Bb(L^2)$ by \reflm(M0-1) (3), 
the last term on the right of \refeq(X1-def) is also of class 
$\Og^{(3)}_{\Lg_{2,1}}(h_2(\lam))$. 
Hence $X(\lam) \in \Og^{(3)}_{\Lg}(h_2(\lam))$ and lemma follows.   
\edpf

\section{Generalities for $H$ with singularities at zero $H$}. 
\lbsec(generalities) 
In this section we record some results which are satisfied 
when $H$ has singularities at zero and which we need in what 
follows. We assume 
$\ax^{\d} V \in (L^1 \cap L^4)$ for $\d=3+\ep$, $\ep>0$,  
if otherwise stated explicitly.  
Then, $\Ker \Mg_0 = S_1 \HL\not=\{0\}$ is of finite rank (recall \reflm(null)). 
We 
let ${\textrm{rank}}\, S_1=n$ and $\{\ph_1, \dots, \ph_n\}$ be 
an orthonormal basis  of $S_1 \HL$. Since $\Mg_0$ is a 
real operator we may take $\ph_1, \dots, \ph_n$ as real valued 
functions. 

If $H$ is singular at zero, then $\Mg(\lam)^{-1}$ 
is singular at $\lam=0$ and for studying the singularities 
we use Feshbach formula and a lemma due to Jensen and 
Nenciu (\cite{JN}): 

\bglm[Feshbach formula] \lblm(FS) 
Let $A=
\begin{pmatrix} a_{11}  & a_{12} \\ a_{21}  & a_{22} \end{pmatrix} $ be 
an operator matrix
in the direct sum of Banach spaces $\Yg= \Yg_1 \oplus \Yg_2$. 
Suppose that $a_{11}$, $a_{22}$ are closed,  
$a_{12}$, $a_{21}$ are bounded and 
$a_{22}^{-1}$ exists. Then $A^{-1}$ exists if and only if 
$d= (a_{11}- a_{12}a_{22}^{-1}a_{21})^{-1}$ exists. In this case we have 
\bqn \lbeq(FS-formula)
A^{-1} = \begin{pmatrix}  d & -d a_{12} a_{22}^{-1} \\
-a_{22}^{-1}a_{21} d & a_{22}^{-1}a_{21}d a_{12} a_{22}^{-1} + a_{22}^{-1}
\end{pmatrix}.
\eqn 
\edlm

\begin{lemma}[\cite{JN}] \lblm(JN) 
Let $A$ be a closed operator in a Hilbert space $\Hg$ 
and $S$ a projection. Suppose $A+S$ has bounded inverse. 
Then, $A$ has bounded inverse if and only if 
\bqn \lbeq(JN-B)
B= S - S(A+S)^{-1}S 
\eqn 
has bounded inverse in $S\Hg$ and, in this case, 
$A^{-1}= (A+S)^{-1}+ (A+S)^{-1}SB^{-1}S (A+S)^{-1}$.  
\end{lemma}

The following lemma is of independent interest and we prove it under  
a weaker decay assumption on $V(x)$.  

\bglm \lblm(7-12) Supppose $V \in (\ax^{-\ep}L^1 \cap L^4)$ for an $\ep>0$. 
\ben
\item[{\textrm{(1)}}] Let $\ph \in S_1 \HL$. Then,  
$u(x) = N_0(v\ph)(x)$ satisfies $|u(x)|\leq C \ax^{-2}$,  
$(-\lap + V)u=0$ in the sense of distributions and $\ph(x) = -w(x) u(x)$. 
\item[{\textrm{(2)}}] If  $u$ satisfies 
$|u(x)|\leq C \ax^{-2}$ and $(-\lap + V)u=0$ in the sense of distributions, 
then $\ph(x) = -w(x) u(x)\in S_1\HL$. 
\item[{\textrm{(3)}}] The correspondence $u \leftrightarrow \ph$ is an 
isomorphis between the set of solutions $u$ of $(-\lap + V)u=0$ which satisfy  
$|u(x)|\leq C \ax^{-2}$ and $\ph \in S_1 \HL$. $u \in L^2(\R^4)$ if and only if 
$\la v, \ph \ra=0$. 
\een
\edlm 
\bgpf (1) We have $(U+ vN_0 v)\ph=0$ and $\ph = - w N_0 v\ph$. It follow that 
$\ph= - w u$ and 
$v\ph = -V N_0 (v\ph)$, viz. $(-\lap u)= - Vu$ or $(-\lap + V)u=0$. 
We show $|u(x)|\leq C \ax^{-2}$. 
We remark that $\ax^{-\ep}L^1 \cap L^4\subset \ax^{-\ep/3}L^2$ by 
H\"older's inequality and for any $0<\ep'<\ep$ arbitrary 
close to $\ep$, $V \in \ax^{-\ep'/3}L^p$ for some $p=p(\ep')>2$. 
We express 
\bqn 
u(x) = 
\frac1{4\pi^2} \left(\int_{|x-y|\geq 1} + \int_{|x-y|\leq 1}\right) 
\frac{(v\ph)(y)}{|x-y|^2}dy = \frac1{4\pi^2}(I_1(x) + I_2(x)).
\eqn 
Since $v\ph \in L^1$, $|I_1(x)|\leq \|v\ph\|_1$; 
$\ph \in \HL$ and $v \in L^8$ imply $v\ph \in L^{\frac85}$, hence 
$u \in L^8$ by the HLS inequality and $v\ph = - Vu  \in L^{\frac83}$ 
which implies 
$|I_2(x)|\leq \|v\ph\|_{\frac83} \||x|^{-2}\chi_{\leq a}\|_{\frac85}$ 
by H\"older's inequality. 

This implies $u \in L^\infty(\R^4)$ and we repeat the argument above with 
this new information which implies  
$v\ph = -V u \in (\ax^{-\ep}L^{1}) \cap (\ax^{-\ep'/3}L^p)$ for any 
$\ep'<\ep$ and $p=p(\ep')>2$. Hence, $|I_1(x)|\leq C\ax^{-\min(\ep,2)}$ 
and  for $q=p/(p-1)<2$, by H\"older's inequality, 
$|I_2(x)|\leq \ax^{-\ep'/3} 
\|\ax^{\ep'/3}V\|_{p} \||x|^{-2}\chi_{\leq a}\|_{q}\leq C\ax^{-\ep'/3}$.
Thus, $|u(x)|\leq \ax^{-\min(\frac{\ep'}{3}, 2)}$, which implies 
$v\ph = - Vu\in (\ax^{-\min(\ep+ \frac{\ep'}{3}, 2+\ep)}L^1) 
\cap (\ax^{-\min(\frac{2\ep'}{3}, 2+\frac{\ep'}{3})}L^2)$.  
Then, the repetition of the argument above gives 
$|u(x)|\leq 
C\ax^{-\min\left(
2, \ep+\frac{\ep'}{3}, \frac{2\ep'}{3}
\right)}$. Iterating $k$ times implies 
$|u(x)|\leq C\ax^{-\min
\left(2, \ep+k\frac{\ep'}{3}, 
\frac{(k+1)\ep'}{3}
\right)
}$ and taking $k$ large enough, we obtain $|u(x)|\leq C\ax^{2}$.

\noindent
(2) and (3) Define $\ph = - w u$. Then, $\ph \in \HL$ and $-\lap{u}= -V\ph$ 
implies $u = N (v\ph)$. Hence $U\ph = -v u = - M_v N M_v \ph$ or 
$\Mg_0 \ph =0$. Thus, $u \mapsto \ph = -w u$ and $\ph \mapsto u=N_0(v\ph)$ 
give the desired inverse of each other and they are the isomorphisms. 
Last statement is obvious since $\hat{u}(\xi)= |\xi|^{-2}\widehat{v\ph}(\xi)$  
(see the proof of \reflm(S1P0)). \edpf 

For studying $\Mg(\lam)^{-1}$, we apply \reflm(JN) to the pair 
$(A,S)= (\Mg(\lam), S_1)$.
We first show $\Mg(\lam)+S_1$ is invertible and compute the 
$B$ for this pair. We begin with studying $(\Mg_0 + S_1)^{-1}$. 
Since $\Mg_0$ is selfadjoint and $0$ is isolated eigenvalue of $\Mg_0$ 
(cf. \reflm(null)), $\Mg_0 + S_1$ is invertible in $\HL$. Recall that  
\[
D_0= (\Mg_0 + S_1)^{-1}.
\]
The following is a variant of \reflm(M0-1) 
(and of Lemma 2.7 of \cite{EGG}). 

\bglm \lblm(7-2) Suppose that $V \in (\ax^{-\ep}L^1 \cap L^4)$. Then 
$D_0=M_U - M_w N_0 M_w + \Hg_2$ and $(M_\m D_0 M_\n)(x,y) \in \Lg$\, for 
$\m, \n \in (L^{2}\cap L^8)$. 
\edlm 
\bgpf  
It is shown in the proof of Lemma 2.7 of \cite{EGG} that $D_0$ satisfies 
\bqn \lbeq(7-1)
D_0 = U-wN_0 w-US_1U+(wN_0v-vN_0v S_1)(wN_0v-vN_0v S_1)D_0\,,
\eqn 
where we denoted $M_U, M_v$ and $M_w$ simply by $U$, $v$ and $w$ respectively.
($N_0$ is denoted by $G_0$ in \cite{EGG} and \refeq(7-1) 
is the adjoint of their formula.) Then, the first statement follows 
since $S_1$ is of finite rank and $(M_wN_0 M_v)^2 \in \Hg_2$ by \reflm(HS). 
We multiply \refeq(7-1) by $M_\m$ and $M_\n$. 
We have $\m U \n \in (L^1 \cap L^2)$ and 
$M_{\m U \n}\in \Lg$;  
$M_{\m{w}}N_0M_{{w}\n} \in \Lg$ by \reflm(M0-1); 
\[
(M_{\m U} S_1 M_{U \n})(x,y)= 
\sum_{j=1}^n (U(x)\m(x)\ph_j(x))(U(y)\n(y)\ph_j(y)) \in \Lg
\]
since $\m\ph_j, \n\ph_j \in (L^1\cap L^2)$ for $j=1, \dots, n$
by virtue of \reflm(7-12). Let 
$L_1=M_\m(M_wN_0M_v-M_vN_0M_v S_1)$ and $L_2=(M_wN_0M_v-M_vN_0M_v S_1)D_0$. 
Then, the operator $L$ which arises from the last term on 
the right of \refeq(7-1) is equal to $L= L_1 L_2 M_\n$. 
$L\in M_\m \Hg_2 M_\n$ as remarked above and 
we have $L(x,y) \in \Lg_1$. 
Define $g (x)= \int_{\R^4}|L_2(x,y)||\n(y)|dy$. 
Then $g \in \HL$ and 
$\int_{\R^4}|L(x,y)|dy \leq \int_{\R^4} |L_1(x,y)||g(y)|dy$ 
by virtue of \reflm(important). Since $L_1 \in \Ag$ by \reflm(M0-1) (3), 
$L(x,y)\in \Lg_{2,1}$ and $L \in \Lg$. This completes the proof.  
\edpf

Recall that 
$\Mg(\lam) = \Mg_0 + g_1 (\lam)\lam^2 P+ \lam^2 \Mg_1 + \tilde{\Mg}_2(\lam)$ 
and that $\Mg_1 \in \Hg_2$ and 
$\tilde{\Mg}_2(\lam)\in \Og^{(3)}_{\Hg_2}(h_{2+\ep_1}(\lam))$ 
for any $\ep_1<\d/2$ (cf. \reflm(M0-3)). Let  
\bqn \lbeq(L1tL1-def)
L_1(\lam)= \colon (g_1 (\lam)\lam^2 P+ \lam^2 \Mg_1 + \tilde{\Mg}_2(\lam))D_0.  
\eqn 
Then $L_1(\lam)\in \Og^{(3)}_{\Hg_2}(h_2(\lam))$, 
$\Mg(\lam)+S_1 =(1+ L_1(\lam))(\Mg_0 + S_1)$ and  $1+ L_1(\lam)$ is  
invertible for small $\lam>0$. It follows that so is $\Mg(\lam)+S_1$,  
$(\Mg(\lam)+S_1)^{-1} = D_0 (1+ L_1(\lam))^{-1}$ and  
\bqn \lbeq(NMlam)
(\Mg(\lam)+S_1)^{-1}= D_0 - D_0 L_1(\lam) + D_0 \tilde L_1(\lam), \ 
\ \tilde L_1(\lam)= L_1(\lam)^2 (1+L_1(\lam))^{-1}.
\eqn  

\bglm \lblm(phDMp)
Let $\m \in (L^2 \cap L^8)$ and $\nu\in \HL$. Then, 
$M_{\mu} D_0 \Mg_1 M_{\nu}\in \Lg$ and 
\begin{align}
M_{\mu} D_0 \tilde{\Mg}_2(\lam) M_{\nu} & \in \Og^{(3)}_{\Lg}(h_{2+\ep_1}(\lam)) 
\ \mbox{for an $\ep_1>1$}, \lbeq(phDMp-2) \\ 
M_{\mu} D_0 \tilde{L}_1(\lam) M_{\nu} & \in \Og^{(3)}_{\Lg}(h_{2}(\lam)^2). 
\lbeq(phDMp-3)
\end{align}
These estimates are satisfied if $D_0$ is replaced by 
the identity operator of $\HL$. 
\edlm 
\bgpf 
By virtue of \reflm(M0-3), it is obvious that the lemma holds if $\Lg$ 
is replaced by $\Lg_1$. We prove that the lemma also holds if 
$\Lg$ is replaced by $\Lg_{2,1}$. 

\noindent 
(1) We first prove  $M_{\mu} D_0 \Mg_1 M_{\nu}\in \Lg_{2,1}$. 
For $|z-y|\geq 1$, we have 
$|\log |z-y||\leq C_\d\az^{\c}\ay^{\c}$ for any $0<\c<1/2$. 
Then, 
$\int_{\R^4}|(M_{\mu} D_0 \Mg_1 M_{\nu})(x,y)|dy \leq I_1(x) + I_2(x)$ 
by virtue of \refeq(M1), where 
\begin{align}  
& I_1(x) = C 
\iint_{|z-y|\leq 1}|(M_{\mu} D_0 M_v)(x,z)|
\la \log |z-y|\ra |(v\nu)(y)| dydz, \lbeq(I-1) \\
& I_2(x)= C \iint_{\R^4 \times \R^4}|
(M_{\mu} D_0 M_v)(x,z)|\az^{\c}\ay^{\c}|(v\nu)(y)| dydz .  \lbeq(I-2)
\end{align} 
Let $g(z)=\int_{|z-y|\leq 1} \la\log |z-y|\ra |(v\nu)(y)|dy$. 
Then $\|g\|_\infty 
\leq \|\chi_{\leq a}(|y|)\la \log |y|\ra \|_{4}\|v\|_4 \|\nu\|_2$ 
by H\"older's inequality and $I_1\in L^2$ since  
$(M_{\mu}{D_0}M_v)(x,z)\in \Lg_{2,1}$ by \reflm(7-2). 
\[
I_2(x) = \left(\int_{\R^4} \ay^{\c}|v\nu(y)| dy \right) 
\left(\int_{\R^4} |(M_{\mu}{D_0})(x,z)\az^{\c}v(z)|dz\right) 
\]
and $\|I_2\|\leq \|\ax^{2\c} V\|_1^{1/2}\|\nu\|_2
\|M_{\mu}{D_0}M_{\az^{\c}v}\|_{\Lg_{2,1}}$ which is finite 
for $0<\c<1/2$ by virtue of \reflm(7-2). 
Thus, $M_{\mu} D_0 \Mg_1 M_{\nu}\in \Lg_{2,1}$. 

\noindent 
(2) We prove \refeq(phDMp-2) for $1<\ep_1<3/2$ with $\Lg_{2,1}$ replacing $\Lg$. 
\refeq(M-2a) implies that, for $0\leq j \leq 3$, 
$|\pa_\lam^j ({M_{\mu}}D_0 \tilde{\Mg}_2(\lam)M_{\nu})(x,y)|$ is bounded by  
\[
Ch_{2+\ep_1-j}(\lam) 
\iint_{\R^{4}}|(M_{\mu}D_0(x,z)M_{\az^{3/2} v})(x,z) 
\la \log|z-y| \ra \ay^{3/2} v\nu)(y)| dz.  
\] 
We split the integral on the right into the one over $|y-z|\leq 1$ 
and the other on $|y-z|\geq 1$. Then the former is bounded by 
\refeq(I-1) with 
$\az^{3/2}v(z)$ and $\ay^{3/2}v(y)$ in places of $v(z)$ and $v(y)$ 
respectively and the latter by \refeq(I-2) with 
$\az^{3/2+\c}v(z)$ and $\ay^{3/2+\c}v(y)$ 
in places of $v(z)$ and $v(y)$. Take $\c>0$ such that $3+2\c<\d$. 
Then, the repetition of the argument of (1) implies 
$(M_{\mu} D_0 \tilde{\Mg}_1(\lam)M_{\nu})(x,y)\in 
\Og^{(3)}_{\Lg_{2,1}}(h_{2+\ep_1}(\lam))$.   

\noindent 
(3) Define $L_2(\lam)=D_0 L_1(\lam)(1+L_1(\lam))^{-1}$. We have 
$L_2(\lam)\in \Og^{(3)}_{\Hg_2}(h_2(\lam))$. 
Define $k_{\lam}(y)=|L_2(\lam)||\nu|(y)$. We have $k_\lam\in \HL$ 
and \reflm(important) implies  
\begin{align*} 
& \int_{\R^4}|(M_{\mu} D_0 \tilde{L_1}(\lam) M_{\nu})(x,y)|dy \\
& \leq \int_{\R^4} |(M_{\mu} D_0 (g_1 (\lam)\lam^2 P+ \lam^2 \Mg_1 + 
\tilde{\Mg}_2(\lam))(x,z)| |k_\lam(z)|dz\,. 
\end{align*} 
The right side is bounded by the sum of the quantities 
which are considered above and we obtain 
$\|(M_{\mu} D_0 \tilde{L}_1(\lam)M_{\nu})(x,y)\|_{\Lg_{2,1}}
\leq C h_2(\lam)\lam^2\la \log \lam\ra$. Likewise we obtain 
for the derivatives that 
$\|\pa_\lam^j (M_{\mu} D_0 \tilde{L}_1(\lam)M_{\nu})(x,y)\|_{\Lg_{2,1}}
\leq C h_2(\lam)\lam^{2-j}\la \log \lam\ra$. Combining these together implies 
that 
$(M_{\mu} D_0 \tilde{L_1}(\lam)M_{\nu})(x,y)
\in \Og^{(3)}_{\Lg_{2,1}}(h_{2}(\lam)^2)$. 
The last statement of the lemma is obvious. 
\edpf 

For shortening formulas we denote 
$\Kg(\lam) = (\Mg(\lam)+S_1)^{-1}$. Recall \refeq(NMlam) 
and that $M_{\m}{D_0}M_{\nu} \in \Lg$ for $\m, \nu \in  (L^2 \cap L^8)$ 
(\reflm(7-2)).

\bglm \lblm(7-3) Let $\m, \nu \in  (L^2 \cap L^8)$. Then, 
$M_{\m} \Kg(\lam) M_{\nu} = M_{\m} D_0M_{\nu} + \Og^{(3)}_{\Lg}(h_{2}(\lam))$ 
and $M_v \Kg(\lam) M_v$ is a good producer. 
\edlm 
\bgpf  Let $\Kg_1(\lam)= \Kg(\lam)-D_0$, then \refeq(L1tL1-def) implies that 
\bqn \lbeq(7-3-2)
\Kg_1(\lam)=
-D_0(g_1 (\lam)\lam^2 P + \lam^2 \Mg_1+ \tilde{\Mg}_2(\lam))D_0 + 
\tilde{L}_1(\lam) \in \Og^{(3)}_{\Hg_2}(h_2(\lam))
\eqn 
and $(M_{\m} \Kg_1(\lam)M_{\nu})(x,y) \in \Og^{(3)}_{\Lg_1}(h_2(\lam))$. 
We prove 
$M_{\m}\Kg_1(\lam)M_{\nu}\in \Og^{(3)}_{\Lg_{2,1}}(h_2(\lam))$ 
which finishes the proof. \reflm(phDMp) with $D_0 \nu$ in replace of 
$\nu$ implies that all operators on the right of \refeq(7-3-2) 
except $g_1 (\lam) \lam^2 M_{\m}D_0 P D_0 M_\nu$ are of the class 
$\Og^{(3)}_{\Lg_{2,1}}(h_2(\lam))$ if sandwiched by $M_\m$ and $M_\n$. 
The latter operator is also of class $\Og^{(3)}_{\Lg_{2,1}}(h_2(\lam))$ 
since it can be written as 
$g_1 (\lam) \lam^2 \|V\|_1^{-1} (M_{\m}D_0 v)\otimes (M_{\nu} D_0 v)(y)$
and $M_{\m}D_0 v, M_{\nu} D_0 v\in L^1 \cap L^2$ by virtue of \reflm(7-2).  
This completes the proof. 
\edpf

\section{The case $H$ has singularity of the first kind at zero}

In this section we prove \refth(main-theorem)\, (2a). Thus, we assume 
$\ax^{\d}V \in (L^1\cap L^4)$ 
for some $\d=3+\ep$, $\ep>0$ and $H$ has singularities of the first kind 
at zero.  Without losing generalities we may assume $0<\ep \leq 4$. 
We use the notation of \refdf(GT-1). Then 
$T_1= S_1P{S_1}\vert_{S_1 \HL}$ is of rank one and is 
invertible in $S_1\HL$. Hence, 
${\textrm{rank}}\, T_1 = {\textrm {rank}}\, S_1=1$ and, since 
$\Mg_0$ is a real operator, we may choose a real function $\ph(x)$ 
as the basis vector of $S_1 \HL$ so that $S_1 = \ph \otimes \ph$. 

\bglm \lblm(B-inv) The operator $B(\lam)$ of \refeq(JN-B) for the pair 
$(\Mg(\lam), S_1)$ is invertible for small $\lam>0$ and 
$B(\lam)^{-1} = m(\lam)(\ph \otimes \p)$, where with an $\ep_1>1$, 
$c_1= |\la \ph, v\ra|^2/\|V\|_1>0$ and $c_2= - \la \ph, \Mg_1 \ph\ra$, 
\[
m(\lam)=\lam^{-2}\mu(\lam), \quad 
\m(\lam)= (c_1g_1(\lam)+ c_2+ \Og^{(3)}(h_{\ep_1}(\lam)))^{-1} 
\]
and, $\m(\lam)$ is and Mikhlin multiplier. 
\edlm 
\bgpf We substitute \refeqs(NMlam,L1tL1-def) for $\Kg(\lam)$ in 
$B(\lam)= S_1 - S_1\Kg(\lam)S_1$. Then, the identity 
$D_0 S_1= S_1 D_0 = S_1$ and \reflm(M0-3) yield 
\[
B(\lam) = g_1 (\lam)\lam^2 S_1 P S_1- 
\lam^2 S_1 \Mg_1 S_1+ S_1\Og^{(3)}_{\Hg_2}(h_{2+\ep_1}(\lam))S_1
= \lam^{2}\m(\lam)^{-1}\ph\otimes \ph \,.
\] 
The lemma follows. 
\edpf 

\reflm(JN) and \reflm(B-inv) imply 
that $\Mg(\lam)$ is invertible for small $\lam>0$ and  
\bqn \lbeq(B-2)
\Mg(\lam)^{-1}= \Kg(\lam)+ \Kg(\lam)S_1B(\lam)^{-1}S_1\Kg(\lam), \quad 
\Kg(\lam) = (\Mg(\lam)+S_1)^{-1}.
\eqn

\bglm \lblm(7-4) Let $m(\lam)$ be as in \reflm(B-inv). Then modulo 
a good producer 
\bqn \lbeq(B-f)
M_v  \Mg(\lam)^{-1} M_v  \equiv m(\lam)|v\ph\ra \la v\ph| \,. 
\eqn 
\edlm 
\bgpf By \reflm(7-3) $M_v \Kg(\lam)M_v $ is a good producer.  
\reflm(B-inv) implies 
\bqn \lbeq(B-fa)
M_v \Kg(\lam)S_1B(\lam)^{-1}S_1\Kg(\lam)M_v
=m(\lam)|M_v \Kg(\lam)\ph\ra \la M_v \Kg(\lam)^\ast \ph|\,.
\eqn 
Let $w(\lam,x)= M_v \Kg(\lam)\ph(x)$. 
Since $\Kg(\lam)= (M_U + M_v G_0(\lam)M_v + S_1)^{-1}$ 
and $G_0(\lam)^\ast(x,y)= \overline{G_0(\lam)(x,y)}$, we have 
$\Kg(\lam)^\ast(x,y)= \overline {\Kg(\lam)(x,y)}$ and 
$M_v \Kg(\lam)^\ast \ph(x) = \overline{w(\lam,x)}$ 
(recall $\ph(x)$ is real). By \refeq(NMlam) and $D_0\ph = \ph$ 
\bqn \lbeq(B-3)  
w(\lam)= M_v\ph - g_1(\lam)\lam^2  M_v D_0P \ph 
- \lam^2 M_v D_0 \Mg_1 \ph -M_v D_0(\tilde{\Mg}_2(\lam)- \tilde{L}_1(\lam))\ph.
\eqn 
Observe that the $\lam$-dependence of the first three terms on the right 
is explicit. 
Since $\ph \in \ax^{-2-\d/2} (L^2 \cap L^8)$  
by virtue of \reflm(7-12), we have \\[2pt]
(i) $M_v\ph =v\ph,\, M_v D_0P \ph ,\, 
M_v D_0 \Mg_1 \ph \in (L^1 \cap L^2)$ by \reflm(7-2) and \refeq(phDMp-2); \\[2pt]
(ii) 
$M_v D_0(\tilde{\Mg}_2(\lam)- \tilde{L}_1)\ph\in 
\Og^{(3)}_{L^1\cap L^2} (h_{2+\ep_1}(\lam))$ for an $1<\ep_1<\d/2$ by 
\refeqs(phDMp-2,phDMp-3). \\[2pt]
Thus, substituting \refeq(B-3) in $m(\lam)w(\lam) \otimes \overline{w(\lam)}$, 
we obtain from (i) and (ii) that 
\[
\refeq(B-fa)= m(\lam)|v \ph\ra \la v\ph| + 
\sum \m_{jk}(\lam)|\p_j \ra \la \overline{\p_k}| 
+ \Og^{(3)}_{\Lg}(h_{\ep_1}(\lam))\,.
\]
with Mikhlin multipliers $\m_{jk}(\lam)$ and 
$|\p_j \ra \la \overline{\p_k}|\in \Lg$.
Since $\ep_1>1$, \reflm(Funda) and \refprop(R-theo) imply 
that the last two terms on the right are good producers and 
\refeq(B-f) follows.  
\edpf 

\paragraph{\bf Proof of \refthb(main-theorem) (2a)} 
Since $W_{\pm}$ are isomtries of $L^2$, we may assume $p\not=2$.   
By virtue of \reflm(7-4), it suffices to prove \refth(main-theorem) (2a) 
for 
\bqn  \lbeq(Wleq1)
\W_{\leq{a}}^{(1)} u(x) = 
\int_0^\infty 
(G_0(-\lam)(v\ph\otimes v\ph) \Pi({\lam})u)(x)\m(\lam)\lam^{-1} 
\chi_{\leq {a}}(\lam)d\lam.
\eqn 
By \refeq(WL) we have 
$\W_{\leq{a}}^{(1)}= \W^{(0,0)}_{\leq{a}}(\mu(\lam)v\ph\otimes v\ph)$. 
Since $ v\ph\otimes v\ph\in \Lg$ by virtue of \reflm(7-12) and $\m(\lam)$ is 
Mikhlin multiplier, \reflm(Funda) implies that 
$\W_{\leq{a}}$ is bounded in $L^p$ for $1<p<2$.  

We next prove that $\W_{\leq{a}}^{(1)}$ is unbounded in $L^p$ if $2<p<\infty$. 
Since $\tchi_{\leq{a}}(|D|)$ is bounded in $L^p$ for all $1\leq p \leq \infty$, 
it suffices to show this for $\tchi_{\leq{a}}(|D|)\W_{\leq{a}}^{(1)}$. 
By \refeq(WL) 
\bqn \lbeq(WL-sum)
\tchi_{\leq{a}}(|D|)\W_{\leq{a}}^{(1)}u=
\int_{\R^8} (v\ph)(z)(v\ph)(y) (\t_z K_{a,\leq}^{(0,0)}\t_{-y} \m(|D|)u) dz dy
\eqn 
and \refeq(KT-2) implies that this  
is equal to $R_2u - R_1 u$ where $R_1$ and $R_2$ are 
the operators obtained from \refeq(WL-sum) by replacing $K_{a,\leq}^{(0,0)}$ 
by $(8\pi^2)^{-1}\tilde{T}_1$ and $(8\pi^2)^{-1}Q$ respectively where 
\begin{align}
\tilde{T}_1 u(x) & = \int_{\R^4} (\t_z T_1 \t_{w}u)(x)
\widehat{\tchi_{\leq{a}}}(z)\widehat{\chi_{\leq{a}}}(w)dz dw, \\  
Q u(x)& =\left(
\frac1{|x|^2}\ast \widehat{{\tchi}_{\leq{a}}}\right) 
\otimes \left(\frac1{|y|^2} \ast \widehat{\chi_{\leq {a}}})\right)\m(|D|)u(x).
\end{align} 
Since  that $T_1$ is bounded in $L^p$ for $2<p<\infty$ by \reflm(T),  
Minkowski's inequality implies so is $\tilde{T}_1$ and, hence, $R_1$. 
We show that $R_2$ is unbounded in $L^p$ for $2<p<\infty$. Let  
\[
F(x) =  
\int_{\R^4}\frac{(v\ph \ast \widehat{{\tchi}_{\leq {a}}})(z)}{|x-z|^2}dz, \ \   
G(y) =  \int_{\R^4}\frac{(v\ph \ast 
\widehat{\chi_{\leq{a}}})(z)}{|y-z|^2}dz, \ \ 
\ell(u)= \la u, \m(|D|) G \ra. 
\]
Then, $F \in L^p\setminus\{0\}$ for $2<p<\infty$ and 
$R_2 u(x)= \int_{\R^8} F(x)G(y) \m(|D|)u(y)dy= \ell(u) F(x)$. 
It follows that, if $R_2$ were bounded in $L^p$, 
$\ell$ were bounded functional on $L^p$ and,  
by the Fourier inversion formula and Riesz' theorem, 
it must be that 
\[
\m(|D|)G(x)=\frac1{4\pi^2}
\int_{\R^4}e^{ix\xi}\hat{G}(\xi)\m(|\xi|)d\xi \in L^q, 
\quad 1<q=p/(p-1)<2 
\]
which would imply by Hausdorff-Young's inequality that   
\bqn \lbeq(HY)
\hat{G}(\xi)\mu(|\xi|)= 
\frac{\m(|\xi|)\widehat{v\ph}(|\xi|) \tchi(|\xi|)}
{|\xi|^{2}}\in L^p.
\eqn 
Since $\widehat{v\ph}\in C^{2}$ and 
and $|\m(\lam)|\geq C (|\log \lam|)^{-1}$ with $C>0$  for small $\lam>0$, 
\refeq(HY) could happen only when 
$\widehat{v\ph}(0)=\frac1{4\pi^2}\la v, \ph\ra =0$.
But $S_1 P S_1 \not=0$ and $\la v, \ph\ra\not=0$. 
Hence $R_2$ must be unbounded in $L^p$ for any 
$2<p<\infty$. This completes the proof. 
\qed

\section{The case $H$ has singularities of the second kind}  
\lbsec(second)

In this section we prove \refth(main-theorem) (2b) assuming that  
$\ax^{\d} V \in (L^{1} \cap L^4)$, $\d=4+\ep$ {for an $\ep>0$} 
and that $H$ has singularities of the second kind at zero, 
viz. the projection $S_1$ onto $\Ker \Mg_0$ 
satisfies $S_1 P S_1\vert_{S_1\HL}=0$ which is equivalent to 
$S_1 P = P S_1 =0$ or $\la v, \ph\ra=\widehat{v\ph}(0)=0$ 
for all $\ph \in S_1\HL$. As in \refsec(generalities) we 
let ${\textrm{rank}}\, S_1=n$ and $\{\ph_1, \dots, \ph_n\}$ be 
an orthonormal basis  of $S_1 \HL$.

\bglm \lblm(S1P0) Let $\ph\in S_1\HL\setminus \{0\}$. Then,
$\ph\in \ax^{-2-(\d/2)}(L^2\cap L^8)$. The function $u$ defined 
$u = N_0 v\ph$ is eigenfunction of $H$ with eigenvalue $0$. 
\edlm 
\bgpf \reflm(7-12) implies the first statement. Hence  
$\widehat{v\ph}\in H^{2+\d}(\R^4)$ and $\widehat{v\ph}(0)=0$. 
It follows that $|\xi|^{-2} \widehat{v\ph}(\xi)\in \HL$ and  
$u = N_0 v\ph \in \HL$. \reflm(7-12) implies 
$(-\lap + V)u(x)=0$. This proves the lemma. 
\edpf 

We study $\Mg(\lam)^{-1}$ for $0<\lam<4a$ for an arbitrarily small 
but fixed $0<a<1$. We apply \reflm(JN) for the pair $(A,S)=(\Mg(\lam), S_1)$. 
The following is Lemma 7.4 of \cite{EGG}. 

\bglm \lblm(10-1) If $H$ has singularities 
of the second kind at zero,  then 
$S_1 \Mg_1 S_1\vert_{S_1\HL}$ is non-singular. Let $D_1$ denotes 
$(S_1 \Mg_1S_1\vert_{S_1\HL})^{-1}$. 
\edlm 

In what follows we denote $\Og_{\Hg_2}^{(j)}(\cdot)$  simply 
by $\Og^{(j)}(\cdot)$ for operators $T(\lam)$ in $S_1\HL$. 
We observe that $\ph_j \otimes \ph_k \in \Lg$ for $j,k=1, \dots, n$ 
and $T(\lam) \in \Og^{(j)}(f(\lam))$ if and only if 
\bqn \lbeq(observe)
T(\lam) = \sum_{j,k=1}^n \a_{jk}(\lam) \ph_j \otimes \ph_k, \quad 
a_{jk}(\lam)\in \Og^{(j)}_{\C}(f(\lam))\,.
\eqn

\bglm \lblm(B-inv-2)  The operator 
$B(\lam)$ for $(A,S)=(\Mg(\lam), S_1)$ is invertible and  
\bqn \lbeq(B-inv-2)
B(\lam)^{-1}= \lam^{-2}(D_1 + \tilde{F}(\lam)), \quad 
\tilde{F}(\lam)\in \Og^{(3)}(\lam^2(\log \lam)^2)\,.
\eqn 
\edlm 
\bgpf 
We recall that $\Kg(\lam)=(\Mg+ S_1)^{-1}$, \refeq(L1tL1-def) 
and \refeq(NMlam). 
Then, since $D_0S_1= S_1 D_0 = S_1$ and $S_1 P= PS_1=0$, we have on 
$B(\lam)= S_1 (\lam^2 \Mg_1 + \tilde{\Mg}_2(\lam)- \tilde L_1(\lam))S_1$ or  
\bqn 
B(\lam)= \lam^2 
(S_1+ \lam^{-2}S_1 \tilde{\Mg}_2(\lam)S_1 D_1 - 
\lam^{-2}S_1 \tilde L_1(\lam)S_1 D_1)(S_1\Mg_1 S_1). \lbeq(BS1)
\eqn 
Here $\lam^{-2} S_1\tilde L_1(\lam)S_1 D_1 = \Og^{(3)}(\lam^2\la \log \lam\ra^2)$ 
and $\lam^{-2}S_1 \tilde{\Mg}_2(\lam) S_1 = \Og^{(3)}(h_{2}(\lam))$ 
by virtue of \refeq(M-2a) with $\ep_1=2$ and that 
$v\ph\in \ax^{-2-\d}L^2$ for $\ph \in S_1\HL$, $\d>4$. Hence $B(\lam)$ 
is invertible and $B(\lam)^{-1}$ may be written in the form 
\refeq(B-inv-2). 
\edpf 

\bglm \lblm(Minv-case2)
Modulo a good producer 
$M_v\Mg(\lam)^{-1}M_v \equiv \lam^{-2}M_v S_1 D_1 S_1 M_v$.  
\edlm 
\bgpf \reflm(JN) implies  
$\Mg(\lam)^{-1}= \Kg(\lam)+\Kg(\lam)S_1B^{-1}(\lam)S_1\Kg(\lam)$. 
$M_v\Kg(\lam)M_v$ is a good producer by \reflm(7-3). 
Let 
$\Ga(\lam)= (\lam^2 \Mg_1+ D_0^{-1}\tilde{\Mg}_2(\lam))D_0 -\tilde L_1(\lam)$. 
Then $\Ga(\lam)\in \Og^{(3)}_{\Hg_2}(\lam^2)$ 
and \refeq(NMlam), \reflm(S1P0) and \refeq(B-inv-2) imply  
\begin{align}
& \Kg(\lam)S_1B^{-1}(\lam)S_1\Kg(\lam)= 
\lam^{-2}D_0(1 -\Ga(\lam))S_1 (D_1+\tilde{F}(\lam) )S_1 (1-\Ga(\lam))
\notag \\ 
& = \lam^{-2}S_1 D_1S_1 + \lam^{-2} S_1\tilde{F}(\lam)S_1 
- \lam^{-2}D_0 \Ga(\lam)S_1 (D_1+ \tilde{F}(\lam))S_1 \notag \\
& - \lam^{-2}S_1 (D_1+ \tilde{F}(\lam))S_1 \Ga(\lam)
+ \lam^{-2}D_0 \Ga(\lam)S_1 (D_1+ \tilde{F}(\lam))S_1 \Ga(\lam) \,. \lbeq(32p)
\end{align} 
We show that, if we sandwich the right of \refeq(32p) by $M_v$, 
all terms become good producers except $\lam^{-2}M_vS_1D_1S_1M_v$, 
which proves the lemma. \\[2pt]
(i) Let $E_1(\lam)=\lam^{-2}M_vS_1 \tilde{F}(\lam)S_1 M_v$. 
By \refeq(B-inv-2), 
$E_1(\lam)= \sum_{j,k=1}^n (\log \lam)^2 \ta_{jk}(\lam) L_{jk}$ 
with $\ta_{jk}(\lam) \in \Og_{\C}^{(3)}(1)$ 
and $L_{jk}=(v\ph_j) \otimes (v\ph_k)\in \Lg$. 
Then, $\W_{\leq{a}}(E_1(\lam))=\sum_{j,k=1}^n 
\W^{(2,2)}_{\leq{a}}(\ta_{jk}(\lam)L_{jk})$ and is a good operator 
by virtue of \reflm(Funda) and $E_1(\lam)$ is a good producer 
(recall \refdf(Wtilde) for $\W_{\leq{a}}(\tilde\Ng(\lam))$).

\noindent 
(ii) Let $E_2(\lam){=} \lam^{-2}M_v D_0 \Ga(\lam) S_1 D_1S_1 M_v $ 
and $c_{jk}= \la \ph_j, S_1\ph_k\ra$, $j,k=1, \dots, n$. 
Then, 
\bqn \lbeq(E-1) 
E_2(\lam) = \sum_{j,k=1}^n c_{jk} 
\lam^{-2}M_v D_0 \Ga(\lam)\ph_j \otimes (v\ph_k)\,. 
\eqn 
$\{c_{jk}\}$ is non-singular by \reflm(10-1) and  
$v\ph_k\in (L^1\cap L^2)$. We shall show 
\bqn 
\lam^{-2}M_v D_0 \Ga(\lam)\ph_j \in (L^1\cap L^2) + 
\Og^{(3)}_{L^1 \cap L^2}(\lam^2 \la \log\lam \ra^2)\,, \lbeq(Gamma-est)
\eqn 
which will imply that  
$E_2(\lam)= \Lg + \Og^{(3)}_{\Lg}(\lam^2 \la \log\lam \ra^2)$ 
and  $E_2(\lam)$ is a good producer by virtue of 
\refprop(R-theo) which holds actually with 
$\Og^{(2)}_{\Lg}(\lam^2 \la \log\lam \ra^2)$ in place of 
$\Og^{(2)}_{\Lg}(\lam^2 \la \log\lam \ra^2)$.  
We separately examine the operators on the right of  
\[ 
\lam^{-2}M_v D_0 \Ga(\lam)\ph_j=
M_v D_0 \Mg_1 \ph_j   +  \lam^{-2}M_v D_0^{-1}\tilde{\Mg}_2(\lam)\ph_j-  
\lam^{-2}M_v D_0 \tilde L_1(\lam)\ph_j. 
\] 
and the following (a), (b) and (c) will jointly prove \refeq(Gamma-est).

\noindent 
(a) $(M_v D_0 \Mg_1)\ph_j \in  (L^1\cap L^2)$ by the first of \refeq(phDMp-2). 

\noindent 
(b) \reflm(M0-3) implies \refeq(phDMp-2) 
with $\m=\n=v$ and $\ep_1=2$ and, \refeq(phDMp-2) 
remains to hold if $D_0$ is replaced  
by $D_0^{-1}=\Mg_0 + S_1$ (see \refeq(7-1)). Hence,  
$\lam^{-2}M_v D_0^{-1}\tilde{\Mg}_2(\lam)\ph_j \in 
\Og^{(3)}_{L^1\cap L^2}(h_2(\lam))$.  

\noindent 
(c) $M_v \lam^{-2}\tilde L_1(\lam) \ph_j \in 
\Og^{(3)}_{L^1\cap L^2}(\lam^2 \la \log\lam \ra^2)$
by the remark below \refeq(phDMp-3).

\noindent 
(iii) $E_3(\lam){=}
\lam^{-2}M_v D_0 \Ga(\lam) S_1 \tilde{F}(\lam)S_1 M_v $
is given by \refeq(E-1) with 
$a_{jk}(\lam)$ of \refeq(B-inv-2) in place of $c_{jk}$. Then 
(a), (b) and (c) above imply that 
$E_3(\lam)\in \Og^{(3)}_{\Lg}(\lam^2 \la \log\lam \ra^2)$ and it 
is a good producer by virtue of \refprop(R-theo).

\noindent 
(iv) $E_4(\lam){=} \lam^{-2} 
M_v S_1 (D_1+ \tilde{F}(\lam))S_1\Ga(\lam)M_v
{=} \sum (c_{jk}+ a_{jk}(\lam))(v\ph_j) \otimes 
\lam^{-2}M_v \Ga(\lam)^\ast \ph_k$ is 
the sum of (almost) the adjoints of operators studied in (ii) and (iii). 
Then, the argument similar to the one used for 
proving \refeq(Gamma-est) impies that 
$\lam^{-2}M_v \Ga(\lam)^\ast \ph_k$ which is equal to 
$M_v(D_0 \Mg_1  + \lam^{-2}D_0 \tilde{\Mg}_2(\lam)^\ast D_0^{-1} + 
\lam^{-2}\tilde L_1(\lam)^\ast)\ph_k$ 
satisfies  
\bqn \lbeq(Gamma-ast)
\lam^{-2}M_v \Ga(\lam)^\ast \ph_k\in (L^1 \cap L^2)+ 
\Og^{(3)}_{L^1\cap L^2}(\lam^2 \la \log\lam \ra^2). 
\eqn 
Hence, $E_4(\lam)$ is a good producer as in (iv) 
by virtue of \refprop(R-theo). 

\noindent 
(v) Finally we need show that  
$\lam^{-2}M_v D_0 \Ga(\lam)S_1 (D_1+ \tilde{F}(\lam))S_1 \Ga(\lam)M_v$ 
which is equal to  
\[
\sum (c_{jk}+ a_{jk}(\lam)
|\lam^{-2}vD_0\Ga(\lam) \ph_j\ra \la  M_v \Ga(\lam)^\ast \ph_k|\,
\]
is a good producer. However, this is obvious from estimates 
\refeq(Gamma-est) and \refeq(Gamma-ast) and \refprop(R-theo). 
This completes the proof of the lemma. 
\edpf 

\paragraph{\bf Proof of \refthb(main-theorem) (2b)} 
For shortening formulas we denote $f_j = v\ph_j, \ \ j=1, \dots, n$. 
\reflm(7-12) implies $f_j \in \ax^{-2-\d}(L^1 \cap L^4)$ 
and $PS_1=S_1P=0$ does  
\bqn \lbeq(fj)
\int_{\R^4}f_j(x) dx =0, \ \ j=1, \dots, n. 
\eqn 

\noindent 
(i) We have $M_vS_1 D_1 S_1 M_v = \sum_{j,k=1}^n c_{jk} f_j \otimes f_k $ 
and thanks to \reflm(Minv-case2)   
\begin{align}
\W_{\leq{a}} u(x) & \equiv \int_0^\infty (G_0(-\lam) 
\lam^{-2}M_vS_1 D_1 S_1 M_v \Pi(\lam)u)(x) \chi_{\leq a}(\lam) \lam d\lam 
\notag \\
& = \sum_{j,k=1}^n c_{jk}\int_{\R^4\times \R^4} f_j(z)f_k(w)
(\t_z K_{a}^{(0,0)}\t_{-w} u)(x) dz dw \lbeq(chuto-1)
\end{align} 
modulo a good operator. Then \refeq(WL) implies 
\bqn 
\W_{\leq{a}} u(x) \equiv 
\sum_{j,k=1}^n c_{jk}\W^{(0,0)}_{\leq{a}}(f_j \otimes f_k)u(x)
\lbeq(chuto-2) 
\eqn 
and $\W_{\leq {a}}$ is bounded in $L^p$ for $1<p<2$ by virtue of \reflm(Funda). 

\noindent 
(ii) We next show that the cancellation property \refeq(fj) 
widens the range of $p$ to $1<p<4$ 
for which $\W_{\leq {a}}$ is bounded in $L^p$. 
By the result in (i) we have only to show that 
$\W_{\leq {a}}$ is bounded in $L^p$ for $2<p<4$. 
Recall that 
$\tilde{\W}^{(0,0)}_{a,\leq}$ and $\tilde{\W}^{(0,0)}_{a,\geq}$ 
are defined by \refeq(WL) with $K^{(0,0)}_{a,\leq}$ and 
$K^{(0,0)}_{a,\geq}$ respectively in place $K_{a}^{(0,0)}$.  
Substituting 
$K_{a}^{(0,0)}=K^{(0,0)}_{a,\leq}+ K^{(0,0)}_{a,\geq}$ in \refeq(chuto-1),  
we have 
\bqn
\W^{(0,0)}_{\leq a}(f_j \otimes f_k)= \tilde{\W}^{(0,0)}_{a,\leq}(f_j \otimes f_k)
+ \tilde{\W}^{(0,0)}_{a,\geq}(f_j \otimes f_k)
{=} {\W}_{jk,\leq a}+ {\W}_{jk,\geq a}  \lbeq(chuto-3)
\eqn 
where definitions should be obvious. By virtue of \reflm(Kgeq)  
\begin{align}
\W_{jk,\geq a}u(x) & \equiv 
\frac1{4\pi^2}\int_{\R^4\times \R^4}
f_j(z)f_k(w) \t_z ((\widehat{\mu_{1,a}} 
\otimes \n^{(0,0)}_{a})\t_{-w} u)(x) dz dw \notag \\
& = \frac1{4\pi^2}(f_j\ast \widehat{{\m}_{1,a}})(x)
\la {f}_k \ast \n^{(0,0)}_a, u\ra    \lbeq(chu-3)
\end{align}
modulo a good operator.  Since $f_j \in \ax^{-2-\d}(L^1 \cap L^4)$ 
we have $f_j\ast \widehat{\m_{1,a}} \in L^p$ for $1 \leq p \leq \infty$ 
from \refeq(mu-1-2). Recalling the definition of $\n^{(0,0)}_a$ 
in \reflm(mu-nu) and observing that 
$\Fg(\xi_j |\xi|^{-2})(x)$ is homogenous of order $3$ and 
$\widehat{\chi_{\leq a}}\in \Sg(\R^4)$, we see that  
\[
(\nabla_y\n^{(0,0)}_a)(y)
= \frac{-i}{4\pi^2}\int_{\R^4} e^{-iy\xi}\chi_{\leq a}(\xi)
\xi |\xi|^{-2} d\xi \in L^q, \quad \frac43< q\leq \infty.
\]
Then, since $\int_{\R^4} f_k(x)dx=0$, Taylor's formula and  
Minkowki's inequality imply 
\[
(f_k \ast \n^{(0,0)}_a)(y)= - \int_0^1 
\left(\int_{\R^4}f_k(z)(z\cdot \nabla_y\n^{(0,0)}_a)(y-\th z) dz \right)d\th 
\in L^q, \quad \frac43< q\leq \infty.
\]
It follows that $\W_{jk,\geq{a}}$ is 
bounded in $L^p$ for $1\leq p<4$ for $j,k=1,\dots, n$. 

We next show that $\W_{jk,\leq{a}}$, $j,k=1,\dots, n$, are 
bounded in $L^p$ for $2<p<4$. We denote slightly formally the 
integral kernel of $T_1$ by $(x^2+y^2-i0)^{-1}|y|^{-2}$ 
and define 
\begin{align*} 
& K_{a,\leq,1}^{(0,0)}(x,y)= -\left(\frac{1}{8\pi^3(x^2-y^2+i0)y^2} \right) 
(\widehat{\tchi_{\leq a}}(x)\otimes \widehat{\chi_{\leq a}}(y)), \\
& K_{a,\leq,2}^{(0,0)}(x,y) =  \left(\frac{1}{8\pi^3 x^2 y^2}\right) 
(\widehat{\tchi_{\leq a}}(x)\otimes \widehat{\chi_{\leq a}}(y)), \\
& \W_{jk,\leq {a}}^{(\ell)}u(x)=\int_{\R^4\times \R^4}
f_j(z)f_k(w) (\t_z K^{(0,0)}_{a,\leq, \ell}\t_{-w} u)(x) dz dw, \quad 
\ell=1,2 
\end{align*}
so that 
$\W_{jk,\leq{a}}= \W_{jk,\leq {a}}^{(1)}+ \W_{jk,\leq {a}}^{(2)}$ 
by virtue of the first equation of \refeq(KT-2). 
By virtue of \reflm(T) $\W_{jk,\leq{a} }^{(1)}$ is bounded in $L^p$ for 
$2<p<\infty$. By the cancellation property \refeq(fj)  
we have as previously that, for $4/3<q \leq \infty$, 
\bqn \lbeq(wjka2)
\W_{jk,\leq{a} }^{(2)}(x,y)=\frac{1}{8\pi^3}
\left(
\frac1{x^2}\ast (\widehat{\tchi_{\leq {a}}}\ast f_j)
\right) 
\otimes 
\left(
\frac1{y^2}\ast (\widehat{\chi_{\leq {a}}}\ast f_k)\right) \in 
L^q \otimes L^q\,.
\eqn 
Hence $\W_{jk,\leq{a} }^{(2)}$ is bounded in $L^p$ 
for $4/3< p<4$ and $\W_{jk,\leq{a}}$ is bounded in $L^p$ for $2<p<4$. 

\noindent 
(iii) Next we show that if all $\ph \in S_1\HL$ satisfy the extra 
cancellation property 
\bqn \lbeq(but)
\int_{\R^4} x_j v(x)\ph(x) dx=0, \ 1\leq j \leq 4,\ \ph \in S_1\HL, 
\eqn 
then $\W_{\leq{a}}$ is a good operator. It suffices by virtue of  
the result (i) to prove that $\W_{\leq{a}}$ is bounded in $L^p$ for 
$2<p<\infty$. In view of \refeq(chuto-2) and \refeq(chuto-3), 
we do this for $\W_{jk,\leq {a}}^{(\ell)}$, $\ell=1,2$ and $1\leq j,k \leq n$. 
If \refeq(fj) and \refeq(but) are satisfied, then Taylors formula produces 
\[
(f_k \ast \n^{(0,0)}_{a})(y) = \frac12 \int_0^1 (1-\th) \left(
\int_{\R^4}
f_k (z)\la z, (\nabla_y^2 \n^{(0,0)}_{a})(y-\th z)z\ra  dz 
\right) d\th  
\]
and, since $\xi_j \xi_l |\xi|^{-2}$ is homogeneous of order $0$,  
\[
(\pa_j \pa_l \n^{(0,0)}_{a})(y)
= \frac{-1}{4\pi^2}\int_{\R^4} e^{-iy\xi}\chi_{\leq a}(\xi)
\xi_j \xi_l |\xi|^{-2} d\xi \in L^q, \quad 1< q \leq \infty.  
\]
It follows that $f_k \ast \n^{(0,0)}_{a}\in L^q$, $k=1, \dots, n$, 
for $1<q\leq \infty$ and \refeq(chu-3) implies that 
$\W_{jk,\geq{a}}$ are good operators. 
Under conditions \refeq(fj) and \refeq(but), \refeq(wjka2) 
can be likewise improved  for  all $ L^q$ for $1<q\leq \infty$ 
and $\W_{jk,\leq{a} }^{(2)}$ becomes good operators. 
Since $\W_{jk,\leq{a} }^{(1)}$ is bounded in $L^p$ for $2<p<\infty$ 
as is shown above, $\W_{jk, \leq{a}}$ is also bounded in $L^p$ for $2<p<\infty$. 

\noindent 
(iv) We finally show that $\W_{\leq{a}}$ is unbounded in 
$L^p$  for any $4<p<\infty$ unless \refeq(but) is 
satisfied for all $\ph \in S_1 \HL$. Of course it suffices to show this 
for an $a>0$ small enough. 
Since $\tchi_{\geq a}(|D|)$ is a good operator, \refeq(chuto-2) 
and \refeq(chuto-3) imply  that it suffices to prove this for 
\[
\sum_{j,k=1}^n c_{jk} \W_{jk,\geq {a}}
= \sum_{k=1}^n \p_{k,a} \otimes (f_k \ast \n^{(0,0)}_a), \quad 
\p_{k,a}= \sum_{j=1}^n c_{jk}(f_j\ast \widehat{\m_{1,a}})\,,
\]
where $\p_{k,a}\in L^p$ for $1\leq p \leq \infty$ and $1\leq k \leq n$ 
as is shown above. 

We first show that 
$\p_{1,a}, \dots, \p_{n,a}$ are linearly independent in $L^p(\R^4)$ 
for any $1<p<\infty$ when $a>0$ is sufficiently small. 
Suppose the contrary. Then, since $(c_{jk})$ is non-singular, 
$f_1\ast \widehat{{\m}_{1,a}}, \dots, f_n\ast \widehat{{\m}_{1,a}}$ 
are linearly dependent and hence, via Fourier transform, so are  
$\widehat{f_1}(\xi)\tchi_{\geq a_m}(\xi)|\xi|^{-2}, \dots,
\widehat{f_n}(\xi)\tchi_{\geq a_m}(\xi)|\xi|^{-2}$.
Then, for any decreasing sequence $a_1, a_2 \dots \to 0$,  
there exist null sets $N_1, N_2, \dots$ such that 
for any $m=1,2, \dots$ there exists 
$(\a_{m1}, \dots, \a_{mn} )\in {\mathbb S}^{n-1}$ such that 
$\sum {\a_{mj}}\widehat{f_j}(\xi)\tchi_{\geq a_m}(\xi)|\xi|^{-2} =0$ 
for $\xi \notin N_m$. Set $N= \cup_{m=1}^\infty N_m$. Then $N$ is still a 
null set and 
\bqn \lbeq(Nset)
\sum {\a_{mj}} \widehat{f_j}(\xi)=0, \quad  \xi \not\in N \ \mbox{and} \ 
|\xi|\geq a_m.
\eqn  
Let, for $m=1,2, \dots$, $S_m$ the set of 
$(\a_{m1}, \dots, \a_{mn} )\in {\mathbb S}^{n-1}$ 
for which \refeq(Nset) is satisfied. Then $S_1 \subset S_2\supset \dots$ 
and they are non-empty compact subset.  
Hence $S= \cap_{m=1}^\infty S_m\not=\emptyset$ and, 
for $(\a_{1}, \dots, \a_{n} )\in S$,   
$\sum {\a_{j}} \widehat{f_j}(\xi)=0$ for all $\xi\not\in N$ and 
$f_1, \dots, f_n$ must be linearly dependent. But, this is a contradiction 
since $\ph_1, \dots, \ph_n$ are orthonormal and $\ph_j + w N_0 f_j=0$, 
$j=1, \dots, n$. 
Thus $\p_{1,a}, \dots, \p_{n,a}$ are linearly independent 
for some $a>0$. Then, if $\W_{\leq{a}}$ were bounded in $L^p$ for a 
$4<p<\infty$, the linear functional $\la u, (v\ph_k)\ast \n^{(0,0)}_{a}\ra$ 
must be continuous in $L^p$ for all $k=1, \dots, n$ by the Hahn-Banach theorem 
and hence   
$(v\ph_k)\ast \n^{(0,0)}_{a}\in L^q$, $q=p/(p-1)$ by 
the Riesz theorem. It follows 
$(v\ph)\ast \n^{(0,0)}_{a}\in L^q$ for all $\ph\in S_1\HL$. 
Then, since $1<q<4/3$, it must be by the Hausdorff-Young inequality that 
\[
(2\pi)^{-2}\Fg((v\ph)\ast \n^{(0,0)}_{a})(\xi)= 
 \widehat{v\ph}(\xi)\widehat{\n^{(0,0)}_{a}}(\xi)= 
\widehat{v\ph}(\xi)|\xi|^{-2}\chi_{\leq a}(\xi) \in L^p. 
\]
But this is impossible for any $4<p<\infty$ if $\ph\in S_1\HL$ 
does not satisfy \refeq(but) for a $j$ because 
$\pa_j \widehat{v\ph}(0)\not=0$ and 
$|\widehat{v\ph}(\xi)||\xi|^{-2}\geq C|\xi|^{-1}$ for a constant 
$C>0$ in an open conic subset set $\{|\xi|\leq a \colon \xi_j >\ep |\xi|\}$ 
for a $a>0$ and $\ep>0$. Thus, $\W_{\leq {a}}$ is unbounded in $L^p(\R^4)$ 
for any $4<p\leq \infty$  
\qed

\section{The case $H$ has singularities of the third kind}

In this section we assume 
$\ax^{\d} V(x)\in (L^1 \cap L^4)$, $\d=4+\ep$ for an $\ep>0$ and that 
$H$ has singularities of the third kind:  
$S_1 P S_1\vert_{S_1\HL}$ is singular and $S_1 P S_1\vert_{S_1\HL}\not=0$. 
Let $S_2$ be the orthogonal projection in $S_1 \HL$ onto 
${\Ker} (S_1 P S_1\vert_{S_1\HL})$  and denote $S_2 S_1$ in $\HL$ 
again $S_2$, viz. we consider $S_2$ is an orthogonal projection in $\HL$ 
which vanishes on $(S_2\HL)^{\perp}$. It is shown in  Corollary 7.3 of 
\cite{EGG} that ${\textrm {rank}}\, (S_1 \ominus S_2) =1$.
We take the orthonormal basis $\{\ph_1, \dots, \ph_n\}$ of $S_1\HL$ such that 
$\ph_1$ spans $(S_1 \ominus S_2)\HL$ and $\{\ph_2, \dots, \ph_n\}$ does  
$S_2\HL$. We denote $S_1 \ominus S_2= S_2^\perp$. We have 
\bqn \lbeq(cancellation-3)
\int_{\R^4} v \ph_1 dx \not=0, \quad 
\int_{\R^4} v \ph_2 dx = \cdots = 
\int_{\R^4} v \ph_n dx =0.  
\eqn 
If we define $u_j(x) = N_0 v\ph_j(x)$,  
$u_2, \dots, u_n$ are eigenfunctions of 
$H$ with the eigenvalue $0$. 

We study $\Mg(\lam)^{-1}$ for small $\lam>0$ by using \reflm(JN), however,  
$B(\lam)^{-1}$ via the Feshbach formula. Recall that  
$B(\lam) = S_1 - S_1 \Kg(\lam) S_1$ and that $\Kg(\lam)=(\Mg(\lam)+ S_1)^{-1}$ 
has the expression \refeq(NMlam). Let  
\begin{gather*}
\ta_{11}= S_2^\perp(g_1(\lam)P + \Mg_1)S_2^\perp, \ \  
\ta_{12} = S_2^\perp \Mg_1 S_2, \ \ 
\ta_{21} = S_2 \Mg_1 S_2^\perp, \\  
\ta_{22} = S_2 \Mg_1 S_2, \ \ 
L_2(\lam) = S_1(\tilde{\Mg}_2(\lam)+ \tilde{L}_1)S_1\,. 
\end{gather*}
Then, in the decomposition $S_1\HL= S_2^\perp \HL \oplus S_2 \HL$ we have 
\begin{gather} \lbeq(B-matrix)
B(\lam) = B_1(\lam) +L_2(\lam), \quad 
B_1(\lam)= \lam^2 
\begin{pmatrix} \ta_{11}  & \ta_{12} \\ \ta_{21} & \ta_{22} 
\end{pmatrix}\,.
\end{gather}
Note that $\ta_{11}$ is one dimensional and 
$\ta_{12}, \ta_{21}$ and $\ta_{22}$ are $\lam$-independent.

\bglm We have for small $0<\lam$ that  
\begin{gather}
\ta_{11}= 
(g_1(\lam)|\la \ph_1, v_1\ra|^2  + \la \ph_1, \Mg_1 \ph_1\ra)|\ph_1\ra \la \ph_1|
\not = 0 \,.   \lbeq(ta11) \\
L_2(\lam) = \sum_{j,k=2}^n e_{jk}(\lam) |\ph_j\ra \la \ph_k|, 
\quad e_{jk}(\lam)=\Og^{(3)}(\lam^4\la \log \lam\ra^2)   \lbeq(L2lam4)
\end{gather}
\edlm 
\bgpf \refeq(ta11) is obvious. \reflm(7-12) implies 
$v\ph_j=\ax^{-2-\d}L^2$ for $j=1, \dots, n$ and, 
by virtue of \refeq(M-2a),   
$\la \ph_j, \tilde{\Mg}_2(\lam)\ph_k\ra= \Og^{(3)}(h_{4}(\lam))$. 
Hence  $\la \ph_j |\tL_1| \ph_k\ra = \Og^{(3)}(\lam^4\la \log \lam \ra^2)$.
\edpf  

It is shown in Lemma 7.4 of \cite{EGG} that  
$\ta_{22} = S_2 \Mg_1 S_2$ is invertible in $S_2 \HL$.  
It follows from \refeq(ta11) that 
$\ta_{11}- \ta_{12}\ta_{22}^{-1}\ta_{21} 
= g_1(\lam)(|\la \ph_1, v_1\ra|^2  + cg_1(\lam)^{-1}) |\ph_1\ra \la \ph_1|$ 
with a constant $c$ and $\ta_{11}- \ta_{12}\ta_{22}^{-1}\ta_{21}$ is invertible 
for small $\lam>0$ with the inverse   
\bqn 
d(\lam)= d_1(\lam)|\ph_1\ra \la \ph_1|, \quad 
d_1(\lam)= 
\frac{1|}{g_1(\lam)(|\la \ph_1, v_1\ra|^2  + cg_1(\lam)^{-1})}\,. 
\lbeq(d1-def)
\eqn 
$d_1(\lam)$ is a Mikhlin multiplier. Then, $B_1(\lam)$ is invertible 
by \reflm(FS) and 
\begin{gather} 
B_1(\lam)^{-1}= \lam^{-2}({S_2(S_2 \Mg_1 S_2)^{-1}S_2}+ {d_1(\lam)}Q)\,, 
\lbeq(B1-inv) \\
Q=\begin{pmatrix} 
|\ph_1\ra \la \ph_1| & -|\ph_1\ra \la \ph_1|\ta_{12}\ta_{22}^{-1} \\
-|\ta_{22}^{-1}\ta_{21}\ph_1\ra \la \ph_1 & 
|\ta_{22}^{-1}\ta_{21}\ph_1\ra \la \ph_1| \ta_{12}\ta_{22}^{-1} 
\end{pmatrix}.
\lbeq(B1-inv-a)
\end{gather}
Note that $Q$ is $\lam$-independent, ${\rm rank}\, Q=2$ 
and $Q= (\ph_1 \oplus \tph) \otimes  (\ph_1 \oplus \tph)$ 
with $\tph=- \ta_{22}^{-1}\ta_{21}\ph_1$. 
Then, \refeq(L2lam4) and \refeq(B1-inv) implies  
$B(\lam)= (1 +L_2(\lam)B_1(\lam)^{-1})B_1(\lam)$ is invertible and 
$B(\lam)^{-1}$ is given by 
\bqn
 B_1(\lam)^{-1} + L_3(\lam), \ 
L_3(\lam)= - B_1(\lam)^{-1}L_2(\lam)B_1(\lam)^{-1}
+ \Og^{(3)}(\lam^2\la \log\lam\ra^4). \lbeq(L3def)
\eqn
Hece $\Mg(\lam)^{-1}$ exists by \reflm(JN) 
and $\Mg(\lam)^{-1}= \Kg(\lam) + \Kg(\lam)S_1 B(\lam)^{-1}S_1 \Kg(\lam)$. 

\bglm \lblm(last-reduction)
Modulo a good producer  
$M_v \Mg(\lam)^{-1}M_v \equiv M_vS_1 B_1(\lam)^{-1} S_1 M_v$\,.  
\edlm 
\bgpf We have 
$M_v\Mg(\lam)^{-1}M_v= M_v\Kg(\lam)M_v+ 
M_v\Kg(\lam)S_1 B(\lam)^{-1}S_1 \Kg(\lam)M_v$. By \reflm(7-3) 
again $M_v \Kg(\lam) M_v$ is a good producer. 
Substituting $B(\lam)^{-1}$ by \refeq(L3def), we see that the second term on 
the right is equal to $ E_1(\lam) + E_2(\lam)$ where 
\[ 
E_1(\lam)=M_v\Kg(\lam)S_1 B_1 (\lam)^{-1}S_1 \Kg(\lam)M_v, \quad 
E_2(\lam) = M_v\Kg(\lam)S_1 L_3 (\lam)S_1 \Kg(\lam)M_v\,. 
\] 
\noindent 
(i) We first show that $E_2(\lam)$ is a good producer. 
We obtain by combining \refeq(L2lam4), \refeq(B1-inv) and \refeq(L3def) 
that 
$S_1 L_3(\lam)S_1= 
\sum_{j,k=1}^n (\log\lam)^2 g_{jk}(\lam) |\ph_j\ra \la \ph_k|$ 
with Mikhlin multipliers 
$g_{jk}(\lam)\in \Og^{(3)}_{\C}(1)$,  $j,k=1, \dots, n$.
We then recall \refeq(B-3) which implies that for $j=1, \dots, n$  
\begin{gather} 
M_v \Kg(\lam)\ph_j= \p_{j0} + 
g_1 (\lam)\lam^2 \p_{j1}(x) + \lam^2 \p_{j2}(x) + \p_{j3}(\lam,x), 
\lbeq(vNlamph) \\ 
\p_{j0},\, \p_{j1},\, \p_{j2} \in (L^1\cap L^2), \quad 
\p_{j3} \in \Og^{(3)}_{L^1\cap L^2}(\lam^4 (\log\lam)^2)\,.  
\lbeq(vNlamph-1)
\end{gather}
Since the integral kernel of $\Kg(\lam)^\ast$ is the complex conjugate of 
$\Kg(\lam)$, $M_v\Kg(\lam)^\ast \ph_k$, $k=1, \dots, n$ is expressed 
similarly. Hence  
$E_2(\lam)= \sum (\log{\lam})^2 g_{jk}(\lam)(\Lg + \Og^{(3)}_{\Lg}(h_2(\lam)))$ 
and \reflm(Funda) for $(j,\ell)=(2,2)$ and \refprop(R-theo) imply that 
$E_2(\lam)$ is a good producer. 

\noindent 
ii) Define $B_2(\lam)= S_1 B_1 (\lam)^{-1}S_1$. 
By virtue of \refeqs(B1-inv,B1-inv-a) we have  
\bqn \lbeq(S1-coeff)
B_2(\lam)= \sum_{j,k=1}^n \lam^{-2} f_{jk}(\lam)|\ph_j \ra \la \ph_k|, 
\quad f_{jk}(\lam) \in \Og^{(3)}_{\C}(1).
\eqn 
Substituting $\Kg(\lam)$ by \refeq(NMlam) and using 
$D_0 S_1= S_1 D_0 = S_1$, we express 
$E_1(\lam)= E_{11}(\lam)+  E_{12}(\lam) +  E_{13}(\lam) +  E_{14}(\lam)$ where 
\begin{gather*}
E_{11}= M_v (D_0 - D_0 L_1(\lam)) B_2(\lam)(D_0 - D_0 L_1(\lam))M_v, \ 
E_{12}= M_v \Kg(\lam) B_2(\lam) \tilde L_1(\lam)M_v,  \\
E_{13}= M_v D_0 \tilde L_1(\lam) B_2(\lam) \Kg(\lam) M_v, \quad 
E_{14}= M_v D_0 \tilde L_1(\lam)  B_2(\lam) \tilde L_1(\lam)M_v . 
\end{gather*}

\noindent 
(a) We first show that $E_{12}(\lam)$ is a good producer. 
$E_{12}= \sum \lam^{-2}f_{jk}(\lam)
(M_v \Kg(\lam)\ph_j) \otimes (M_v \tilde L_1(\lam)^\ast \ph_k)$ 
by \refeq(S1-coeff). We can apply \refeqs(vNlamph,vNlamph-1) to 
$M_v \Kg(\lam)\ph_j$ and we have 
$M_v \tilde L_1(\lam)^\ast \ph_k \in \Og^{(3)}_{\Hg\cap \Lg}(\lam^4(\log\lam)^2)$ 
by virtue of \refeqs(L1tL1-def,NMlam). It follows  
$E_{12}(\lam) \in \Og^{(3)}_{\Lg}(\lam^2(\log\lam)^2)$ and 
$E_{12}(\lam)$ is a good producer by virtue of \refprop(R-theo). 
Similar argument implies $E_{13}(\lam)$ and $E_{14}(\lam)$ are both 
good producers. 

\noindent 
(b) We have 
$E_{11}(\lam)=M_v B_2(\lam)M_v+ E_3(\lam)$ where $E_3(\lam)$ is defined by 
\[
E_3(\lam)= - M_vB_2 (\lam) L_1(\lam)M_v - M_vD_0 L_1(\lam)B_2 (\lam)M_v 
+ M_vD_0 L_1(\lam) B_2 (\lam)L_1(\lam)M_v. 
\] 
We prove  $E_3(\lam)$ is a good producer to finish the proof. 
Recalling \refeq(L1tL1-def) and that $v\ph \in \ax^{-2-\d}(L^1\cap L^4)$ 
for $\ph \in S_1 \HL$, we obtain as previously that, for $j=1, \dots, n$, 
\begin{gather*}
\lam^{-2} M_vD_0 L_1(\lam) \ph_j(x)= g_1 (\lam)\tp_{j1}(x) + \tp_{j2}(x) + 
\tp_{j3}(\lam,x),\\ 
\tp_{j1}, \ \tp_{j2} \in (L^1\cap L^2), \  
\tp_3(\lam) \in \Og^{(3)}_{L^1\cap L^2}(h_{2}(\lam)). 
\end{gather*} 
An obvious modification of the argument shows that 
similar expressions are satisfied by 
$\lam^{-2} M_v L_1(\lam)^\ast\ph_k$ for $k=1, \dots, n$. Combining these  
with \refeq(S1-coeff) produces the expression for $E_3(\lam)$: 
\[
E_3(\lam)= \sum_{j,k=1}^n (\log\lam)^2 h_{jk}(\lam)\Lg_{jk} 
+ \Og^{(3)}_{\Lg}(\lam^2 \la \log\lam \ra^2) 
\]
with $\Lg_{jk} \in \Lg$ and $h_{jk}(\lam)\in \Og_{\C}^{(3)}(1)$ 
for $1\leq j,k \leq n$. Thus, 
$E_3(\lam)$ is a good producer by \reflm(Funda) and \refprop(R-theo). 
\edpf 

\paragraph{\bf Proof of \refthb(main-theorem) (2c)}  
In view of \reflm(last-reduction), it suffices to prove that 
the operator $Z$ defined by 
\bqn \lbeq(Z)
Zu(x)= \int_0^\infty G_0(-\lam) 
M_{v}S_1 B_1(\lam)^{-1}S_1 M_v \Pi(\lam)u(x) 
\chi_{\leq a}(\lam) \lam d\lam
\eqn 
is bounded in $L^p$ for $1<p <2$ and unbounded for $2<p<\infty$. 
We substitute \refeq(B1-inv) for $B_1(\lam)^{-1}$, which makes  
$Z= Z_1+Z_2$ where $Z_1$ and $Z_2$ are produced by 
$\lam^{-2}S_2(S_2 \Mg_1S_2)^{-1}S_2$ and 
$\lam^{-2}d_1(\lam)(v(\ph_1+\tph))\otimes (v(\ph_1+\tph))$. 

We may repeat the argument of the previous section \refsec(second) 
to $Z_1$ with obvious modifications, which proves that $Z_1$ 
is bounded in $L^p$ for $1<p<4$ and unbounded 
for $4<p<\infty$ in general and, it becomes a good operator if all 
$\ph \in S_2\HL$ satisfy the extra cancellation property 
$\la v, y_j\ph\ra=0$, $j=1,\dots, 4$. 
 
The operator $Z_2$ is the same as the one defined by \refeq(Wleq1) if 
$\ph$ and $\m(\lam)$ are replaced by $\ph_1+\tph$ and $d_1(\lam)$ respectively. 
Then, the repetition of the argument below \refeq(Wleq1) implies that 
$Z_1$ is bounded in $L^p$ for $1<p<2$ and is unbounded for $2<p<\infty$. 
This completes the proof of \refth(main-theorem). 
\qed











\end{document}